\pdfoutput=1

\documentclass{article}
\usepackage[margin=1in]{geometry}
\usepackage[utf8]{inputenc}
\usepackage[T1]{fontenc}
\usepackage{hyperref}
\usepackage{url}
\usepackage{booktabs}
\usepackage{longtable}
\usepackage{amsmath}
\usepackage{amssymb}
\usepackage{amsfonts}
\usepackage{amsthm}
\usepackage{nicefrac}
\usepackage{microtype}
\usepackage{graphicx}
\usepackage{float}
\usepackage[section]{placeins}
\usepackage{tikz}
\usepackage{algorithm}
\usepackage{algpseudocode}
\usepackage{todonotes}

\theoremstyle{plain}
\newtheorem{theorem}{Theorem}
\newtheorem{lemma}[theorem]{Lemma}
\newtheorem{proposition}[theorem]{Proposition}
\newtheorem{corollary}[theorem]{Corollary}
\theoremstyle{definition}

\theoremstyle{remark}
\newtheorem{remark}[theorem]{Remark}
\newcommand{\dfn}[1]{#1}

\title{Algorithms for Soft Tree Containment on Tree-Child Networks}

\author{Erdem Ugurlu (Bilkent University) \\ Louxin Zhang (National University of Singapore)}
\date{}

\begin{document}
\maketitle
\begin{abstract}
Phylogenetic networks generalize evolutionary trees by allowing lineages to merge. The Tree
Containment problem asks whether a network $N$ displays a tree $T$ on the same taxa. When the
polytomies of $T$ reflect uncertainty rather than simultaneous divergence, the relevant notion is soft
display, in which $N$ may refine the polytomies of $T$. We take it in its asymmetric reading. Soft Tree
Containment is NP-hard on reticulation-visible networks even at reticulation indegree three, and it
was open for a non-binary tree and a tree-child network with multifurcating tree nodes or reticulations
of indegree above two.

We solve this case. A cluster characterization reduces soft display to realizing all clusters of $T$
by one switching, and in a tree-child network the vertices that can realize a cluster lie on a single
chain. The per-cluster requirements then form a constraint network of width one, which arc consistency
decides without search, and a $2$-SAT instance when $N$ is binary. This yields, to our
knowledge, the first polynomial-time algorithm for soft Tree Containment on single-labelled tree-child
networks of arbitrary reticulation indegree, and a linear-time algorithm when $T$ is binary. The same
method decides the symmetric variant of the problem, in which the polytomies of $N$ may be resolved as
well.
\end{abstract}

\section{Introduction}

Evolutionary relationships among species, or more generally among taxa, are traditionally represented by phylogenetic trees. A phylogenetic tree is a rooted tree whose leaves are bijectively labelled by taxa and whose internal branching vertices represent ancestral speciation events. Its branching structure describes the evolutionary descent of taxa from common ancestors. However, phylogenetic trees model only divergent evolution. Reticulate evolutionary events, such as hybridization, recombination, and horizontal gene transfer~\cite{treangen2011,huson2010,chan2013}, involve the merging of distinct evolutionary lineages and cannot be represented by trees alone. Phylogenetic networks~\cite{huson2010} generalize phylogenetic trees to accommodate such events. A rooted phylogenetic network (RPN) is a rooted directed acyclic graph whose leaves represent taxa, tree nodes represent divergence events, and reticulations, with two or more parents, represent the merging of evolutionary lineages.

Given an RPN $N$ and a phylogenetic tree $T$ on the same set of taxa, the Tree Containment Problem asks whether $N$ displays $T$. Informally, this means that a subdivision of $T$ can be obtained from $N$ by selecting one incoming edge at each reticulation and removing redundant vertices and edges. Tree Containment is important for verifying phylogenetic network models~\cite{gunawanLuZhang2016}, since different genes or genomic regions may have distinct tree-like evolutionary histories, and a proposed species network should be consistent with these histories. In general, Tree Containment is NP-hard~\cite{kanj2008}, even when both $N$ and $T$ are binary~\cite{vijwscanwidth,companion}. Consequently, considerable research has focused on developing efficient algorithms for restricted classes of phylogenetic networks.

Van Iersel, Semple, and Steel~\cite{viss2010} established polynomial-time solvability of Tree Containment for binary tree-child networks and binary level-$k$ networks. Subsequent work extended efficient algorithms to the broader class of reticulation-visible networks, including certain non-binary networks~\cite{gunawan2018,weller2018}. Janssen and Murakami~\cite{jm2020,jm2021} developed a different approach based on cherry-picking sequences, which progressively reduce a network by removing leaves and reticulation edges. Their linear-time algorithm addresses the more general Network Containment Problem, of which Tree Containment is a special case, and accommodates reticulations of arbitrary indegree.

An important distinction in Tree Containment concerns the interpretation of multifurcations, or polytomies, in the input tree. Under firm display, the multifurcations of $T$ must be preserved exactly. This interpretation treats a multifurcation as representing a genuine simultaneous divergence event. In practice, however, polytomies often reflect uncertainty about the branching order rather than simultaneous divergence. Such polytomies may be resolved into finer branching structures, motivating the notion of soft display. Under asymmetric soft display, $N$ displays $T$ if it displays a refinement of $T$. Thus, the polytomies of $T$ may be resolved, whereas those of $N$ are not. The two notions coincide when $T$ is binary.

The algorithms of~\cite{viss2010,weller2018}, which allow certain non-binary networks, adopt the firm interpretation of display. Janssen and Murakami~\cite{jm2020,jm2021} define containment through refinements of the query network, but their linear-time algorithm is stated for semi-binary networks, in which every tree node has outdegree two. In this setting, the relevant notions of display coincide. These algorithms therefore do not directly resolve the general Soft Tree Containment Problem for non-binary trees and non-binary tree-child networks.

Bentert et al.~\cite{bmw2018,bmw2021} initiated the algorithmic study of Soft Tree Containment and demonstrated that its complexity depends strongly on reticulation indegree. For reticulation-visible networks on $n$ taxa with maximum reticulation indegree two, Soft Tree Containment can be solved in $O(n^3)$ time~\cite{bmw2018}. In contrast, the problem becomes NP-hard when reticulation indegree is unrestricted. These results establish an important complexity boundary, but leave open the tractability of Soft Tree Containment on tree-child networks with unbounded reticulation indegrees.

For binary input trees, Tree Containment on tree-child networks can be solved efficiently even when reticulations have arbitrary indegree~\cite{jm2020}. The distinction between firm and asymmetric soft display arises only when the input tree is non-binary. For such trees, previously known polynomial-time algorithms for soft display require reticulation indegree at most two. It has therefore remained open whether Soft Tree Containment is polynomial-time solvable on tree-child networks when $T$ is non-binary and $N$ may contain multifurcating tree nodes or reticulations of indegree greater than two. This paper studies Tree Containment for these unsolved cases.

Our approach is based on a characterization of soft display in terms of clusters. It reduces Soft Tree Containment to finding a single choice of incoming edges at the reticulations that simultaneously realizes all clusters of $T$. To exploit the structure of tree-child networks, we use the lineage-taxon encoding, including the lineage-taxon sequences (LTSs) introduced by Zhang et al.~\cite{zhang2023} and their extension to multifurcating trees, the lineage-taxon hypersequences (LTHs)~\cite{phylofusion}. These encodings were previously used for firm display, while PhyloFusion handles soft display heuristically \cite{phylofusion}. By combining the encoding with a canonical ordering of taxa, we identify principal paths and rank intervals that allow the cluster conditions to be evaluated efficiently. For binary input trees, the encoding alone suffices to decide containment. For non-binary trees, the cluster conditions lead to a $2$-SAT formulation in the binary-network case and a tractable constraint-satisfaction formulation in the general case.

Throughout the paper, phylogenetic networks are assumed to be single-labelled. Our main contributions are summarized as follows.

\begin{itemize}

\item Cluster characterization of soft display.  We prove that a network $N$ softly displays a tree $T$ if and only if there exists a switching $f$, which retains exactly one incoming edge at each reticulation, such that
$\mathcal{C}(T)\subseteq \mathcal{C}(T(N,f))$,
where $T(N,f)$ is the tree displayed by the switching and $\mathcal{C}$ denotes the set of clusters (Theorem~\ref{thm:star}). This characterization holds for arbitrary phylogenetic networks and forms the basis of our algorithms.

\item Linear-time algorithm for binary trees.  When $T$ is binary and $N$ is a tree-child network with arbitrary tree-node outdegrees and reticulation indegrees, we develop an $O(|N|+|T|)$-time algorithm (Theorem~\ref{thm:case1}). This improves the $O(|N|d^2)$ bound of Weller's Theorem~2, where $d$ is the maximum reticulation indegree.

\item  2-SAT-based algorithm for binary tree-child networks. When $T$ is non-binary and $N$ is a binary tree-child network with $k$ reticulations, we reduce Soft Tree Containment to a single $2$-SAT instance and obtain an $O(|N|+|T|+nk^2)$-time algorithm (Corollary~\ref{cor:binarycase}). This improves the previously known $O(n^3)$ bound~\cite{bmw2018}, particularly when $k$ is small relative to $n$.

\item Polynomial-time algorithm for general tree-child networks. For a non-binary tree $T$ and a tree-child network $N$ with arbitrary tree-node outdegrees and reticulation indegrees, we develop an $O(|N|+|T|+nd_{\Sigma}^2)$-time algorithm, where $d_{\Sigma}$ is the total reticulation indegree of $N$ (Theorem~\ref{thm:case3}). To our knowledge, this is the first polynomial-time algorithm for Soft Tree Containment on tree-child networks without degree restrictions.

\item Extension to symmetric soft display. We extend our approach to the symmetric notion of soft display studied by Bentert et al.~\cite{bmw2018,bmw2021}, in which multifurcations of both $N$ and $T$ may be resolved. For arbitrary phylogenetic trees and single-labelled tree-child networks without degree restrictions, we obtain an $O(|N|+|T|+n|N|(d_{\Sigma}+1))$-time algorithm (Theorem~\ref{thm:symcase}).

\end{itemize}

All these results concern networks and trees defined on the same set of taxa. In a companion paper~\cite{companion}, we show that allowing $T$ to omit taxa of $N$ makes Tree Containment NP-complete on tree-child networks, even when both inputs are binary. Table~\ref{tab:landscape} compares our results with previous work.

The remainder of the paper is organized as follows. Section~\ref{sec:prelim} introduces the necessary definitions, including tree-child networks, the notions of display, and the lineage-taxon encoding. Section~\ref{sec:char} establishes the cluster characterization of soft display. Sections~\ref{sec:case1} and~\ref{sec:case3} develop the algorithms, for a binary and for a non-binary tree. Section~\ref{sec:sym} extends the approach to symmetric soft display, and Section~\ref{sec:discussion} discusses related approaches and open problems. Proofs omitted from the main text are provided in the appendix, in the order in which the corresponding results appear.

\begin{table}[H]
\centering
\small
\caption{Summary of previous results and those of this paper for Tree Containment. Firm display requires the multifurcations of $T$ to be preserved exactly. Asymmetric soft display allows edge contractions after removing reticulation edges, whereas symmetric soft display additionally permits node splits. Both notions coincide with firm display when $T$ is binary and the tree nodes of $N$ are binary.
Here, $n$ denotes the number of taxa in $N$ and $T$; $k$ is the number of reticulations in $N$; $d_{\Sigma}$ is the total reticulation indegree of $N$; and $|N|$ and $|T|$ denote the sums of the numbers of vertices and edges in $N$ and $T$, respectively. An empty cell indicates that no restriction is imposed on the corresponding parameter.\\
}
\label{tab:landscape}
\begin{tabular}{@{}lllcccl@{}}
\toprule
 & & \multicolumn{3}{c}{$N$} & & \\
\cmidrule(lr){3-5}
 & Display & Class & Tree nodes & In-deg. & $T$ & Time \\
\midrule
\cite{gunawan2018}            & firm & reticulation-visible & binary & $2$     & binary     & linear \\
\cite{weller2018}             & firm & reticulation-visible & binary &      & binary     & linear \\
\cite{jm2020,jm2021}          & soft\,$=$\,firm & tree-child & binary &      & binary     & linear \\
\cite{bmw2018}                & soft & reticulation-visible & binary & $2$     &     & $O(n^{3})$ \\
\cite{bmw2018,bmw2021}        & soft & reticulation-visible & binary & $\ge 3$ &  & NP-hard \\
\cite{bruchholdweller} & soft (sym.) &  &  &  &  & FPT (scanwidth) \\
\midrule
Thm.~\ref{thm:case1} & soft\,$=$\,firm & tree-child &     &      & binary     & $O(|N|{+}|T|)$ \\
Cor.~\ref{cor:binarycase} & soft & tree-child           & binary & $2$     &  & $O(|N|{+}|T|{+}nk^{2})$ \\
Thm.~\ref{thm:case3} & soft & tree-child           &     &      &  & $O(|N|{+}|T|{+}nd_{\Sigma}^{2})$ \\
\bottomrule
\end{tabular}
\end{table}

\section{Preliminaries}\label{sec:prelim}

Let $X$ be a finite set of taxa.
A RPN on $X$ is a rooted directed
acyclic graph (DAG) with no parallel edges, whose root is the
unique vertex of indegree $0$ and has outdegree $1$.
All nodes other than the root belong to one of the following
three classes:
\begin{enumerate}
\item[(i)] {\em leaves}, each of indegree $1$ and outdegree $0$,
    bijectively labelled by the taxa in $X$;
\item[(ii)] {\em tree nodes}, each of indegree $1$ and outdegree
    at least $2$; or
\item[(iii)] {\em reticulations}, each of indegree at least $2$
    and outdegree $1$.
\end{enumerate}
We use $\ell_x$  to denote the leaf that is labeled with $x\in X$. A node is  {\it internal} if it is a tree node or a reticulation.  A node is {\it binary} if its outdegree or indegree is $2$;  it is a {\it polytomy} if it has outdegree at least $3$.

Let $N$ be an RPN on $X$. We denote its vertex and edge sets by $V(N)$ and $E(N)$, respectively, and define $|N|=|V(N)|+|E(N)|$. We also denote the set of reticulations of $N$ by $R(N)$.

For a directed edge $(u,v)\in E(N)$, we call $u$ its {\it tail} and $v$ its {\it head}. An edge is called a {\it tree edge} if its head is a tree node or a leaf, and a {\it reticulation edge} if its head is a reticulation. An edge is internal if its head is not a leaf.

Let $e=(u,v)\in E(N)$ be an edge whose endpoints are both tree nodes. The {\it contraction} of $e$ in $N$ is the operation that identifies $u$ and $v$ into a single vertex: every outgoing edge $(v,x)$ is replaced by $(u,x)$, and the edge $e$ and vertex $v$ are deleted. Note that the contraction of $e$ may lead to parallel edges if both $u$ and $v$ are the parents of a reticulation.

Let $u$ be a polytomy of $N$ with outdegree $d_{\mathrm{out}}(u)\geq 3$. A {\it split} of $u$ is performed by selecting an outgoing edge $(u,v)$ and a subset
\[
E'\subseteq \{(u,x)\in E(N):x\neq v\}
\]
such that
\[
1\leq |E'|\leq d_{\mathrm{out}}(u)-2.
\]
A new tree node $w$ is inserted into the edge $(u,v)$, and every edge $(u,x)\in E'$ is replaced by $(w,x)$. The resulting graph is again an RPN, denoted by ${\rm split}(N, (u, v), E')$.

For $u,v\in V(N)$, we write $u\preceq v$ (equivalently, $v\succeq u$) if $u=v$ or there exists a directed path from $v$ to $u$. Thus, the root is the greatest vertex under $\preceq$. We write $u\prec v$ if $u\preceq v$ and $u\neq v$. In this case, $v$ is called an {\it ancestor} of $u$, and $u$ a {\it descendant} of $v$; $u$ is {\it reachable}  from $v$. In particular, $u$ is a {\it child} of $v$ if $(v,u)\in E(N)$.

For each reticulation $r\in R(N)$, we use $\mathrm{par}(r)$ to denote the set of parents of $r$ and $d_r$ to denote its indegree. We define
\[
d(N):=\max_{r\in R(N)}d_r,
\]
called the {\it maximum reticulation indegree}, with the convention that $d(N)=1$ if $R(N)=\varnothing$. We also define
\[
d_{\Sigma}(N):=\sum_{r\in R(N)}d_r,
\]
called the {\it total reticulation indegree}, where the empty sum is taken to be $0$.

\subsection{Trees, tree-child and reticulation-visible networks}

A {\it rooted phylogenetic tree} is an RPN with no reticulations. An RPN is {\it tree-child}~\cite{crv2009} if every non-leaf vertex has at least one child that is either a leaf or a tree node.

\begin{lemma} \label{lem:committed}
Let $N$ be a tree-child network. For every non-leaf vertex $w$ of $N$, there exists a leaf $\ell$ reachable from $w$ along a path consisting entirely of tree edges such that every directed path from the root to $\ell$ passes through $w$.
\end{lemma}

\begin{proof}
By the tree-child property, every non-leaf vertex has a child of indegree one, connected to it by a tree edge. Starting at $w$, repeatedly follow such an edge. Since $N$ is finite and acyclic, this process terminates at a leaf $\ell$. Every vertex on the resulting path, except possibly $w$, has indegree one, with its unique parent being the preceding vertex on the path. Consequently, every directed path from the root to $\ell$ must pass through $w$.
\end{proof}

A reticulation $r$ is {\it visible} if there exists a leaf $\ell$ such that every directed path from the root to $\ell$ passes through $r$. An RPN is {\it reticulation-visible} if all its reticulations are visible.

Every rooted phylogenetic tree is tree-child, and every tree-child network is reticulation-visible, as follows immediately from Lemma~\ref{lem:committed}.

\subsection{Vertex clusters}

Let $N$ be an RPN on $X$. For each vertex $u\in V(N)$, its \dfn{cluster} $L_N(u)\subseteq X$ is the set of taxa labelling the leaves reachable from $u$. We simply write $L(u)$ when $N$ is clear from the context. In particular, the cluster of the root is $X$. The \dfn{cluster set} of $N$ is defined as
\[
\mathcal{C}(N)=\{L_N(u):u\in V(N)\}.
\]

Any two clusters of a rooted phylogenetic tree are either disjoint or nested. A family of sets satisfying this property is called \dfn{laminar}. A laminar family of nonempty subsets of $X$ is \dfn{maximal} if no additional nonempty subset of $X$ can be included without violating laminarity. A laminar family of nonempty subsets of $X$ is maximal if and only if it is the set of clusters of a binary phylogenetic tree on $X$.

A rooted phylogenetic tree is uniquely determined, up to isomorphism, by its cluster set~\cite[Theorem~3.5.2]{semplesteel2003}. For two rooted phylogenetic trees $T$ and $T'$ on $X$, we say that $T'$ \dfn{refines} $T$ if
\[
\mathcal{C}(T)\subseteq\mathcal{C}(T').
\]
Equivalently, $T$ can be obtained from $T'$ by contracting internal edges.

For a rooted phylogenetic tree $R$ on $X$ and a nonempty subset $Y\subseteq X$, the \dfn{restriction} $R|_Y$ is the rooted phylogenetic tree obtained by taking the minimal subtree connecting the leaves labelled by $Y$ and their least common ancestor, suppressing all vertices of indegree one and outdegree one, and adding a conventional root of outdegree one if necessary. Its cluster set satisfies
\[
\mathcal{C}(R|_Y)
=
\{C\cap Y:C\in\mathcal{C}(R)\}\setminus\{\emptyset\}.
\]

\subsection{Tree display}

Let $N$ be an RPN on $X$. A {\it switching} $f$ of $N$ selects exactly one incoming edge at each reticulation $r\in R(N)$ and deletes all other incoming edges of $r$. Let $f(r)$ denote the parent of $r$ such that $(f(r), r)$ is retained. Writing $\mathrm{par}(r)$ for the set of parents of $r$, a switching can be represented as a function
\[
f\in\prod_{r\in R(N)}\mathrm{par}(r).
\]
Thus, the total number of switchings of $N$ is  $\prod_{r\in R(N)}|\mathrm{par}(r)|$.

Let $f(N)$ denote the directed graph obtained from $N$ by applying a switching $f$. Since every vertex other than the root has exactly one incoming edge in $f(N)$, the graph $f(N)$ is a rooted directed tree. However, it may contain unlabelled leaves and vertices of indegree one and outdegree one. By deleting all vertices and edges that do not lie on a path from the root to a labelled leaf, and subsequently suppressing all vertices of indegree one and outdegree one, we obtain a rooted phylogenetic tree on $X$, denoted by $T(N,f)$.

Let $T$ be a phylogenetic tree on $X$.
$T$ is {\it displayed} by $N$ if there exists a switching $f$ such that $T(N,f)$ is isomorphic to $T$~\cite[Definition~1]{bmw2021}. Display is also referred to as hard display, or firm-display, in the literature (see, for example, \cite{phylofusion}).

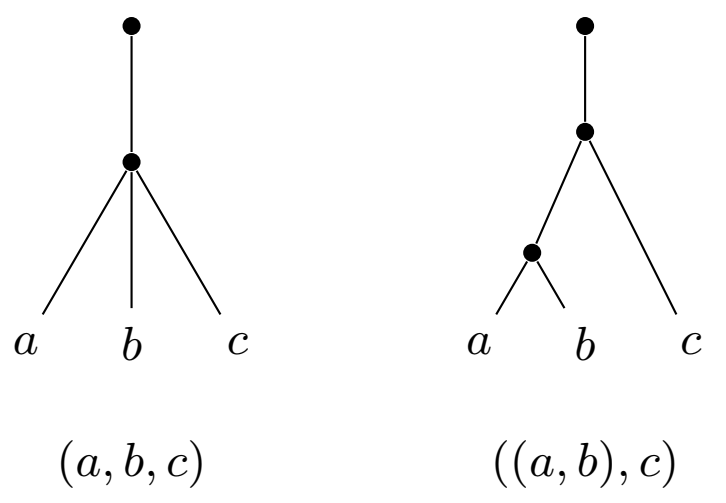
\begin{figure}[t!]
\centering
\begin{tikzpicture}[
  int/.style={circle, fill, inner sep=1.2pt},
  lf/.style={font=\small},
  capt/.style={font=\small}
]
\begin{scope}
  \node[int] (rho1) at (0,2.6) {};
  \node[int] (v1)   at (0,1.7) {};
  \node[lf] (a1) at (-0.7,0.5) {$a$};
  \node[lf] (b1) at (0,0.5) {$b$};
  \node[lf] (c1) at (0.7,0.5) {$c$};
  \draw (rho1)--(v1);
  \draw (v1)--(a1); \draw (v1)--(b1); \draw (v1)--(c1);
  \node[capt] at (0,-0.3) {$(a,b,c)$};
\end{scope}
\begin{scope}[xshift=3cm]
  \node[int] (rho2) at (0,2.6) {};
  \node[int] (r2)   at (0,1.9) {};
  \node[int] (q2)   at (-0.35,1.1) {};
  \node[lf] (a2) at (-0.7,0.5) {$a$};
  \node[lf] (b2) at (0,0.5) {$b$};
  \node[lf] (c2) at (0.7,0.5) {$c$};
  \draw (rho2)--(r2);
  \draw (r2)--(q2); \draw (r2)--(c2);
  \draw (q2)--(a2); \draw (q2)--(b2);
  \node[capt] at (0,-0.3) {$((a,b),c)$};
\end{scope}
\end{tikzpicture}
\caption{Asymmetry of soft display. As an RPN,  the polytomy $(a,b,c)$ 
does not softly display $T=((a,b),c)$. On the other hand, $((a,b),c)$ softly displays $(a,b,c)$. 
\label{fig:asymmetry}
}
\end{figure}

$T$ is {\it softly} displayed by $N$ if there exists a switching $f$ such that $T$ can be obtained from $T(N,f)$ by edge contraction. Equivalently, $N$ softly displays $T$ if it displays a refinement of $T$ 
(Figure~\ref{fig:asymmetry}). The motivation for  this asymmetric version of display is that the polytomies of $T$ may not represent true multifurcation events \cite{phylofusion}.

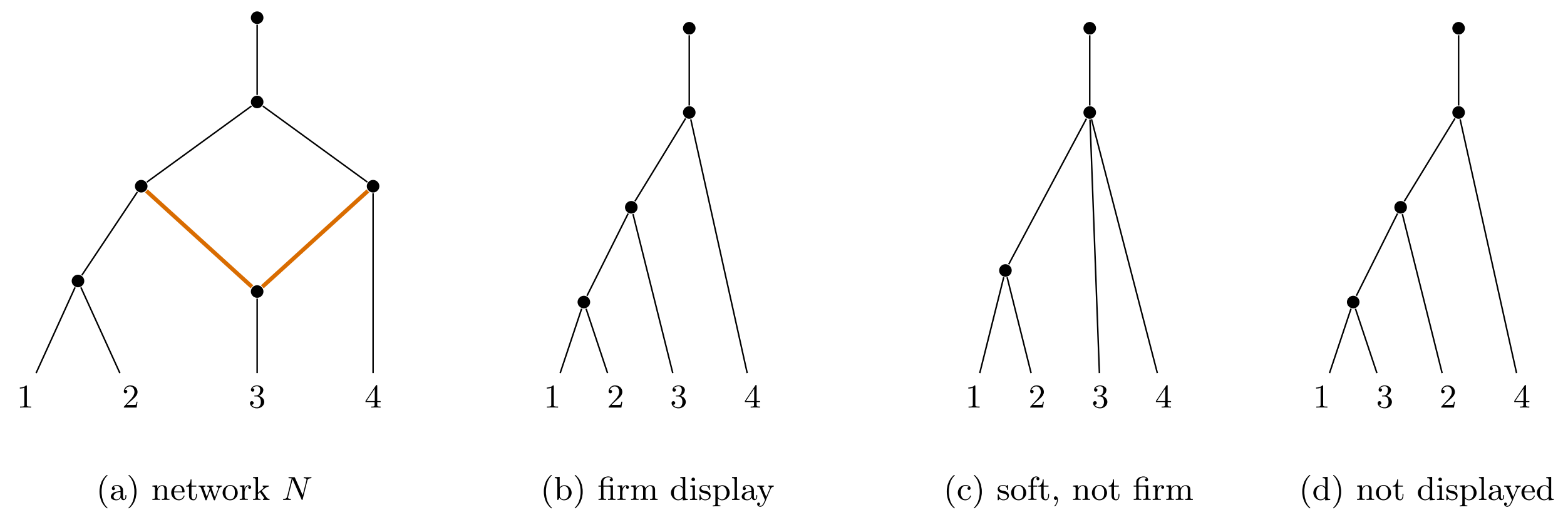
\begin{figure}[b!]
\centering
\resizebox{\textwidth}{!}{%
\begin{tikzpicture}[
  int/.style={circle, fill, inner sep=1.2pt},
  lf/.style={font=\small},
  re/.style={line width=1.1pt, orange!85!black},
  capt/.style={font=\small}
]
\begin{scope}
  \node[int] (rho) at (0,4.1) {};
  \node[int] (s)   at (0,3.3) {};
  \node[int] (g1)  at (-1.1,2.5) {};
  \node[int] (g2)  at (1.1,2.5) {};
  \node[int] (mm)  at (-1.7,1.6) {};
  \node[int] (ret) at (0,1.5) {};
  \node[lf] (La) at (-2.2,0.5) {$1$};
  \node[lf] (Lb) at (-1.2,0.5) {$2$};
  \node[lf] (Lc) at (0,0.5) {$3$};
  \node[lf] (Ld) at (1.1,0.5) {$4$};
  \draw (rho)--(s);
  \draw (s)--(g1); \draw (s)--(g2);
  \draw (g1)--(mm); \draw (g2)--(Ld);
  \draw[re] (g1)--(ret); \draw[re] (g2)--(ret);
  \draw (mm)--(La); \draw (mm)--(Lb); \draw (ret)--(Lc);
  \node[capt] at (-0.5,-0.4) {(a) network $N$};
\end{scope}
\begin{scope}[xshift=4.3cm]
  \node[lf] (ba) at (-1.5,0.5) {$1$};
  \node[lf] (bb) at (-0.9,0.5) {$2$};
  \node[lf] (bc) at (-0.3,0.5) {$3$};
  \node[lf] (bd) at (0.4,0.5) {$4$};
  \node[int] (bq) at (-1.2,1.4) {};
  \node[int] (bp) at (-0.75,2.3) {};
  \node[int] (bt) at (-0.2,3.2) {};
  \node[int] (br) at (-0.2,4.0) {};
  \draw (br)--(bt);
  \draw (bq)--(ba); \draw (bq)--(bb);
  \draw (bp)--(bq); \draw (bp)--(bc);
  \draw (bt)--(bp); \draw (bt)--(bd);
  \node[capt] at (-0.5,-0.4) {(b) firm display};
\end{scope}
\begin{scope}[xshift=8.0cm]
  \node[lf] (ca) at (-1.2,0.5) {$1$};
  \node[lf] (cb) at (-0.6,0.5) {$2$};
  \node[lf] (cc) at (0,0.5) {$3$};
  \node[lf] (cd) at (0.6,0.5) {$4$};
  \node[int] (cq) at (-0.9,1.7) {};
  \node[int] (ct) at (-0.1,3.2) {};
  \node[int] (cr) at (-0.1,4.0) {};
  \draw (cr)--(ct);
  \draw (cq)--(ca); \draw (cq)--(cb);
  \draw (ct)--(cq); \draw (ct)--(cc); \draw (ct)--(cd);
  \node[capt] at (-0.3,-0.4) {(c) soft, not firm};
\end{scope}
\begin{scope}[xshift=11.6cm]
  \node[lf] (da) at (-1.5,0.5) {$1$};
  \node[lf] (dc) at (-0.9,0.5) {$3$};
  \node[lf] (db) at (-0.3,0.5) {$2$};
  \node[lf] (dd) at (0.4,0.5) {$4$};
  \node[int] (dq) at (-1.2,1.4) {};
  \node[int] (dp) at (-0.75,2.3) {};
  \node[int] (dt) at (-0.2,3.2) {};
  \node[int] (dr) at (-0.2,4.0) {};
  \draw (dr)--(dt);
  \draw (dq)--(da); \draw (dq)--(dc);
  \draw (dp)--(dq); \draw (dp)--(db);
  \draw (dt)--(dp); \draw (dt)--(dd);
  \node[capt] at (-0.5,-0.4) {(d) not displayed};
\end{scope}
\end{tikzpicture}%
}
\caption{Tree display in a tree-child network $N$ on $\{1, 2, 3, 4\}$ (panel a). The two possible switchings of $N$ yield
$T(N,f_1)=(((1,2),3),4)$ (panel b) and $T(N,f_2)=((1,2),(3,4))$.
$N$ softly but not firmly displays the multifurcating tree $((1,2),3,4)$ (panel c) through an edge contraction of $T(N, f_1)$ or $T(N, f_2)$.
$N$ does not softly display $(((1,3),2),4)$ (panel d), because  taxon $1$ is always grouped with $2$ in  $T(N, f_1)$ and $T(N, f_2)$ and thus their edge contractions.}
\label{fig:display}
\end{figure}

$T$ is {\it symmetrically softly} displayed by $N$
if there exists a switching $f$ such that $T$ can be obtained from $T(N,f)$ by edge contraction or node splits.
Figure~\ref{fig:display} illustrates three possibilities: display, soft but not firm display, and non-display.

The tree containment problems are formally defined as:
\\

\noindent {\bf (Soft, Symmetric Soft) Tree Containment}\\
\noindent {\bf Instance} A phylogenetic tree $T$   and an  RPN, both  on a taxon set $X$.\\
\noindent {\bf Question} Is $T$ (softly, symmetrically softly) displayed by $N$?\\

Bentert and Weller~\cite[Corollary~3]{bmw2021} proved that symmetric Soft Tree Containment is NP-hard on reticulation-visible networks of maximum reticulation indegree three.

\subsection{Flatten representation of tree-child networks}

In this subsection, we introduce the lineage-taxon encoding of tree-child networks, first developed in \cite{zhang2023},  which forms the basis of our algorithms for the Tree Containment Problem. Given an ordering of the taxa, this encoding assigns to each taxon a sequence that captures the branching structure along a path from its corresponding leaf toward the root of the network. 

\subsubsection{Minimum taxa and  node labels}

Fix an order $\pi=(\pi_1,\dots,\pi_n)$ of $X$, and let $T$ be a
phylogenetic tree on $X$.
For each non-leaf node $v\in V(T)$, let
$m_\pi(v)$
denote the minimum taxon in the cluster $L_T(v)$ with respect to $\pi$. Clearly,
$$m_{\pi}(v)=\min_{c\in \mathrm{Ch}(v)} m_{\pi}(c_i),$$
where $\mathrm{Ch}(v)$ is the set of children of $v$.

Each internal node $v$ with children $c_1,\dots,c_t$
($t\geq 2$) is assigned the label \cite{zhang2023,phylofusion}
\[
\lambda_\pi(v)
=\{m_\pi(c_1),\dots,m_\pi(c_t)\}
\setminus\{m_\pi(v)\}.
\]
Since the children of $v$ have disjoint clusters,
$\lambda_\pi(v)$ consists of exactly $t-1$ taxa.
The root $\rho$ of $T$
is assigned the label $\lambda_\pi(\rho)=\min X$,
which is the minimum taxon of $X$ under $\pi$.

Each taxon $\tau\in X$ appears in the label of exactly
one node $v_\tau$; in particular,
$v_{\pi_1}=\rho$.

Define the {\it principal path} $P(\tau)$ of $\tau$ to be
the directed path from $v_\tau$ to the leaf labeled $\tau$.
The {\it lineage-taxon hypersequence} $\mathrm{LTH}(\tau)$
is the sequence of labels of the internal nodes along
$P(\tau)$, excluding $v_\tau$ and the leaf, read from top to bottom.
The family $\{\mathrm{LTH}(\tau)\}_{\tau\in X}$ is called
the {\it LTH encoding} of $T$. When $T$ and $\pi$ must be named, we write $\mathrm{LTH}_{T,\pi}(\tau)$ for $\mathrm{LTH}(\tau)$. 
If $T$ is binary, the label of each node contains exactly one taxon. 
Throughout, singleton label is written without braces.

\begin{lemma}\label{lem:host}
(\cite{zhang2023,phylofusion})
Let $\pi$ be an order of $X$ and $T$ a phylogenetic tree on $X$. Then every taxon $\tau\neq\pi_1$ occurs exactly once in the LTH encoding of $T$, and $\pi_1$ does not occur in it.
\end{lemma}

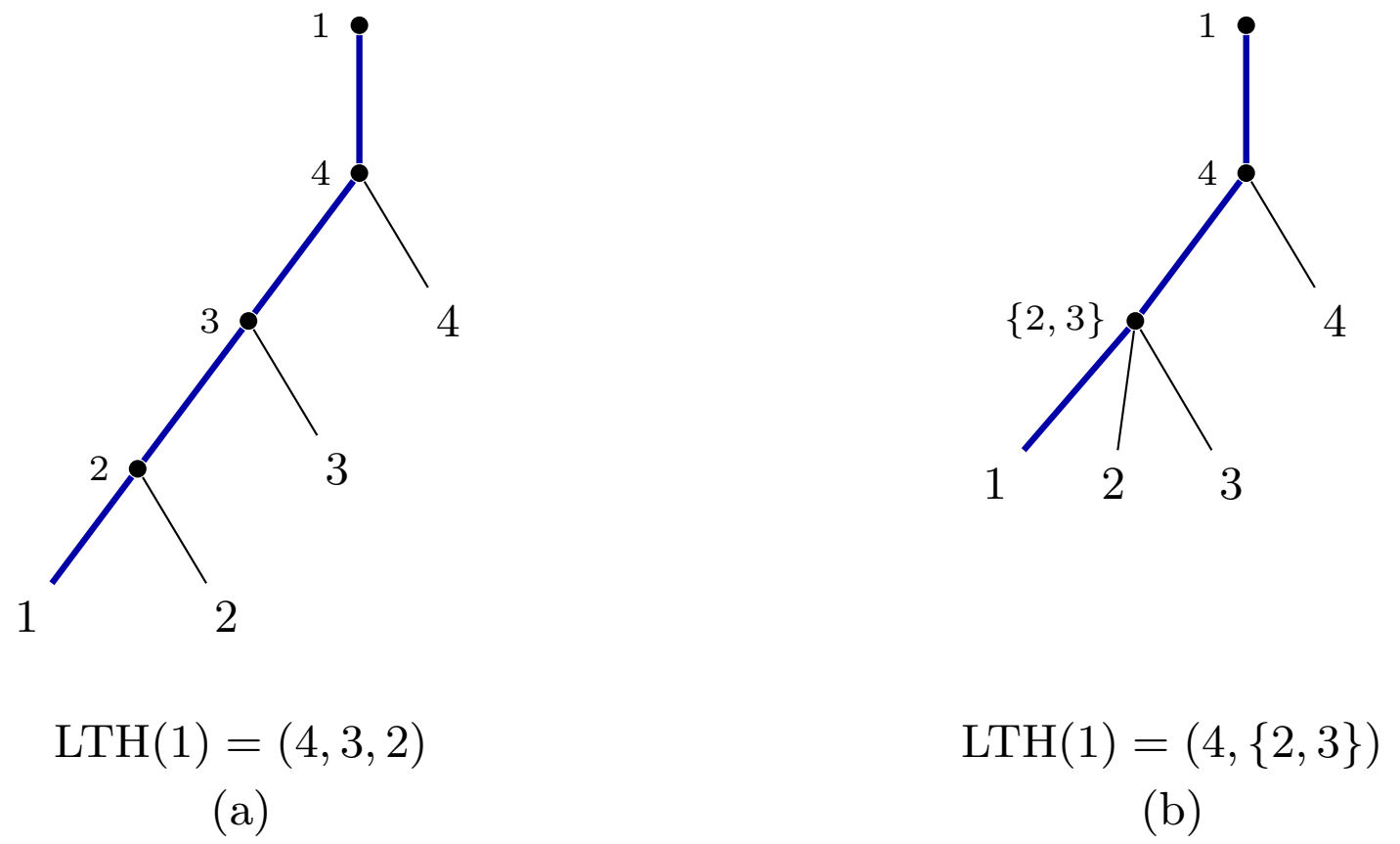
\begin{figure}[H]
\centering
\begin{tikzpicture}[
  int/.style={circle, fill, inner sep=1.2pt},
  leaf/.style={font=\small},
  pp/.style={line width=1.2pt, blue!65!black},
  lab/.style={font=\scriptsize}
]
\begin{scope}
  \node[int, label={[lab]left:$1$}] (rt) at (0,4) {};
  \node[int, label={[lab]left:$4$}] (r) at (0,3) {};
  \node[int, label={[lab]left:$3$}] (a) at (-0.75,2) {};
  \node[int, label={[lab]left:$2$}] (b) at (-1.5,1) {};
  \node[leaf] (l1) at (-2.25,0) {$1$};
  \node[leaf] (l2) at (-0.9,0) {$2$};
  \node[leaf] (l3) at (-0.15,1) {$3$};
  \node[leaf] (l4) at (0.6,2) {$4$};
  \draw[pp] (rt)--(r)--(a)--(b)--(l1);
  \draw (r)--(l4);  \draw (a)--(l3);  \draw (b)--(l2);
  \node[font=\small, align=center] at (-0.8,-1.1) {$\mathrm{LTH}(1)=(4,3,2)$\\ [2pt](a)};
\end{scope}
\begin{scope}[xshift=6cm]
  \node[int, label={[lab]left:$1$}]       (Rt) at (0,4) {};
  \node[int, label={[lab]left:$4$}]       (R) at (0,3) {};
  \node[int, label={[lab]left:$\{2,3\}$}] (A) at (-0.75,2) {};
  \node[leaf] (L1) at (-1.7,0.9) {$1$};
  \node[leaf] (L2) at (-0.9,0.9) {$2$};
  \node[leaf] (L3) at (-0.1,0.9) {$3$};
  \node[leaf] (L4) at (0.6,2) {$4$};
  \draw[pp] (Rt)--(R)--(A)--(L1);
  \draw (R)--(L4);  \draw (A)--(L2);  \draw (A)--(L3);
  \node[font=\small, align=center] at (-0.5,-1.1) {$\mathrm{LTH}(1)=(4,\{2,3\})$\\[2pt](b)};
\end{scope}
\end{tikzpicture}
\caption{An illustration of lineage-taxon encoding for phylogenetic trees.  The order is $\pi=1<2<3<4$.  The label of each tree node is beside the node;
the principal path of taxon~$1$ is highlighted.
(a) The lineage-taxon labeling of binary tree $(((1,2),3),4)$. 
(b) The lineage-taxon labeling of non-binary tree 
$((1, 2, 3), 4)$. Here the empty LTH sequences are omitted.
\label{fig:lts}
}
\end{figure}

Take $X=\{1,2,3,4\}$ with $\pi=1<2<3<4$ and  $T=(((1,2),3),4)$ of Fig.~\ref{fig:lts}(a).
All internal vertices have minimum taxon $1$, the root is labelled $1$, and the three branching vertices
carry, from the bottom up, the labels $2$, $3$, $4$. Reading the branching labels along the principal
path of $1$ (the whole left spine, top to bottom) gives $\mathrm{LTH}(1)=(4,3,2)$. The other taxa branch off
immediately, which leaves their sequences empty. Contracting the edge between the two lowest branching
vertices gives $T'=((1,2,3),4)$ (Fig.~\ref{fig:lts}(b)): the labels $2$ and $3$ merge into the
set-valued label $\{2,3\}$, and thus 
\[\mathrm{LTH}(1)=(4,\{2,3\}),\qquad \mathrm{LTH}(2)=\mathrm{LTH}(3)=\mathrm{LTH}(4)=\epsilon,\]
where $\epsilon$ denotes the empty sequence.

\begin{lemma}\label{lem:faithful}
(\cite{zhang2023,phylofusion})
With  a fixed order $\pi$ of $X$, every rooted phylogenetic tree on $X$ is bijectively represented  by its
encoding $\{\mathrm{LTH}(\tau)\}_{\tau\in X}$.
\end{lemma}

\subsubsection{Encoding tree-child networks}

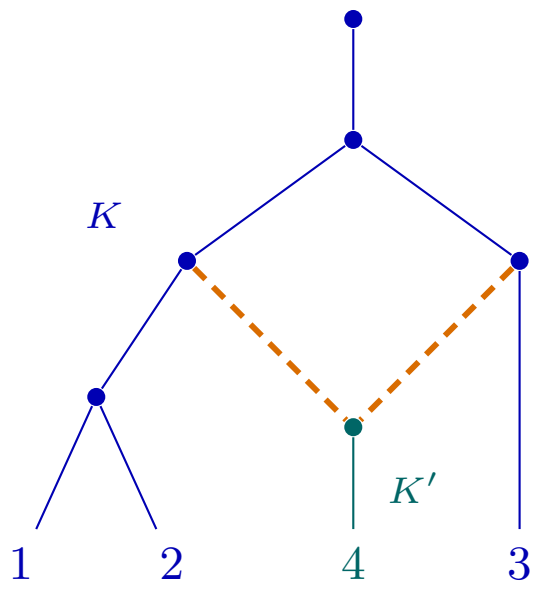
\begin{figure}[!ht]
\centering
\begin{tikzpicture}[
  int/.style={circle, fill, inner sep=1.2pt},
  lf/.style={font=\small},
  c0/.style={blue!70!black},
  c1/.style={teal!80!black},
  re/.style={line width=1.1pt, orange!85!black, densely dashed},
  cc/.style={font=\scriptsize, text=black}
]
  \node[int, c0] (rho) at (0,4.1) {};
  \node[int, c0] (s)  at (0,3.3) {};
  \node[int, c0] (g1) at (-1.1,2.5) {};
  \node[int, c0] (g2) at (1.1,2.5) {};
  \node[int, c0] (mm) at (-1.7,1.6) {};
  \node[lf,  c0] (l1) at (-2.2,0.5) {$1$};
  \node[lf,  c0] (l2) at (-1.2,0.5) {$2$};
  \node[lf,  c0] (l3) at (1.1,0.5) {$3$};
  \node[int, c1] (ret) at (0,1.4) {};
  \node[lf,  c1] (l4)  at (0,0.5) {$4$};
  \draw[c0] (rho)--(s);
  \draw[c0] (s)--(g1); \draw[c0] (s)--(g2);
  \draw[c0] (g1)--(mm); \draw[c0] (g2)--(l3);
  \draw[c0] (mm)--(l1); \draw[c0] (mm)--(l2);
  \draw[c1] (ret)--(l4);
  \draw[re] (g1)--(ret); \draw[re] (g2)--(ret);
  \node[cc, c0] at (-1.65,2.8) {$K$};
  \node[cc, c1] at (0.40, 1.00) {$K'$};
\end{tikzpicture}
\caption{
An illustration of tree components and canonical orders
for a tree-child network $N$ on $\{1,2,3,4\}$.
Removing the reticulation edges (dashed) partitions $N$
into two tree components, $K$ and $K'$.
The component $K$ contains leaves 1, 2, and 3,
while $K'$ contains only leaf 4.
An order of $\{1,2,3,4\}$ is canonical if and only if
4 appears last, such as $\pi=1<3<2<4$.
}
\label{fig:components}

\end{figure}

While any order $\pi$ suffices  for encoding a phylogenetic tree, encoding a tree-child network $N$  requires additional constraints on the taxon order \cite{zhang2023}. Removing all reticulation edges from a network $N$ yields a forest of directed subtrees, which are  rooted at either the root of $N$ or a reticulation.
These rooted subtrees are called {\it tree components} (Figure~\ref{fig:components}).
Clearly, every vertex $v\in V(N)$  belongs to a unique tree component, which we denote by $\mathrm{comp}(v)$.

Consider a topological order of the vertices of $N$, in which every vertex appears after its parents. Let $r_0,r_1,\ldots,r_k$ denote the root and reticulations of $N$, listed in this order, with $r_0$ being the root. An order $\pi$ of $X$ is called {\it canonical} if every taxon in $\mathrm{comp}(r_i)$ precedes every taxon in $\mathrm{comp}(r_j)$ whenever $i<j$ (Figure~\ref{fig:components}). Since a topological order of $N$ can be computed in linear time, a canonical order of $X$ can also be constructed in linear time (Algorithm~\ref{alg:canonical}).

\begin{algorithm}[H]
\caption{Computing a canonical order of a tree-child network.}
\label{alg:canonical}
\begin{algorithmic}
\Procedure{CanonicalOrder}{$N$} \Comment{\texttt{$N$: a tree-child network on $X$}}
  \State compute a topological order of $V(N)$, and let $r_0,r_1,\ldots,r_k$ be the list \\
  \;\;\;\;\;\; of the root  and the reticulations of $N$ in this order;
  \State $\pi\gets$ the empty sequence
  \For{$i=0,1,\ldots,k$}
    \State append to $\pi$ the taxa of the leaves of $\mathrm{comp}(r_i)$, in the order\\
     \;\;\;\;\;\;\;\;\;\;\;in which a depth-first search from $r_i$
    along tree edges reaches them;
  \EndFor
  \State \Return $\pi$
\EndProcedure
\end{algorithmic}
\end{algorithm}

As in the tree case, let $m_\pi(v)$ denote the minimum taxon in the cluster $L(v)$ of a vertex $v$ with respect to a canonical order $\pi$ of $X$ for $N$.

\begin{lemma}\label{lem:canonical}
For a canonical order $\pi$ of $X$ for a tree-child network $N$,
the following statements hold.
\begin{enumerate}
\item[(i)] For every vertex $v\in V(N)$, $m_\pi(v)$ is the label of a leaf belonging to $\mathrm{comp}(v)$.
\item[(ii)] For every reticulation $r$ of $N$ and every parent $p$ of $r$, $m_\pi(p)<_\pi m_\pi(r)$.
\end{enumerate}
\end{lemma}

\begin{proof}
We first note that if a reticulation $r$ lies below a vertex $v$, then $\pi$ lists all taxa of
$\mathrm{comp}(v)$ before all taxa of $\mathrm{comp}(r)$. Indeed, the root of $\mathrm{comp}(v)$ is an ancestor of $r$, so it comes before $r$ in the topological order, and $\pi$ is canonical.

(i) By Lemma~\ref{lem:committed}, $v$ reaches a leaf along tree edges, and this leaf is in
$\mathrm{comp}(v)$. Let $\ell$ be a leaf below $v$ that is not in $\mathrm{comp}(v)$, and let $r$ be the last
reticulation on a path from $v$ to $\ell$. Then $\ell$ is in $\mathrm{comp}(r)$ and $r$ lies below $v$, so the
taxa of $\mathrm{comp}(v)$ precede the taxon of $\ell$. Hence $m_{\pi}(v)$ is a taxon of $\mathrm{comp}(v)$.

(ii) Since $r$ lies below its parent $p$, the taxa of $\mathrm{comp}(p)$ precede those of $\mathrm{comp}(r)$.
By (i), $m_\pi(p)$ is a taxon of $\mathrm{comp}(p)$ and $m_\pi(r)$ is a taxon of $\mathrm{comp}(r)$. Hence
$m_\pi (p)<_\pi m_\pi (r)$.
\end{proof}

Let $N$ be a tree-child network on $X$ and $\pi$ a canonical
order of $X$ for $N$. We encode $N$ in a manner analogous
to the tree case. Each tree node $v\in V(N)$ with children
$c_1,\dots,c_t$ is assigned the label
\[
  \lambda_\pi(v)
  =\{m_\pi(c_1),\dots,m_\pi(c_t)\}\setminus\{m_\pi(v)\},
  \qquad m_\pi(v)=\min_{1\le i\le t}m_\pi(c_i),
\]
where the second equality holds because $L(v)=L(c_1)\cup\dots\cup L(c_t)$.

The root of $N$ is assigned the label $\pi_1$, the minimum
taxon of $X$ under $\pi$. By the definition of a canonical
order, $\pi_1$ belongs to the tree component containing
the root.

For each taxon $\tau\in X$, start at the leaf $\ell_\tau$
and repeatedly move to its parent as long as the current
vertex has indegree one and its parent has minimum taxon
$\tau$. Let $\mathrm{top}(\tau)$ denote the vertex at which this
process terminates. The vertices visited form a directed
path from $\mathrm{top}(\tau)$ to $\ell_\tau$, called the
{\it principal path} of $\tau$ in $N$, denoted by
$\mathrm{path}(\tau)$.
By Lemma~\ref{lem:canonical}, this path lies entirely within
the tree component containing $\ell_\tau$.

Only $\mathrm{top}(\tau)$ can be a reticulation on
$\mathrm{path}(\tau)$. We define the
{\it lineage-taxon hypersequence} (LTH)
$\mathrm{LTH}_{N,\pi}(\tau)$ to be the sequence of labels
of the tree nodes on $\mathrm{path}(\tau)$, read from
$\mathrm{top}(\tau)$ down to $\ell_\tau$ \cite{phylofusion}.
The family
\[
  \{\mathrm{LTH}_{N,\pi}(\tau)\}_{\tau\in X}
\]
is called the {\it LTH encoding} of $N$.

The next lemma locates the occurrences of each taxon in the encoding of $N$. Its proof is presented in Appendix~\ref{app:prelim}.

\begin{lemma} \label{lem:dichotomy}
Let $\pi$ be a canonical order of $X$ for a tree-child network $N$, and let $\tau \in X$. We have the following facts.

(i)  The vertices of $N$ with minimum taxon $\tau$ are exactly the vertices of $\mathrm{path}(\tau)$, and
$\mathrm{top}(\tau)$ is the only vertex of $\mathrm{path}(\tau)$ that can be a reticulation.

 (ii) If $\mathrm{top}(\tau)$ is a tree node or a leaf, then $\tau$ occurs in the encoding of $N$ exactly once, in
the label of the parent of $\mathrm{top}(\tau)$.

 (iii) If $\mathrm{top}(\tau)$ is a reticulation $r$, then $\tau$ occurs in the encoding of $N$ exactly in the labels
of the parents of $r$.

 (iv) If $\mathrm{top}(\tau)$ is the root, which happens exactly when $\tau=\pi_1$, then $\tau$ does not occur in the
encoding of $N$.

(v) In particular, distinct reticulations have distinct minimum taxa, none of which is $\pi_1$.
\end{lemma}

For a switching $f$ of $N$ and a vertex $v\in V(N)$,
let $L_f(v)$ denote the set of taxa reachable from $v$
in $f(N)$.

\begin{lemma}\label{lem:minsurvive}
Let $f$ be a switching of $N$ and $\pi$ a canonical
order of $X$ for $N$. Then, for every vertex
$v\in V(N)$,
\[
    \min_\pi L_f(v)=m_\pi(v).
\]
\end{lemma}

\begin{proof}
Let $t=m_\pi(v)$. Consider a directed path from $v$
to the leaf $\ell_t$ obtained by repeatedly following
a child with minimum taxon $t$.
For every edge $(u,u')$ on this path, we have
$m_\pi(u)=m_\pi(u')=t$.
By Lemma~\ref{lem:canonical}(ii), none of these edges
is a reticulation edge. Hence, the entire path is
preserved under every switching $f$, implying that
$t\in L_f(v)$.
Since $L_f(v)\subseteq L(v)$ and
$t=\min_\pi L(v)$, we conclude that
$\min_\pi L_f(v)=t$.
\end{proof}

\begin{figure}[H]
\centering
\begin{tikzpicture}[
  int/.style={circle, fill, inner sep=1.2pt},
  lf/.style={font=\small},
  re/.style={line width=1.1pt, orange!85!black},
  lab/.style={font=\scriptsize},
  clab/.style={font=\scriptsize, text=orange!75!black}
]
\begin{scope}
  \node[int] (rho) at (0,4.1) {};
  \node[int, label={[lab]left:$3$}]    (s)  at (0,3.3) {};
  \node[int, label={[clab]left:$4_r$}] (g1) at (-1.1,2.5) {};
  \node[int, label={[clab]right:$4_r$}](g2) at (1.1,2.5) {};
  \node[int, label={[lab]left:$2$}]    (mm) at (-1.7,1.6) {};
  \node[int, label={[lab]right:$r$}] (ret) at (0,1.4) {};
  \node[lf] (l1) at (-2.2,0.5) {$1$};
  \node[lf] (l2) at (-1.2,0.5) {$2$};
  \node[lf] (l4) at (0,0.5) {$4$};
  \node[lf] (l3) at (1.1,0.5) {$3$};
  \draw (rho)--(s);
  \draw (s)--(g1); \draw (s)--(g2);
  \draw (g1)--(mm); \draw (g2)--(l3);
  \draw[re] (g1)--(ret); \draw[re] (g2)--(ret);
  \draw (mm)--(l1); \draw (mm)--(l2); \draw (ret)--(l4);
  \node[font=\small] at (-0.3,-0.4) {(a) tree-child network $N$};
\end{scope}
\begin{scope}[xshift=4.8cm]
    \node[anchor=west, align=left, font=\small] at (0,2.6)
      {$\mathrm{LTH}_{N,\pi}(1)=(\,3,\ {\color{orange!75!black}4_r},\ 2\,)$\\[3pt]
       $\mathrm{LTH}_{N,\pi}(3)=(\,{\color{orange!75!black}4_r}\,)$\\[3pt]
       $\mathrm{LTH}_{N,\pi}(2)=(\,)$\\[3pt]
       $\mathrm{LTH}_{N,\pi}(4)=(\,)$}; 
       
    \node[font=\small] at (1.7,-0.4) {(b) encoding of $N$};
  \end{scope}
\end{tikzpicture}
\caption{An illustration of the encoding of a tree-child network. Here, $N$ is the network of
Figure~\ref{fig:components} and $\pi=1<2<3<4$; the label of each tree node is beside the node. Both
parents of the reticulation $r$ carry the element $4=m_\pi(r)$ (orange, written $4_r$). (a) The labels of
$N$. (b) The encoding of $N$. If a switching $f$ keeps the left edge into $r$, deleting $4_r$ from the
right parent of $r$ gives the encoding of $T(N,f)=(((1,2),4),3)$ (Lemma~\ref{lem:rowlocality}).}
\label{fig:nlth}
\end{figure}
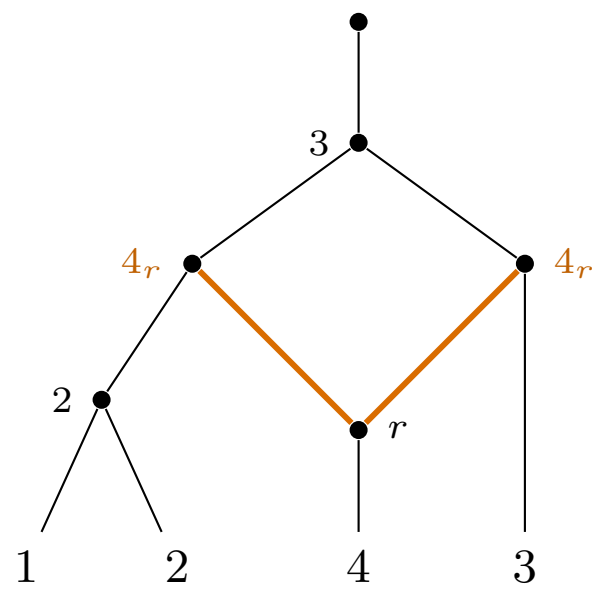

\section{The cluster characterization}\label{sec:char}

All three algorithms are based on one fact: for any RPN $N$ and a phylogenetic tree $T$, $N$ softly displays $T$ if and only if there exists a switching
of $N$ that realizes all clusters of $T$. %
In this section, we prove this cluster characterization of soft display for trees and networks.

For $X'\subseteq X$, a phylogenetic tree $T$ on $X'$ is soft-diplayed by $N$ if $T$ can be obtained from
$T(N,f)|_{X'}$ by contracting internal edges for some switching $f$. For $X'=X$, this is the definition of
Section~\ref{sec:prelim}.

\begin{lemma} \label{lem:expansion}
Let $N$ be an RPN on $X$ and $T$ a phylogenetic tree on $X'\subseteq X$. Then $N$
softly displays $T$ if and only if some switching $f$ makes $T$ a contraction of $T(N,f)|_{X'}$ by
internal edges.
\end{lemma}

\begin{proof}
This lemma follows directly from the definition of soft display.
\end{proof}

\begin{theorem}[Cluster characterization]\label{thm:star}
Let $N$ be an RPN on $X$ and let $T$ be a phylogenetic tree on $X'\subseteq X$. Then
\begin{equation}
  N \text{ softly displays } T \qquad\Longleftrightarrow\qquad
  \mathcal{C}(T)\subseteq\mathcal{C}\bigl(T(N,f)|_{X'}\bigr) \ \text{ for some switching } f ,
  \label{a-st}
\end{equation}
where for $X'=X$ the condition becomes $\mathcal{C}(T)\subseteq\mathcal{C}(T(N,f))$.
\end{theorem}

\begin{proof}
Fix a switching $f$ and let $T'=T(N,f)|_{X'}$. By Lemma~\ref{lem:expansion}, it suffices to show that
$T$ is a contraction of $T'$ by internal edges if and only if $\mathcal{C}(T)\subseteq\mathcal{C}(T')$.
Contracting an edge $(u,v)$ deletes $v$ and does not change the clusters of the other vertices, so if $T$
is a contraction of $T'$, then $\mathcal{C}(T)\subseteq\mathcal{C}(T')$. Conversely, assume
$\mathcal{C}(T)\subseteq\mathcal{C}(T')$, and contract every edge $(u,v)$ of $T'$ with $L(v)\notin\mathcal{C}(T)$.
This is allowed: since $\mathcal{C}(T)$ contains all singletons and $X'$, the vertex $v$ is neither a leaf
nor the child of the root, so $u$ and $v$ are tree nodes. The resulting tree has cluster set $\mathcal{C}(T)$,
so it is isomorphic to $T$, as a phylogenetic tree is determined by its cluster set (Section~\ref{sec:prelim}).
\end{proof}

\section{Display of a binary phylogenetic tree in a tree-child network}\label{sec:case1}

In this section, we let $X$ be a set of taxa.  We assume that $T$ is a binary phylogenetic tree  and $N$ a tree-child network, both on $X$.

\begin{lemma}\label{lem:binarycollapse}
If $T$ is binary, then $N$ softly displays $T$ if and only
if $T(N,f) = T$ for some switching $f$, that is, if and only if $N$ firmly displays $T$.
\end{lemma}

\begin{proof}
If $T(N,f)=T$ for a switching $f$, then $N$ softly displays $T$. Conversely, suppose that $T$ is obtained from $T(N,f)$ by contracting internal edges, for some switching $f$. Contracting an internal edge $(u,v)$ merges two tree nodes, each with at least two children, into a node with at least three children, and the node containing $u$ keeps at least three children under later contractions. Since every tree node of the binary tree $T$ has exactly two children, no edge is contracted, and $T(N,f)=T$.
\end{proof}

By Lemma~\ref{lem:binarycollapse},   we simply investigate whether $N$ displays $T$ if $T$ is binary.

In a tree, distinct children of a node have disjoint clusters, and hence distinct minimum taxa. In a network, two children of a tree node may have common descendants. The next lemma shows that, under a canonical order, they still have distinct minimum taxa.

\begin{lemma}\label{lem:distinctmin}
Let $\pi$ be a canonical order of $X$ for a tree-child network $N$. Then distinct children of a tree node $v$
of $N$ have distinct minimum taxa. Hence $|\lambda_\pi(v)|=\mathrm{outdeg}(v)-1$, and each element of
$\lambda_\pi(v)$ is $m_\pi(c)$ for exactly one child $c$ of $v$.
\end{lemma}

\begin{proof}
Suppose that two distinct children $u'$ and $u''$ of $v$ have the same minimum taxon $t$. By
Lemma~\ref{lem:dichotomy}(i), both lie on $\mathrm{path}(t)$, so one of them, say $u'$, is a strict ancestor of
the other on this path. Then $u''\neq\mathrm{top}(t)$, so $u''$ is not a reticulation, and its only parent is the vertex
preceding it on $\mathrm{path}(t)$. This parent is $v$. Hence $v$ lies on $\mathrm{path}(t)$ below $u'$, and
$v\neq u'$ because $(v,u')$ is an edge. So $N$ has a path from $u'$ to $v$, which together with the edge
$(v,u')$ forms a directed cycle, a contradiction.

Hence the minimum taxa of the children of $v$ are pairwise distinct. Since $\lambda_\pi(v)$ is obtained from them
by removing $m_\pi(v)$, it has $\mathrm{outdeg}(v)-1$ elements, and each of them is the minimum taxon of exactly one
child of $v$.
\end{proof}

Algorithm~\ref{alg:case1} below never constructs a tree $T(N,f)$. It compares encodings instead, which suffices by Lemma~\ref{lem:faithful}: $T(N,f)=T$ if and only if the two trees have the same encoding. The next lemma
shows how the encoding of $T(N,f)$ is obtained from the encoding of $N$. For instance, in Figure~\ref{fig:nlth}, if
$f$ removes the edge from the right parent of $r$, then $4_r$ is deleted from the label of this parent, the label
becomes empty, and $\mathrm{LTH}_{T(N,f),\pi}(3)=\epsilon$, while $\mathrm{LTH}_{T(N,f),\pi}(1)=(3,4,2)$.

Zhang et al.~\cite{zhang2023} introduced this encoding for networks in which tree nodes are binary. They showed that deleting one reticulation edge leaves each sequence a subsequence of the original one. This also holds for tree-child networks with non-binary tree nodes.

\begin{lemma}\label{lem:rowlocality}
Let $\pi$ be a canonical order of $X$ for a tree-child network $N$, and let $f$ be a switching of $N$. For every
taxon $\tau$, $\mathrm{LTH}_{T(N,f),\pi}(\tau)$ is obtained from $\mathrm{LTH}_{N,\pi}(\tau)$ as follows: from the
label of each tree node $v$ on $\mathrm{path}(\tau)$, delete the minimum taxa of the reticulation children of $v$
whose incoming edge from $v$ is removed by $f$, and then remove the labels that have become empty.
\end{lemma}

\begin{proof}
Remove from $N$ the reticulation edges that $f$ does not keep, and call the resulting graph $G$. In $G$, every vertex
other than the root has exactly one incoming edge, and every vertex still reaches a leaf along tree edges
(Lemma~\ref{lem:committed}). So $G$ is a tree whose leaves are the leaves of $N$, and $T(N,f)$ is obtained from $G$
by suppressing the vertices with one parent and one child. By Lemma~\ref{lem:minsurvive}, every vertex has the same
minimum taxon in $G$ as in $N$, and suppressing vertices does not change the cluster of any remaining vertex.

Now let $v$ be a tree node on $\mathrm{path}(\tau)$. In $G$, $v$ keeps its tree children, including the child
attaining $m_\pi(v)$ (Lemma~\ref{lem:canonical}(ii)), and loses the reticulation children whose incoming edge from
$v$ is removed. By Lemma~\ref{lem:distinctmin}, each element of $\lambda_\pi(v)$ is the minimum taxon of exactly one
child of $v$, so the label of $v$ loses exactly the minimum taxa of the lost children. If the label becomes empty,
then $v$ has a single child in $G$ and is suppressed. Otherwise $v$ remains a tree node of $T(N,f)$, and its label
there is the label of $v$ after the deletion.

Finally, every reticulation has one parent and one child in $G$, and so it is suppressed. Hence the tree nodes of
$T(N,f)$ with minimum taxon $\tau$ are the tree nodes on $\mathrm{path}(\tau)$ whose labels do not become empty, in
the same order from top to bottom (Lemma~\ref{lem:dichotomy}(i)). By Lemma~\ref{lem:dichotomy}(i) applied to the
tree $T(N,f)$, their labels form $\mathrm{LTH}_{T(N,f),\pi}(\tau)$.
\end{proof}

Let $r$ be a reticulation of $N$ and $v$ a parent of $r$. By Lemma~\ref{lem:canonical}(ii),
$m_\pi(r)\neq m_\pi(v)$, so $m_\pi(r)\in\lambda_\pi(v)$. Taxon $m_\pi(r)$ is said to be
{\it controlled} by the reticulation edge $(v,r)$. An element of a label that is not controlled by any
reticulation edge is called {\it forced}. By Lemma~\ref{lem:distinctmin}, each element of $\lambda_\pi(v)$ is
$m_\pi(c)$ for exactly one child $c$ of $v$, and it is controlled if and only if $c$ is a reticulation. By
Lemma~\ref{lem:dichotomy}(ii)--(iii), a taxon $\tau\neq\pi_1$ occurs only in the labels of the parents of
$\mathrm{top}(\tau)$, as the minimum taxon of the child $\mathrm{top}(\tau)$. Hence its occurrences are all forced if
$\mathrm{top}(\tau)$ is a tree node or a leaf, and all controlled if $\mathrm{top}(\tau)$ is a reticulation; no taxon occurs
both as a forced and as a controlled element. In Figure~\ref{fig:nlth}, the two occurrences of $4_r$ are
controlled, and all other elements are forced.

Given a canonical order $\pi$  of $X$ for a tree-child network $N$, 
Algorithm~\ref{alg:encode} computes the encoding of $N$ with respect to $\pi$. It computes the minimum taxa
bottom-up, by $m_\pi(v)=\min_\pi\{m_\pi(c): c\ \text{a child of}\ v\}$ (Section~\ref{sec:prelim}), and it
tags an element of $\lambda_\pi(v)$ as controlled if the child contributing it is a reticulation, and as
forced otherwise. 
It runs in a linear time, since the third loop visits each vertex $v$ only for $\tau=m_\pi(v)$ (Lemma~\ref{lem:dichotomy}(i)).

\begin{algorithm}[H]
\caption{Encoding a tree-child network with respect to a canonical order.}
\label{alg:encode}
\begin{algorithmic}
\Procedure{Encode}{$N,\pi$} \Comment{{\tt { $\pi$: a canonical order for the input tree-child network $N$}}}
  \For{each vertex $v$ of $N$, in reverse topological order}
    \State $m_\pi(v)\gets$ (the taxon of $v$ if $v$ is a leaf, and $\min_{c\in {\rm Ch}(v)}m_\pi(c)$ otherwise)
  \EndFor
  \For{each tree node $v$ of $N$}
    \State $\lambda_\pi(v)\gets\varnothing$
    \For{each child $c$ of $v$ with $m_\pi(c)\neq m_\pi(v)$}
      \State $\lambda_\pi(v)\gets \lambda_\pi(v) \cup \{m_\pi(c)\}$
      \If{$c$ is a reticulation}
         tag $m_\pi(c)$ as controlled by the edge $(v,c)$
      \Else
          \;\; tag it as forced
      \EndIf
    \EndFor
  \EndFor
  \For{each taxon $\tau\in X$}
    \State $\mathrm{path}(\tau)\gets(\ell_\tau)$
    \While{the first vertex of $\mathrm{path}(\tau)$ has indegree one and its parent $p$ satisfies $m_\pi(p)=\tau$}
      \State put $p$ in front of $\mathrm{path}(\tau)$
    \EndWhile
    \State $\mathrm{LTH}_{N,\pi}(\tau)\gets$ the labels $\lambda_\pi(x)$ of the tree nodes $x$ on $\mathrm{path}(\tau)$, from top to bottom
  \EndFor
  \State \Return $\{\mathrm{LTH}_{N,\pi}(\tau)\}_{\tau\in X}$, with the elements of all labels tagged
\EndProcedure
\end{algorithmic}
\end{algorithm}

We now describe Algorithm~\ref{alg:case1}. 
Recall that a switching keeps exactly one incoming
edge of each reticulation, and we identify a switching with the set of reticulation edges it keeps. For a taxon
$\tau$, we call $\mathrm{LTH}_{N,\pi}(\tau)$ and $\mathrm{LTH}_{T,\pi}(\tau)$ the {\it rows} of $\tau$ in $N$ and in
$T$. Let $v_1,\dots,v_p$ be the tree nodes on $\mathrm{path}_N(\tau)$, from top to bottom. Then,  
 $$\mathrm{LTH}_{N,\pi}(\tau)=(\lambda_\pi(v_1),\dots,\lambda_\pi(v_p)).$$
 Since $T$ is binary, each label of $T$ is a single taxon, and we write 
 $$\mathrm{LTH}_{T,\pi}(\tau)=(w_1,\dots,w_s).$$

By Lemmas~\ref{lem:faithful} and~\ref{lem:rowlocality}, a switching $f$ satisfies $T(N,f)=T$ if and only if the
following holds for every taxon $\tau$: when each label $\lambda_\pi(v_j)$ loses the elements controlled by the
edges that $f$ does not keep, the labels that do not become empty are $\{w_1\},\dots,\{w_s\}$, in this order. Forced
elements are never lost.

For each taxon $\tau$, the algorithm reads the row of $\tau$ in $N$ from top to bottom and keeps the index $q$ of the
first entry of $(w_1,\dots,w_s)$ not yet matched. When it decides that the label of $v_j$ becomes $\{w_q\}$, we say
that it matches $w_q$ at $v_j$, and it increases $q$ by one. At a tree node $v_j$:
\begin{itemize}
  \item if $\lambda_\pi(v_j)$ has two or more forced elements, the algorithm answers `No', since these elements are
    never lost and each label of $T$ has only one element;
  \item if $\lambda_\pi(v_j)$ has exactly one forced element, the algorithm answers `No' unless this element is
    $w_q$; then it matches $w_q$ at $v_j$ and keeps no edge from $v_j$ to a reticulation;
  \item if $\lambda_\pi(v_j)$ has no forced element and $w_q=m_\pi(r)$ for a reticulation child $r$ of $v_j$, the
    algorithm keeps the edge $(v_j,r)$ and matches $w_q$ at $v_j$; otherwise it keeps no edge from $v_j$, and the
    label of $v_j$ becomes empty.
\end{itemize}

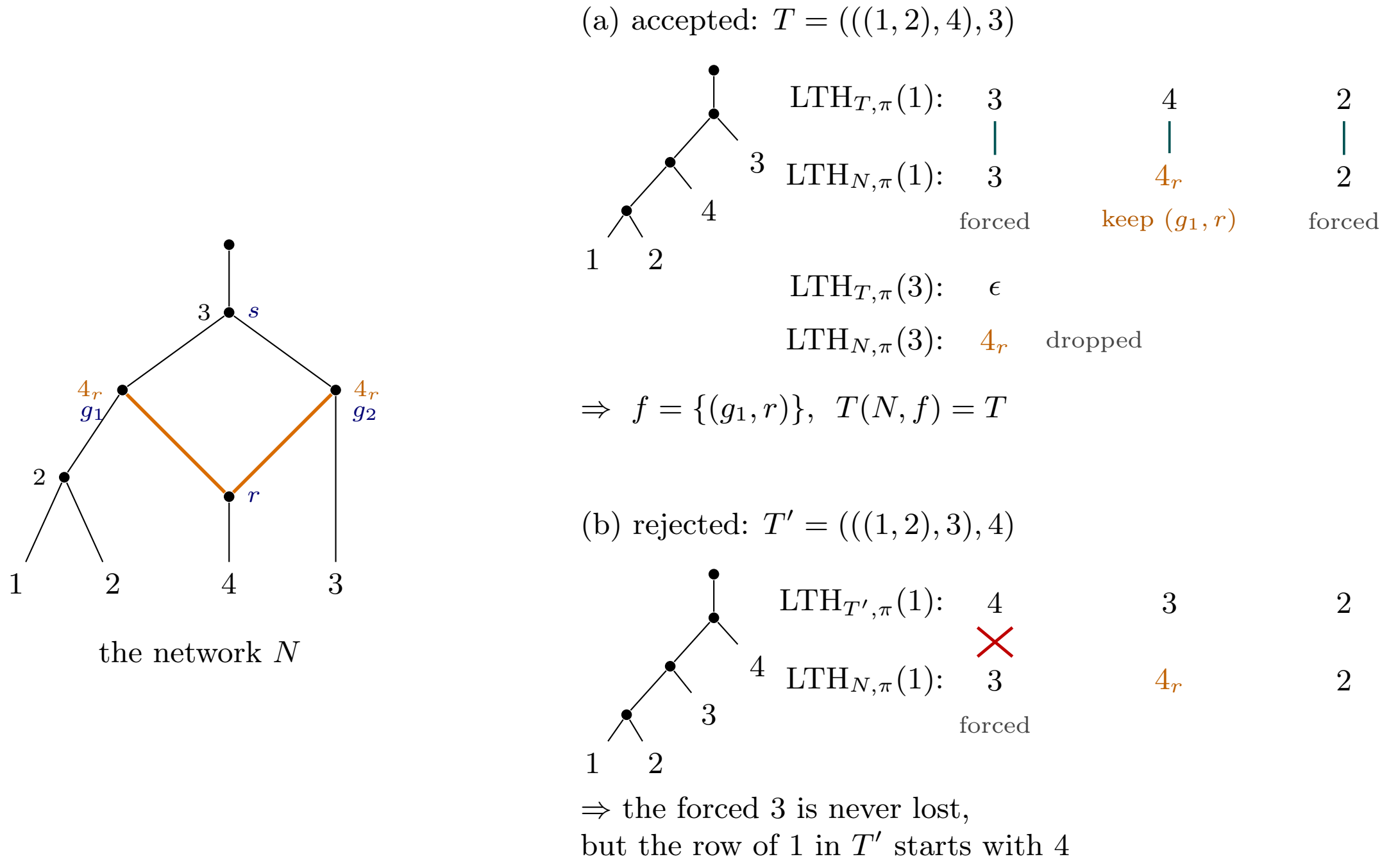
\begin{figure}[!ht]
\centering
\begin{tikzpicture}[
  int/.style={circle, fill, inner sep=1.1pt},
  lf/.style={font=\small},
  nm/.style={font=\scriptsize, text=blue!45!black},
  lab/.style={font=\scriptsize},
  clab/.style={font=\scriptsize, text=orange!75!black},
  re/.style={line width=1.0pt, orange!85!black},
  ent/.style={font=\small},
  ctrl/.style={font=\small, text=orange!75!black},
  rowlab/.style={font=\small, anchor=east},
  tag/.style={font=\scriptsize, text=gray!55!black},
  match/.style={draw=teal!65!black, line width=0.7pt},
  capt/.style={font=\small}
]
\begin{scope}[yshift=0.7cm]
  \node[int] (rho) at (0,2.0) {};
  \node[int, label={[lab]left:$3$}, label={[nm]right:$s$}] (s) at (0,1.3) {};
  \node[int, label={[clab]left:$4_r$}, label={[nm]below left:$g_1$}] (g1) at (-1.1,0.5) {};
  \node[int, label={[clab]right:$4_r$}, label={[nm]below right:$g_2$}] (g2) at (1.1,0.5) {};
  \node[int, label={[lab]left:$2$}] (mm) at (-1.7,-0.4) {};
  \node[int, label={[nm]right:$r$}] (ret) at (0,-0.6) {};
  \node[lf] (l1) at (-2.2,-1.5) {$1$};
  \node[lf] (l2) at (-1.2,-1.5) {$2$};
  \node[lf] (l4) at (0,-1.5) {$4$};
  \node[lf] (l3) at (1.1,-1.5) {$3$};
  \draw (rho)--(s);
  \draw (s)--(g1); \draw (s)--(g2);
  \draw (g1)--(mm); \draw (g2)--(l3);
  \draw[re] (g1)--(ret); \draw[re] (g2)--(ret);
  \draw (mm)--(l1); \draw (mm)--(l2); \draw (ret)--(l4);
  \node[capt] at (-0.3,-2.2) {the network $N$};
\end{scope}
\begin{scope}[xshift=5.0cm, yshift=2.6cm]
  \node[capt, anchor=west] at (-1.5,2.4) {(a) accepted: $T=(((1,2),4),3)$};
  \node[int] (art) at (0,1.9) {};
  \node[int] (as)  at (0,1.45) {};
  \node[int] (ag)  at (-0.45,0.95) {};
  \node[int] (am)  at (-0.9,0.45) {};
  \node[lf] (a1) at (-1.25,-0.05) {$1$};
  \node[lf] (a2) at (-0.6,-0.05) {$2$};
  \node[lf] (a4) at (-0.05,0.45) {$4$};
  \node[lf] (a3) at (0.45,0.95) {$3$};
  \draw (art)--(as); \draw (as)--(ag); \draw (as)--(a3);
  \draw (ag)--(am); \draw (ag)--(a4);
  \draw (am)--(a1); \draw (am)--(a2);
  \node[rowlab] at (2.5,1.6) {$\mathrm{LTH}_{T,\pi}(1)$:};
  \node[ent] (ta1) at (2.9,1.6) {$3$};
  \node[ent] (ta2) at (4.7,1.6) {$4$};
  \node[ent] (ta3) at (6.5,1.6) {$2$};
  \node[rowlab] at (2.5,0.8) {$\mathrm{LTH}_{N,\pi}(1)$:};
  \node[ent] (na1) at (2.9,0.8) {$3$};
  \node[ctrl] (na2) at (4.7,0.8) {$4_r$};
  \node[ent] (na3) at (6.5,0.8) {$2$};
  \draw[match] (na1)--(ta1);
  \draw[match] (na2)--(ta2);
  \draw[match] (na3)--(ta3);
  \node[tag] at (2.9,0.35) {forced};
  \node[tag, text=orange!70!black] at (4.7,0.35) {keep $(g_1,r)$};
  \node[tag] at (6.5,0.35) {forced};
  \node[rowlab] at (2.5,-0.35) {$\mathrm{LTH}_{T,\pi}(3)$:};
  \node[ent] at (2.9,-0.35) {$\epsilon$};
  \node[rowlab] at (2.5,-0.9) {$\mathrm{LTH}_{N,\pi}(3)$:};
  \node[ctrl] at (2.9,-0.9) {$4_r$};
  \node[tag, anchor=west] at (3.3,-0.9) {dropped};
  \node[capt, anchor=west] at (-1.5,-1.6) {$\Rightarrow\ f=\{(g_1,r)\}$, \ $T(N,f)=T$};
\end{scope}
\begin{scope}[xshift=5.0cm, yshift=-2.6cm]
  \node[capt, anchor=west] at (-1.5,2.4) {(b) rejected: $T'=(((1,2),3),4)$};
  \node[int] (brt) at (0,1.9) {};
  \node[int] (bs)  at (0,1.45) {};
  \node[int] (ba)  at (-0.45,0.95) {};
  \node[int] (bm)  at (-0.9,0.45) {};
  \node[lf] (b1) at (-1.25,-0.05) {$1$};
  \node[lf] (b2) at (-0.6,-0.05) {$2$};
  \node[lf] (b3) at (-0.05,0.45) {$3$};
  \node[lf] (b4) at (0.45,0.95) {$4$};
  \draw (brt)--(bs); \draw (bs)--(ba); \draw (bs)--(b4);
  \draw (ba)--(bm); \draw (ba)--(b3);
  \draw (bm)--(b1); \draw (bm)--(b2);
  \node[rowlab] at (2.5,1.6) {$\mathrm{LTH}_{T',\pi}(1)$:};
  \node[ent] at (2.9,1.6) {$4$};
  \node[ent] at (4.7,1.6) {$3$};
  \node[ent] at (6.5,1.6) {$2$};
  \node[rowlab] at (2.5,0.8) {$\mathrm{LTH}_{N,\pi}(1)$:};
  \node[ent] at (2.9,0.8) {$3$};
  \node[ctrl] at (4.7,0.8) {$4_r$};
  \node[ent] at (6.5,0.8) {$2$};
  \draw[draw=red!75!black, line width=0.9pt] (2.72,1.05)--(3.08,1.35);
  \draw[draw=red!75!black, line width=0.9pt] (3.08,1.05)--(2.72,1.35);
  \node[tag] at (2.9,0.35) {forced};
  \node[capt, anchor=west, align=left] at (-1.5,-0.7)
    {$\Rightarrow$ the forced $3$ is never lost,\\ but the row of $1$ in $T'$ starts with $4$};
\end{scope}
\end{tikzpicture}
\caption{Algorithm~\ref{alg:case1} on the network $N$ of Figure~\ref{fig:nlth}, with $\pi=1<2<3<4$. The rows of $N$
are $\mathrm{LTH}_{N,\pi}(1)=(3,4_r,2)$ and $\mathrm{LTH}_{N,\pi}(3)=(4_r)$. The element $4_r$ is controlled by the
edges into $r$, and all other elements are forced.
(a)~For $T=(((1,2),4),3)$, the row of $1$ in $T$ is $(3,4,2)$. The algorithm matches the forced elements $3$ and $2$,
and it matches $4$ at $g_1$ by keeping the edge $(g_1,r)$. The row of $3$ in $T$ is empty, so no edge from $g_2$ is
kept. The algorithm returns `Yes' and $f=\{(g_1,r)\}$, and $T(N,f)=T$.
(b)~For $T'=(((1,2),3),4)$, the row of $1$ in $T'$ is $(4,3,2)$. The first tree node $s$ on $\mathrm{path}(1)$ has
the forced element $3\neq 4$, so the algorithm answers `No', and $N$ does not display $T'$.}
\label{fig:case1match}
\end{figure}

The algorithm answers `No' if some entry of $(w_1,\dots,w_s)$ is not matched. It matches each entry at the first
tree node where this is possible. For instance, if two parents of a reticulation $r$ lie on $\mathrm{path}(\tau)$, it
may keep the edge from the upper parent although the lower one could also be used. Theorem~\ref{thm:case1} shows that
this never loses a solution, and that the edges kept in all rows form a switching $f$ with $T(N,f)=T$.
Figure~\ref{fig:case1match} shows an accepting and a rejecting run on the network of
Figure~\ref{fig:nlth}.

\begin{algorithm}[H]
\caption{Tree Containment for a tree-child network and a binary phylogenetic tree.}
\label{alg:case1}
\begin{algorithmic}%
\State {\bf Input}: A  tree-child network $N$ and a binary tree $T$ on the same taxon set $X$.
\State {\bf Output}: `Yes' and a switching $f$ with $T(N,f)=T$ if $N$ displays $T$, and `No' otherwise.\\
\State $\pi\gets$ \Call{CanonicalOrder}{$N$}; \Comment{{\tt Algorithm~\ref{alg:canonical}}}
\State compute $\{\mathrm{LTH}_{T,\pi}(\tau)\}_{\tau\in X}$ by \Call{Encode}{$T,\pi$} and $\{\mathrm{LTH}_{N,\pi}(\tau)\}_{\tau\in X}$ by \Call{Encode}{$N,\pi$}; \Comment{{\tt Algorithm~\ref{alg:encode}}}
\State $f \gets \varnothing$; \Comment{{\tt the reticulation edges kept so far}}
\For{each taxon $\tau$}
  \State $w \gets \mathrm{LTH}_{T,\pi}(\tau)$;\quad $q \gets 1$; \Comment{{\tt row of $\tau$ in $T$; $q$: next unmatched position}}
  \For{each tree node $v$ on $\mathrm{path}(\tau)$, from top to bottom}
    \If{$\lambda_\pi(v)$ has two or more forced elements}
      \State return `No'; \Comment{{\tt $v$ has at least three children in every $T(N,f)$}}
    \ElsIf{$\lambda_\pi(v)$ has exactly one forced element $y$}
      \State {\bf if} {$q>|w|$ \textbf{ or } $y\ne w_q$} {\bf then} {return `No';}
      \State $q \gets q+1$; \Comment{{\tt every controlled element of $\lambda_\pi(v)$ is dropped}}
    \ElsIf{$q\le|w|$ \textbf{ and } $w_q=m_\pi(r)$ for a reticulation child $r$ of $v$}
      \State $f \gets f \cup \{(v,r)\}$;\quad $q \gets q+1$; \Comment{{\tt keep $(v,r)$; drop the other controlled elements}}
    \EndIf
  \EndFor
  \State {\bf if} {$q \le |w|$} {\bf then} {return `No';} \Comment{{\tt row of $\tau$ not fully matched}}
\EndFor
\State return `Yes' and $f$. \Comment{{\tt $f$ is a switching and $T(N,f)=T$ (Theorem~\ref{thm:case1})}}
\end{algorithmic}
\end{algorithm}

\begin{theorem}[Correctness of Algorithm~\ref{alg:case1}]\label{thm:case1}
For a binary phylogenetic tree $T$ and a tree-child network $N$ on $X$, Algorithm~\ref{alg:case1} returns
`Yes' if and only if $N$ displays $T$, and then $T(N,f)=T$ for the switching $f$ it returns. It runs in
$O(|N|+|T|)$ time.
\end{theorem}
\begin{proof}
Each tree node $v$ of $N$ lies on exactly one principal path, $\mathrm{path}(m_\pi(v))$
(Lemma~\ref{lem:dichotomy}(i)). So the algorithm processes $v$ only once, in the row of $m_\pi(v)$, and $f$ contains
an edge $(v,r)$ only if the algorithm added it while processing $v$. Within a row, the algorithm matches the entries
$w_1,w_2,\dots$ in this order, and it matches an entry only at a tree node whose label contains it.

Suppose first that the algorithm returns `Yes' and $f$. Let $r$ be a reticulation. By Lemma~\ref{lem:dichotomy},
$r=\mathrm{top}(m_\pi(r))$ and $m_\pi(r)\neq\pi_1$. So $m_\pi(r)$ occurs in the encoding of $N$ only in the labels of the
parents of $r$, where it is controlled, and it occurs exactly once in the encoding of $T$ (Lemma~\ref{lem:host}), say
as the entry $w_k$ of the row of some taxon $\tau$. Since the algorithm returned `Yes', it matched $w_k$ at a tree node $v_j$ on
$\mathrm{path}(\tau)$. As $m_\pi(r)$ is not forced, this happened in the third case above, and since distinct
reticulations have distinct minimum taxa (Lemma~\ref{lem:dichotomy}), the algorithm added the edge $(v_j,r)$ to $f$.
It added no other incoming edge of $r$, because it adds an edge $(v,r)$ only when it matches an entry equal to
$m_\pi(r)$, and $w_k$ is the only such entry. Hence $f$ keeps exactly one incoming edge of each reticulation, and $f$
is a switching.

Now fix a taxon $\tau$. If the algorithm matched $w_k$ at $v_j$, then either $w_k$ is the only forced element of
$\lambda_\pi(v_j)$ and $f$ keeps no edge from $v_j$, or $\lambda_\pi(v_j)$ has no forced element and the only edge
from $v_j$ in $f$ is the edge $(v_j,r)$ with $m_\pi(r)=w_k$. In both cases, the label of $v_j$ becomes $\{w_k\}$ when it loses the
elements controlled by the edges that $f$ does not keep. If the algorithm matched nothing at $v_j$, then
$\lambda_\pi(v_j)$ has no forced element and $f$ keeps no edge from $v_j$, so the label of $v_j$ becomes empty. Since
the algorithm matched all of $w_1,\dots,w_s$, in this order, the labels that do not become empty are
$\{w_1\},\dots,\{w_s\}$, in this order, and so $\mathrm{LTH}_{T(N,f),\pi}(\tau)=(w_1,\dots,w_s)=\mathrm{LTH}_{T,\pi}(\tau)$ by
Lemma~\ref{lem:rowlocality}. This holds for every taxon, and so $T(N,f)=T$ by Lemma~\ref{lem:faithful}.

Conversely, suppose that $N$ displays $T$. By Lemma~\ref{lem:binarycollapse}, there is a switching $f^\ast$ with
$T(N,f^\ast)=T$. Fix a taxon $\tau$. By Lemma~\ref{lem:rowlocality}, as $T(N,f^\ast)=T$, when the labels lose the
elements controlled by the edges that $f^\ast$ does not keep, the labels that do not become empty are those of some
tree nodes $v_{j_1},\dots,v_{j_s}$ with $j_1<\dots<j_s$, and the label of $v_{j_k}$ becomes $\{w_k\}$. Since forced
elements are never lost, each $v_j$ with a forced element is some $v_{j_k}$, and then $w_k$ is its only forced element.
If $v_{j_k}$ has no forced element, then $w_k$ is a controlled element of $\lambda_\pi(v_{j_k})$.

For $0\le j\le p$, let $c_j$ be the number of indices $k$ with $j_k\le j$. We show by induction on $j$ that the
algorithm does not answer `No' at $v_1,\dots,v_j$, and that it matches at least $c_j$ entries there. This is clear for $j=0$. Let $j\ge1$, and let $g\ge c_{j-1}$ be the number of entries that the algorithm
matches at $v_1,\dots,v_{j-1}$.
\begin{itemize}
  \item Suppose that $\lambda_\pi(v_j)$ has a forced element. Then $v_j=v_{j_k}$ for some $k$, with the single forced
    element $w_k$, and $c_{j-1}=k-1$ because $j_k=j$. Since $w_k$ is forced, $\mathrm{top}(w_k)$ is not a reticulation, and so $w_k$ occurs in
    the encoding of $N$ only in $\lambda_\pi(v_j)$ (Lemma~\ref{lem:dichotomy}(ii)). Hence $w_k$ was not matched at
    $v_1,\dots,v_{j-1}$, and, as the entries are matched in order, $g\le k-1$. So $g=k-1$, that is, $q=k$ at $v_j$,
    the only forced element of $\lambda_\pi(v_j)$ is $w_q$, and the algorithm matches $w_k$ at $v_j$.
  \item Suppose that $\lambda_\pi(v_j)$ has no forced element. Then the algorithm does not answer `No' at $v_j$. If
    $v_j=v_{j_k}$ and $g=k-1$, then $q=k$, $w_k$ is a controlled element of $\lambda_\pi(v_j)$, and the algorithm
    matches $w_k$ at $v_j$. Otherwise, either $v_j$ is none of $v_{j_1},\dots,v_{j_s}$, and $c_j=c_{j-1}\le g$, or
    $v_j=v_{j_k}$ with $g\ge k=c_j$.
\end{itemize}
In both cases the algorithm matches at least $c_j$ entries at $v_1,\dots,v_j$. For $j=p$, it matches at least
$c_p=s$ entries, that is, all of $w_1,\dots,w_s$, and it does not answer `No' after the row. As this holds for every
taxon, the algorithm returns `Yes'.

For the running time, Algorithms~\ref{alg:canonical} and~\ref{alg:encode} take $O(|N|+|T|)$ time. The main loop
processes each tree node $v$ once and scans its label and its children once, and it reads each row of $T$ once, so it
takes $O(|N|+|T|)$ time.
\end{proof}

The theorem extends the known linear-time bounds for firm containment on tree-child
networks~\cite{gunawan2018,weller2018} to multifurcating tree nodes. It removes the factor $d^{2}$ from the $O(|N|d^{2})$ bound of~\cite[Theorem~2]{weller2018}, where $d$ is the maximum reticulation indegree. 

\section{Soft display of a phylogenetic tree in a tree-child network}\label{sec:case2}\label{sec:case3}

In this section, $N$ is a tree-child network on $X$ whose tree nodes and reticulations may have any
degrees, $T$ is a phylogenetic tree on $X$, and $\pi$ is the canonical order of $X$ for $N$ computed by
Algorithm~\ref{alg:canonical}. We write $k$ for the number of reticulations of $N$, and $d_r$, $d$ and
$d_\Sigma$ for the indegree of a reticulation $r$, the maximum reticulation indegree and the total
reticulation indegree (Section~\ref{sec:prelim}). By Theorem~\ref{thm:star}, $N$ softly displays $T$ if and
only if $\mathcal{C}(T)\subseteq\mathcal{C}(T(N,f))$ for some switching $f$. We show that this condition can be
decided in $O(|N|+|T|+nd_\Sigma^{2})$ time. We first turn the condition into one condition per cluster of
$T$. We then show that the switchings satisfying the condition of one cluster are described by constraints
on single reticulations and on pairs of reticulations, and that the resulting constraint network is solved
without search.

Recall that a switching $f$ keeps the edge $(f(r),r)$ at each reticulation $r$, so that the set of
switchings is $\prod_{r\in R(N)}\mathrm{par}(r)$. A {\it box} is a set of switchings of the form
$\prod_{r\in R(N)}B_r$ with $B_r\subseteq\mathrm{par}(r)$ for every $r$, that is, the set of switchings
$f$ with $f(r)\in B_r$ for every reticulation $r$; it is empty if some $B_r$ is empty. For a vertex $v$,
$L_f(v)$ is the set of taxa reachable from $v$ in $f(N)$
(Section~\ref{sec:prelim}).

\subsection{Reduction to clusters}

Let $C\subseteq X$. A switching $f$ {\it realizes} $C$ at a vertex $w$ of $N$ if $L_f(w)=C$, that is,
if $C$ is the cluster of $w$ in $f(N)$. For a vertex $w$, let $S_w(C)$ be the set of switchings that
realize $C$ at $w$, and let $\mathrm{realized}(C)$ be the set of switchings that realize $C$ at some
vertex:
\[
  S_w(C)=\{f: L_f(w)=C\}
  \qquad\text{and}\qquad
  \mathrm{realized}(C)=\bigcup_{w\in V(N)}S_w(C).
\]
The next lemma shows that the clusters of $T(N,f)$ are exactly the sets that $f$ realizes. Hence the
condition $\mathcal{C}(T)\subseteq\mathcal{C}(T(N,f))$ of Theorem~\ref{thm:star} says that $f$ realizes
every cluster of $T$, and it splits into one condition per cluster.

\begin{lemma}\label{thm:decomp}
For every switching $f$ of $N$ and every $C\subseteq X$, $C\in\mathcal{C}(T(N,f))$ if and only if
$L_f(w)=C$ for some vertex $w$ of $N$. Hence $N$ softly displays $T$ if and only if
\[
  \bigcap_{C\in\mathcal{C}(T)}\mathrm{realized}(C)\neq\varnothing .
\]
\end{lemma}

\begin{proof}
In $f(N)$, every vertex other than the root has exactly one incoming edge, so $f(N)$ is a tree rooted at
the root of $N$, and $L_f(w)$ is the cluster of $w$ in this tree. By Lemma~\ref{lem:committed}, every
vertex reaches a leaf along tree edges, which $f$ keeps. So every vertex of $f(N)$ lies on a path from
the root to a leaf, and the deletion step in the definition of $T(N,f)$ (Section~\ref{sec:prelim})
removes nothing: $T(N,f)$ is obtained from $f(N)$ by suppressing the vertices with one parent and one
child. Suppressing a vertex does not change the clusters of the other vertices, and the suppressed
vertex has the same cluster as its child. Hence the cluster set of $T(N,f)$ is $\{L_f(w): w\in V(N)\}$,
which is the first claim. By the first claim, $\mathcal{C}(T)\subseteq\mathcal{C}(T(N,f))$ if and only
if $f\in\mathrm{realized}(C)$ for every $C\in\mathcal{C}(T)$, that is, if and only if
$f\in\bigcap_{C\in\mathcal{C}(T)}\mathrm{realized}(C)$. The second claim now follows from
Theorem~\ref{thm:star}.
\end{proof}

Let $n=|X|$. The clusters of $T$ with one or $n$ elements are the sets $\{x\}$, for $x\in X$, and $X$
itself. Every switching $f$ realizes them, since $L_f(\ell_x)=\{x\}$ and $L_f(\mathrm{root})=X$. Hence only
the clusters $C$ of $T$ with $2\le|C|\le n-1$ have to be considered.

\subsection{Hosts of a cluster}

A vertex $w$ is a {\it host} of $C$ if $S_w(C)\neq\varnothing$, and $W(C)$ denotes the set of hosts of
$C$.

\begin{lemma}\label{cor:hostrow}
Let $C\subseteq X$ be nonempty and $\tau=\min_\pi C$. Every host of $C$ lies on $\mathrm{path}(\tau)$. In
particular, any two hosts of $C$ are comparable.
\end{lemma}

\begin{proof}
Let $w$ be a host of $C$ and $f\in S_w(C)$. By Lemma~\ref{lem:minsurvive},
$m_\pi(w)=\min_\pi L_f(w)=\min_\pi C=\tau$. By Lemma~\ref{lem:dichotomy}(i), $w$ lies on
$\mathrm{path}(\tau)$. Since $\mathrm{path}(\tau)$ is a directed path, any two of its vertices are
comparable.
\end{proof}

\subsection{Realizing a cluster at a vertex}

Fix $C\subseteq X$ and a vertex $w$. If $C\not\subseteq L(w)$, then $S_w(C)=\varnothing$, since
$L_f(w)\subseteq L(w)$ for every $f$. Let $K_0(w)$ be the set of vertices reachable from $w$ along tree
edges; it is $\mathrm{comp}(w)$ if $w$ is a reticulation or the root, and the part of $\mathrm{comp}(w)$ below
$w$ otherwise. Let $K$ be a tree component or $K_0(w)$. Then $K$ contains a leaf (Lemma~\ref{lem:committed}).
Let $L(K)$ be the set of taxa of the leaves in $K$. We say that $K$ is {\it must-in} if $L(K)\subseteq C$,
{\it must-out} if $L(K)\cap C=\varnothing$, and {\it mixed} otherwise; this is the {\it status} of $K$. A reticulation $r$ has the status of the tree component $\mathrm{comp}(r)$,
whose root it is; for instance, $r$ is must-in if $\mathrm{comp}(r)$ is must-in.

A reticulation $r$ is {\it constrained} at $w$ if $w$ is a strict ancestor of $r$, and {\it free} at $w$
otherwise. If $r$ is free at $w$, then no path from $w$ uses an incoming edge of $r$, so the parent of $r$
that a switching keeps does not affect $L_f(w)$.

Let $R(w)$ be the set of vertices that $w$ reaches without entering a tree component that is not
must-in. Formally, $R(w)$ is the set of vertices reachable from $w$ in the subgraph $G_{\mathrm{in}}(C)$ of
$N$ obtained by deleting the incoming edges of the reticulations that are not must-in.

Let $f$ be a switching that realizes $C$ at $w$, and let $r$ be a reticulation constrained at $w$. As we
show in Lemma~\ref{lem:carve}, $f$ keeps a parent of $r$ in $R(w)$ if $r$ is must-in, so that $w$ reaches
the taxa of $\mathrm{comp}(r)$, and a parent of $r$ outside $R(w)$ if $r$ is must-out, so that $w$ does not
reach them. Moreover, $r$ is not mixed, since $w$ reaches either all taxa of $\mathrm{comp}(r)$ or none.
Accordingly, for each reticulation $r$, let
\[
  A_w(r)=
  \begin{cases}
    \mathrm{par}(r)\cap R(w) & \text{if $r$ is constrained at $w$ and must-in},\\[2pt]
    \mathrm{par}(r)\setminus R(w) & \text{if $r$ is constrained at $w$ and must-out},\\[2pt]
    \varnothing & \text{if $r$ is constrained at $w$ and mixed},\\[2pt]
    \mathrm{par}(r) & \text{if $r$ is free at $w$}.
  \end{cases}
\]
\begin{lemma}[Carve]\label{lem:carve}
Let $C\subseteq L(w)$.
\begin{enumerate}
\item[(i)] If $K_0(w)$ is not must-in, then $S_w(C)=\varnothing$.
\item[(ii)] If $K_0(w)$ is must-in, then $S_w(C)=\prod_rA_w(r)$. In particular, $S_w(C)$ is a
box, and it is nonempty if and only if $A_w(r)\neq\varnothing$ for every reticulation $r$.
\end{enumerate}
\end{lemma}

Figure~\ref{fig:carve} shows the sets $S_w(C)$ for two clusters.
\begin{figure}[!h]
\centering
\begin{tikzpicture}[
  int/.style={circle, fill, inner sep=1.2pt},
  lf/.style={font=\small},
  hostv/.style={circle, draw, fill=white, line width=0.9pt, inner sep=1.7pt},
  c0/.style={blue!70!black},
  c1/.style={teal!80!black},
  mo/.style={red!65!black},
  dimmed/.style={gray!55},
  keep/.style={line width=1.1pt, orange!85!black},
  drop/.style={line width=1pt, gray!55, densely dashed},
  re/.style={line width=1pt, orange!85!black, densely dashed},
  cc/.style={font=\scriptsize}
]
\begin{scope}
  \fill[blue!6, rounded corners] (-1.55,0.85) rectangle (1.55,4.0);
  \node[cc, anchor=west, text=blue!45!black] at (-1.5,1.05) {below $w$};
  \node[int, dimmed] (arho0) at (0.7,5.05) {};
  \node[int, dimmed] (arho)  at (0.7,4.55) {};
  \node[int, dimmed] (az)    at (2.5,3.9) {};
  \node[int, dimmed, label={[cc]right:$z'$}] (azp) at (2.05,3.0) {};
  \node[lf, dimmed] (al5) at (1.9,2.15) {$5$};
  \node[int, dimmed, label={[cc]right:$t$}] (at) at (3.25,3.05) {};
  \node[int, dimmed, label={[cc]left:$t'$}] (atp) at (2.9,2.2) {};
  \node[int, dimmed, label={[cc]right:$s$}] (as) at (3.6,2.2) {};
  \node[lf, dimmed] (al6) at (2.85,1.4) {$6$};
  \node[lf, dimmed] (al7) at (3.6,1.4) {$7$};
  \node[hostv, c0, label={[cc]above left:host $w$}] (aw) at (0,3.6) {};
  \node[int, c0, label={[cc, c0]left:$a$}] (aa) at (-0.8,2.8) {};
  \node[lf, c0] (al1) at (-1.2,1.95) {$1$};
  \node[lf, c0] (al2) at (-0.5,1.95) {$2$};
  \node[int, c1, label={[cc, c1]below left:$r$}] (ar) at (0.8,2.8) {};
  \node[int, c1, label={[cc, c1]left:$v$}] (av) at (0.8,2.0) {};
  \node[lf, c1] (al3) at (0.45,1.2) {$3$};
  \node[lf, c1] (al4) at (1.2,1.2) {$4$};
  \draw[dimmed] (arho0)--(arho); \draw[dimmed] (arho)--(aw); \draw[dimmed] (arho)--(az);
  \draw[dimmed] (az)--(azp); \draw[dimmed] (az)--(at); \draw[dimmed] (azp)--(al5);
  \draw[dimmed] (at)--(atp); \draw[dimmed] (atp)--(al6); \draw[dimmed] (as)--(al7);
  \draw[re] (at)--(as); \draw[re] (atp)--(as);
  \draw[c0] (aw)--(aa); \draw[c0] (aa)--(al1); \draw[c0] (aa)--(al2);
  \draw[c1] (ar)--(av); \draw[c1] (av)--(al3); \draw[c1] (av)--(al4);
  \draw[keep] (aw)--(ar);
  \draw[drop] (azp)--(ar);
  \node[cc, c0, align=center] at (-1.15,3.15) {$K_0$\\must-in};
  \node[cc, c1, align=center] at (1.36,2.42) {$K$\\must-in};
  \node[font=\small] at (1.1,0.5) {(a) $C=\{1,2,3,4\}$: $K$ must-in};
\end{scope}
\begin{scope}[xshift=5.9cm]
  \fill[blue!6, rounded corners] (-1.55,0.85) rectangle (1.55,4.0);
  \node[cc, anchor=west, text=blue!45!black] at (-1.5,1.05) {below $w$};
  \node[int, dimmed] (brho0) at (0.7,5.05) {};
  \node[int, dimmed] (brho)  at (0.7,4.55) {};
  \node[int, dimmed] (bz)    at (2.5,3.9) {};
  \node[int, dimmed, label={[cc]right:$z'$}] (bzp) at (2.05,3.0) {};
  \node[lf, dimmed] (bl5) at (1.9,2.15) {$5$};
  \node[int, dimmed, label={[cc]right:$t$}] (bt) at (3.25,3.05) {};
  \node[int, dimmed, label={[cc]left:$t'$}] (btp) at (2.9,2.2) {};
  \node[int, dimmed, label={[cc]right:$s$}] (bs) at (3.6,2.2) {};
  \node[lf, dimmed] (bl6) at (2.85,1.4) {$6$};
  \node[lf, dimmed] (bl7) at (3.6,1.4) {$7$};
  \node[hostv, c0, label={[cc]above left:host $w$}] (bw) at (0,3.6) {};
  \node[int, c0, label={[cc, c0]left:$a$}] (ba) at (-0.8,2.8) {};
  \node[lf, c0] (bl1) at (-1.2,1.95) {$1$};
  \node[lf, c0] (bl2) at (-0.5,1.95) {$2$};
  \node[int, mo, label={[cc, mo]below left:$r$}] (br) at (0.8,2.8) {};
  \node[int, mo, label={[cc, mo]left:$v$}] (bv) at (0.8,2.0) {};
  \node[lf, mo] (bl3) at (0.45,1.2) {$3$};
  \node[lf, mo] (bl4) at (1.2,1.2) {$4$};
  \draw[dimmed] (brho0)--(brho); \draw[dimmed] (brho)--(bw); \draw[dimmed] (brho)--(bz);
  \draw[dimmed] (bz)--(bzp); \draw[dimmed] (bz)--(bt); \draw[dimmed] (bzp)--(bl5);
  \draw[dimmed] (bt)--(btp); \draw[dimmed] (btp)--(bl6); \draw[dimmed] (bs)--(bl7);
  \draw[re] (bt)--(bs); \draw[re] (btp)--(bs);
  \draw[c0] (bw)--(ba); \draw[c0] (ba)--(bl1); \draw[c0] (ba)--(bl2);
  \draw[mo] (br)--(bv); \draw[mo] (bv)--(bl3); \draw[mo] (bv)--(bl4);
  \draw[keep] (bzp)--(br);
  \draw[drop] (bw)--(br);
  \node[cc, c0, align=center] at (-1.15,3.15) {$K_0$\\must-in};
  \node[cc, mo, align=center] at (1.36,2.42) {$K$\\must-out};
  \node[font=\small] at (1.1,0.5) {(b) $C'=\{1,2\}$: $K$ must-out};
\end{scope}
\begin{scope}[xshift=2.9cm, yshift=-2.6cm]
  \fill[orange!22] (0,0.8) rectangle (2.4,1.6);
  \fill[red!12] (0,0) rectangle (2.4,0.8);
  \draw (0,0) rectangle (2.4,1.6);
  \draw (1.2,0)--(1.2,1.6);
  \draw (0,0.8)--(2.4,0.8);
  \node[cc, anchor=east] at (-0.12,1.2) {$f(r){=}w$};
  \node[cc, anchor=east] at (-0.12,0.4) {$f(r){=}z'$};
  \node[cc] at (0.6,1.78) {$f(s){=}t$};
  \node[cc] at (1.8,1.78) {$f(s){=}t'$};
  \node[int, orange!60!black] at (0.6,1.2) {};
  \node[int, orange!60!black] at (1.8,1.2) {};
  \node[int, mo] at (0.6,0.4) {};
  \node[int, mo] at (1.8,0.4) {};
  \node[cc, anchor=west, text=orange!60!black] at (2.52,1.2) {$S_w(C)$};
  \node[cc, anchor=west, mo] at (2.52,0.4) {$S_w(C')$};
  \node[font=\small] at (1.2,-0.45) {(c) the boxes $S_w(C)$ and $S_w(C')$};
\end{scope}
\end{tikzpicture}
\caption{The sets $S_w(C)$ of Lemma~\ref{lem:carve} in a tree-child network $N$ on $\{1,\dots,7\}$ with the
reticulations $r$, with parents $w$ and $z'$, and $s$, with parents $t$ and $t'$. The vertex $w$ and its descendants
are shaded; $K_0(w)=\{w,a,1,2\}$, and $K=\mathrm{comp}(r)=\{r,v,3,4\}$ is the only tree component of $N$ whose
root is constrained at $w$. The reticulation $s$ is free at $w$, so $A_w(s)=\mathrm{par}(s)$ for every cluster.
(a)~For $C=\{1,2,3,4\}$, $K$ is must-in, and $A_w(r)=\mathrm{par}(r)\cap R(w)=\{w\}$, since $z'$ is not a descendant of $w$.
(b)~For $C'=\{1,2\}$, $K$ is must-out, and $A_w(r)=\mathrm{par}(r)\setminus R(w)=\{z'\}$. For $C''=\{1,2,3\}$, $K$
is mixed, so $A_w(r)=\varnothing$ and $S_w(C'')=\varnothing$. (c)~The boxes
$S_w(C)=\{w\}\times\mathrm{par}(s)$ and $S_w(C')=\{z'\}\times\mathrm{par}(s)$.}
\label{fig:carve}
\end{figure}

\subsection{Candidate hosts}

Let $C\subseteq X$ be nonempty and $\tau=\min_\pi C$. The hosts of $C$ lie on $\mathrm{path}(\tau)$
(Lemma~\ref{cor:hostrow}). In this subsection, we use the order $\pi$ to find a set of vertices of
$\mathrm{path}(\tau)$ that contains all hosts of $C$ and can be computed quickly.

For a taxon $x$, let $\mathrm{rank}(x)$ be the position of $x$ in $\pi$. For a vertex $v$, let $I(v)$ be
the set of ranks of the taxa in $L(K_0(v))$, which are the taxa of the leaves below $v$ in the tree
$\mathrm{comp}(v)$. In $\pi$, the taxa of $\mathrm{comp}(v)$ occupy consecutive positions, in the order in
which a depth-first search of $\mathrm{comp}(v)$ reaches them (Algorithm~\ref{alg:canonical}). Such a
search reaches the leaves below $v$ consecutively. Hence $I(v)$ is an interval of integers.

Let $\mathrm{comp}(\tau)$ be the tree component containing the leaf $\ell_\tau$; it contains
$\mathrm{path}(\tau)$ (Section~\ref{sec:prelim}). Let $I_C$ be the interval of integers from the least to
the greatest rank of a taxon in $C\cap L(\mathrm{comp}(\tau))$. A vertex $v$ of $\mathrm{path}(\tau)$ is a
{\it candidate host} of $C$ if $I(v)=I_C$, and $\widehat W(C)$ denotes the set of candidate hosts of $C$.

\begin{lemma}\label{lem:signature}\samepage
Let $C\subseteq X$ be nonempty and $\tau=\min_\pi C$.
\begin{enumerate}
\item[(i)] Every host $w$ of $C$ satisfies $L(K_0(w))=C\cap L(\mathrm{comp}(\tau))$. Hence
$W(C)\subseteq\widehat W(C)$, and if $C$ has a host, then $|C\cap L(\mathrm{comp}(\tau))|=|I_C|$.
\item[(ii)] If $|C\cap L(\mathrm{comp}(\tau))|=|I_C|$, then $K_0(v)$ is must-in for every candidate
host $v$ of $C$.
\item[(iii)] The candidate hosts of $C$ are consecutive vertices of $\mathrm{path}(\tau)$, and there are
at most $d_\Sigma+3$ of them.
\end{enumerate}
\end{lemma}

The lemma is proved in Appendix~\ref{app:case3}, together with the following one, which describes the
tables that make the lemma usable in constant time.

\begin{lemma}\label{lem:preproc}
In $O(|N|+|T|+nk)$ time, one can compute the following tables.
\begin{enumerate}
\item[(T1)] For every vertex $v$: the interval $I(v)$ and an interval $J(v)$ of integers such that
$u\in K_0(v)$ if and only if $J(u)\subseteq J(v)$. For every tree component $K$: the number $|L(K)|$. A
dictionary that returns, for a taxon $\tau$ and an interval $I$, the vertices $v$ of $\mathrm{path}(\tau)$
with $I(v)=I$, in $O(1)$ time plus $O(1)$ time per vertex returned.
\item[(T2)] For every cluster $C$ of $T$ and every tree component $K$ with $C\cap L(K)\neq\varnothing$:
the number $|C\cap L(K)|$ and the least and the greatest rank of a taxon in $C\cap L(K)$.
\end{enumerate}
With these tables, the status of a tree component with respect to a cluster $C$, the interval $I_C$, the
condition of Lemma~\ref{lem:signature}(ii), and the set $\widehat W(C)$ are obtained in $O(1)$ time, plus
$O(1)$ time per candidate host.
\end{lemma}

\subsection{Constraints on pairs of reticulations}

For a set $A$ of switchings and reticulations $r\neq r'$, let
\[
  \mathrm{proj}_r(A)=\{f(r): f\in A\}
  \qquad\text{and}\qquad
  \mathrm{proj}_{r,r'}(A)=\{(f(r),f(r')): f\in A\}.
\]
We show that $\mathrm{realized}(C)$ is determined by these projections. Fix a cluster $C$ with at least
one host and let $\tau=\min_\pi C$. By Lemma~\ref{cor:hostrow}, the hosts of $C$ form a chain
$w_0\succ w_1\succ\dots\succ w_m$, and so do the candidate hosts. An {\it initial segment} of a chain
consists of its first $i$ elements, for some $i\ge0$, and a {\it final segment} of its last $i$ elements.
For a reticulation $r$ and a parent $p$ of $r$, let
\[
  H_{r,p}=\{\,w\in W(C): p\in A_w(r)\,\}.
\]
By Lemma~\ref{lem:carve}, $S_w(C)=\prod_rA_w(r)$ at every host $w$, so a switching $f$ lies in $S_w(C)$ if
and only if $w\in H_{r,f(r)}$ for every reticulation $r$.

\begin{lemma}\label{lem:monoreach}
Let $w\succeq w'$ be candidate hosts of $C$. Then $w'\in R(w)$ and $R(w')\subseteq R(w)$. Hence, for every
vertex $p$, the candidate hosts $w$ with $p\in R(w)$ form an initial segment of $\widehat W(C)$, and the
hosts $w$ with $p\in R(w)$ form an initial segment of $W(C)$.
\end{lemma}

\begin{proof}
Both $w$ and $w'$ lie on $\mathrm{path}(\tau)$, and the vertices of $\mathrm{path}(\tau)$ other than
$\mathrm{top}(\tau)$ are not reticulations (Lemma~\ref{lem:dichotomy}(i)). So the part of $\mathrm{path}(\tau)$ from
$w$ to $w'$ consists of tree edges, which belong to $G_{\mathrm{in}}(C)$. Hence $w'\in R(w)$, and every
vertex reachable from $w'$ in $G_{\mathrm{in}}(C)$ is reachable from $w$. Thus, if $p\in R(w')$ for a
candidate host $w'$, then $p\in R(w)$ for every candidate host $w\succeq w'$, which is the last claim.
\end{proof}

\begin{lemma}\label{lem:nostraddle}
Let $r$ be a reticulation. If some host $w_0$ of $C$ satisfies $w_0\preceq r$, then no host of $C$ is a strict
ancestor of $r$.
\end{lemma}

\begin{proof}
Let $w$ be a host of $C$, and suppose that $w$ is a strict ancestor of $r$. By
Lemmas~\ref{cor:hostrow} and~\ref{lem:dichotomy}(i), $m_\pi(w_0)=m_\pi(w)=\tau$. Since
$L(w_0)\subseteq L(r)\subseteq L(w)$, the taxon $\tau$ lies in $L(r)$, and $\min_\pi L(r)$ is not smaller
than $\min_\pi L(w)=\tau$; hence $m_\pi(r)=\tau$. By Lemma~\ref{lem:dichotomy}(i), $r$ lies on
$\mathrm{path}(\tau)$, and since $r$ is a reticulation, $r=\mathrm{top}(\tau)$. Then $r$ is an ancestor of every
vertex of $\mathrm{path}(\tau)$, including $w$, which contradicts that $w$ is a strict ancestor of $r$.
\end{proof}

\begin{lemma}\label{lem:bound}
Let $r$ be a must-in reticulation. Then either $r$ is free at every host of $C$, or $r$ is constrained at
every host of $C$. In the second case, we call $r$ {\it bound}.
\end{lemma}

\begin{proof}
Suppose that $r$ is free at some host $w$, and let $f\in S_w(C)$. We claim that $w\preceq r$.
This is clear if $w=r$. Otherwise $r\not\preceq w$, as $r$ is free at $w$. Since $\mathrm{comp}(r)$ is must-in and has a
leaf, some taxon of $C$ labels a leaf $x$ of $\mathrm{comp}(r)$, and $x$ is reachable from $w$, because
$C=L_f(w)\subseteq L(w)$. If $w\notin\mathrm{comp}(r)$, then a path from $w$ to $x$ enters
$\mathrm{comp}(r)$ at its root $r$, as every other vertex of $\mathrm{comp}(r)$ has its only parent in
$\mathrm{comp}(r)$; this gives $r\preceq w$, a contradiction. Hence
$w\in\mathrm{comp}(r)$, and so $w\preceq r$. By Lemma~\ref{lem:nostraddle}, no host of $C$ is a strict
ancestor of $r$, that is, $r$ is free at every host.
\end{proof}

At a host $w$, Lemma~\ref{lem:carve} thus gives $A_w(r)=\mathrm{par}(r)$ if $r$ is free at every host,
$A_w(r)=\mathrm{par}(r)\cap R(w)$ if $r$ is bound, and $A_w(r)=\mathrm{par}(r)\setminus R(w)$ if $r$ is
must-out and constrained at $w$. A mixed reticulation is free at every host, since $A_w(r)\neq\varnothing$
at a host.

\begin{theorem}[Two-decomposition]\label{thm:2decomp}
Let $C$ be a cluster with at least one host. For every reticulation $r$ and every parent $p$ of $r$, the
set $H_{r,p}$ consists of consecutive hosts of the chain $W(C)$. Consequently, a switching $f$ belongs to
$\mathrm{realized}(C)$ if and only if $f(r)\in\mathrm{proj}_r(\mathrm{realized}(C))$ for every reticulation
$r$ and $(f(r),f(r'))\in\mathrm{proj}_{r,r'}(\mathrm{realized}(C))$ for all reticulations $r\neq r'$.
\end{theorem}

The theorem is proved in Appendix~\ref{app:case3}. The second claim follows from the first: the sets
$H_{r,f(r)}$, $r\in R(N)$, are sets of consecutive hosts, and if any two of them intersect, then all of
them share a host $w$, for which $f\in S_w(C)$.

\subsection{Width one}

The constraint network $\mathcal{N}$ has a variable for each reticulation $r$, with domain
$\mathrm{par}(r)$, and, for each cluster $C$ of $T$ with $2\le|C|\le n-1$, the unary constraint
$\mathrm{proj}_r(\mathrm{realized}(C))$ on each variable $r$ and the binary constraint
$\mathrm{proj}_{r,r'}(\mathrm{realized}(C))$ on each pair of variables $r\neq r'$. A {\it solution} is a
switching $f$ with $f(r)\in\mathrm{proj}_r(\mathrm{realized}(C))$ and
$(f(r),f(r'))\in\mathrm{proj}_{r,r'}(\mathrm{realized}(C))$ for all these constraints. If every cluster of
$T$ has a host, then by Theorem~\ref{thm:2decomp} the solutions of $\mathcal{N}$ are exactly the switchings
in $\bigcap_{C\in\mathcal{C}(T)}\mathrm{realized}(C)$, the clusters $X$ and $\{x\}$ being realized by every
switching.

{\it Arc consistency} is enforced on $\mathcal{N}$ by removing from the domain of a variable $r$ every
value $p$ that violates a unary constraint on $r$, or for which some binary constraint on $r$ and another
variable $r'$ contains no pair $(p,p')$ with $p'$ in the current domain of $r'$, until no value can be
removed. A removed value belongs to no solution, so the solutions do not change.

Fix a topological order of the vertices of $N$, in which every vertex appears after its parents, and say
that a vertex is {\it deeper} than another if it appears later in this order. For two parents $p,p'$ of a
reticulation, let $p\wedge p'$ be the deeper of the two, and for switchings $f,g$ let $f\wedge g$ be the
switching with $(f\wedge g)(r)=f(r)\wedge g(r)$ for every reticulation $r$.

\begin{lemma}[Min-closure]\label{lem:minclosed}
For every cluster $C$ and all $f,g\in\mathrm{realized}(C)$, $f\wedge g\in\mathrm{realized}(C)$.
\end{lemma}

\begin{theorem}[Width one]\label{thm:minclosed}
After arc consistency has been enforced on $\mathcal{N}$, either some domain is empty, and then
$\mathcal{N}$ has no solution, or every domain is nonempty, and then the switching that assigns to each
reticulation the deepest value of its domain is a solution.
\end{theorem}

Both are proved in Appendix~\ref{app:case3}. Thus, once every cluster has a host, arc consistency decides
whether $\bigcap_C\mathrm{realized}(C)\neq\varnothing$ and, if so, produces a switching in this
intersection.

\begin{remark}\label{rem:minvs2dec}
Lemma~\ref{lem:minclosed} alone yields a polynomial algorithm: each $\mathrm{realized}(C)$ is the union
of at most $d_\Sigma+3$ boxes (Lemmas~\ref{lem:carve} and~\ref{lem:signature}), hence a constraint closed
under $\wedge$, and a network of constraints closed under one semilattice operation is solved by generalized
arc consistency~\cite{jeavonscooper1995,dalmau1999}. Section~\ref{sec:sym} proceeds in this way.
Theorem~\ref{thm:2decomp} gives in addition the binary network $\mathcal{N}$, whose constraints hold at
most $d_\Sigma^{2}$ pairs in all, and the $2$-SAT instance of Corollary~\ref{cor:binarycase}. It does not
follow from Lemma~\ref{lem:minclosed}: the relation $\{(x_1,x_2,x_3)\in\{0,1\}^{3}: x_3=x_1\wedge x_2\}$ is
closed under the coordinate-wise minimum, but $(1,1,0)$ satisfies all its one- and two-variable
projections without belonging to it.
\end{remark}
\subsection{The algorithm}

\begin{lemma}\label{lem:eligsets}
After the tables of Lemma~\ref{lem:preproc} have been built, the following can be computed for a cluster
$C$ in $O(d_\Sigma^{2})$ time: whether $C$ has a host; if so, the hosts of $C$; for every reticulation $r$
and every parent $p$ of $r$, the first and the last host in $H_{r,p}$; and the projections
$\mathrm{proj}_r(\mathrm{realized}(C))$ and $\mathrm{proj}_{r,r'}(\mathrm{realized}(C))$ for all
reticulations $r\neq r'$.
\end{lemma}

The lemma is proved in Appendix~\ref{app:case3}, where the computation is given as
Algorithm~\ref{alg:carve}. Algorithm~\ref{alg:case3} decides soft display.

\begin{algorithm}[H]
\caption{Soft Tree Containment for a tree-child network and a phylogenetic tree.}
\label{alg:case3}
\begin{algorithmic}%
\State {\bf Input}: A tree-child network $N$ and a phylogenetic tree $T$ on the same taxon set $X$.
\State {\bf Output}: `Yes' and a switching $f$ with $\mathcal{C}(T)\subseteq\mathcal{C}(T(N,f))$ if $N$ softly displays $T$, and `No' otherwise.\\
\State $\pi\gets$ \Call{CanonicalOrder}{$N$}, and build the tables of Lemma~\ref{lem:preproc}; \Comment{{\tt Algorithms~\ref{alg:canonical} and~\ref{alg:preproc}}}
\State $\mathcal{N}\gets$ the constraint network with a variable $r$ of domain $\mathrm{par}(r)$ for each reticulation $r$, and no constraints;
\For{each cluster $C$ of $T$ with $2\le|C|\le n-1$}
  \State compute the hosts of $C$ and the projections of $\mathrm{realized}(C)$; \Comment{{\tt Lemma~\ref{lem:eligsets}, Algorithm~\ref{alg:carve}}}
  \State {\bf if} $C$ has no host {\bf then} return `No'; \Comment{{\tt $\mathrm{realized}(C)=\varnothing$}}
  \State add the constraints $\mathrm{proj}_r(\mathrm{realized}(C))$ and $\mathrm{proj}_{r,r'}(\mathrm{realized}(C))$ to $\mathcal{N}$;
\EndFor
\State enforce arc consistency on $\mathcal{N}$;
\State {\bf if} some domain is empty {\bf then} return `No'; \Comment{{\tt $\bigcap_C\mathrm{realized}(C)=\varnothing$}}
\State $f(r)\gets$ the deepest value in the domain of $r$, for each reticulation $r$; \Comment{{\tt Theorem~\ref{thm:minclosed}}}
\State compute $T(N,f)$; {\bf if} $\mathcal{C}(T)\not\subseteq\mathcal{C}(T(N,f))$ {\bf then} return `No'; \Comment{{\tt certifies $f$; never happens}}
\State return `Yes' and $f$.
\end{algorithmic}
\end{algorithm}

\begin{theorem}[Correctness of Algorithm~\ref{alg:case3}]\label{thm:case3}
For a tree-child network $N$ and a phylogenetic tree $T$ on $X$, Algorithm~\ref{alg:case3} returns `Yes'
if and only if $N$ softly displays $T$, and then $\mathcal{C}(T)\subseteq\mathcal{C}(T(N,f))$ for the
switching $f$ it returns. It runs in $O(|N|+|T|+nd_\Sigma^{2})$ time, which is
$O(|N|+|T|+nk^{2}d^{2})$.
\end{theorem}

\begin{proof}
If the algorithm returns `Yes' and $f$, then it has verified that
$\mathcal{C}(T)\subseteq\mathcal{C}(T(N,f))$, and $N$ softly displays $T$ by Theorem~\ref{thm:star}.
Conversely, suppose that $N$ softly displays $T$. By Lemma~\ref{thm:decomp}, some switching lies in
$\mathrm{realized}(C)$ for every cluster $C$ of $T$. So every cluster has a host, and the algorithm does not
stop in the loop. By Theorem~\ref{thm:2decomp}, the solutions of $\mathcal{N}$ are the switchings in
$\bigcap_{C\in\mathcal{C}(T)}\mathrm{realized}(C)$, so $\mathcal{N}$ has a solution. By
Theorem~\ref{thm:minclosed}, arc consistency empties no domain, and the switching $f$ chosen by the
algorithm is a solution. Hence $f\in\mathrm{realized}(C)$ for every cluster $C$ of $T$, which gives
$\mathcal{C}(T)\subseteq\mathcal{C}(T(N,f))$ by Lemma~\ref{thm:decomp}, and the algorithm returns `Yes'.

For the running time, Algorithm~\ref{alg:canonical} and the tables take $O(|N|+|T|+nk)$ time
(Lemma~\ref{lem:preproc}). Since every internal node of $T$ other than the root has at least two children,
$T$ has at most $2n$ vertices and at most $2n$ clusters, and each cluster is processed in $O(d_\Sigma^{2})$
time (Lemma~\ref{lem:eligsets}), for $O(nd_\Sigma^{2})$ in total. The network has $O(k^{2})$ constraints,
and the constraint on $r,r'$ holds at most $d_rd_{r'}$ pairs. Arc consistency is enforced in time linear
in the total size $\sum_{r\neq r'}d_rd_{r'}\le d_\Sigma^{2}$ of the constraints by the algorithm
AC-4~\cite{mohrhenderson1986}. Computing $T(N,f)$ takes $O(|N|)$ time. For the final test, number the
leaves of $T(N,f)$ in the order of a depth-first search, so that the cluster of each vertex of $T(N,f)$
consists of the leaves with numbers in an interval, and compute for each cluster $C$ of $T$, bottom-up in
$T$, its size and the least and the greatest number of a leaf in $C$. Then $C$ is a cluster of $T(N,f)$ if
and only if the greatest number minus the least number is $|C|-1$ and the interval between them is the
interval of some vertex of $T(N,f)$; sorting the $O(n)$ intervals of both trees together by radix sort
answers all these tests in $O(|N|+|T|)$ time. Since $k\le d_\Sigma$, the total is
$O(|N|+|T|+nd_\Sigma^{2})$, and $d_\Sigma\le kd$ gives the second bound.
\end{proof}

\begin{corollary}\label{cor:binarycase}
Let $N$ be binary. For each reticulation $r$, write $\mathrm{par}(r)=\{p^{0}_r,p^{1}_r\}$, let $x_r$ be a
Boolean variable, and identify a switching $f$ with the assignment that sets $x_r=i$ when $f(r)=p^{i}_r$.
Then, for every cluster $C$ with a host, $\mathrm{realized}(C)$ is the set of models of the clauses of at
most two literals that are satisfied by every element of $\mathrm{realized}(C)$. Hence $N$ softly displays
$T$ if and only if every cluster of $T$ has a host and the $2$-SAT instance formed by these clauses, over
all clusters $C$ of $T$ with $2\le|C|\le n-1$, is satisfiable. On a binary network,
Algorithm~\ref{alg:case3} runs in $O(|N|+|T|+nk^{2})$ time.
\end{corollary}

\begin{proof}
A subset of $\{0,1\}$ is the set of models of at most one clause of one literal, and a subset of
$\{0,1\}^{2}$ is the set of models of at most four clauses of two literals, one excluding each missing
pair. Hence $\mathrm{proj}_r(\mathrm{realized}(C))$ and $\mathrm{proj}_{r,r'}(\mathrm{realized}(C))$ are the
sets of models of clauses of at most two literals that every element of $\mathrm{realized}(C)$ satisfies,
and by Theorem~\ref{thm:2decomp} the common models of these clauses are exactly the elements of
$\mathrm{realized}(C)$. Every other clause of at most two literals satisfied by all of
$\mathrm{realized}(C)$ keeps the set of common models unchanged. This proves the first claim. The second
claim follows from Lemma~\ref{thm:decomp}, because the models of the instance are the switchings in
$\bigcap_C\mathrm{realized}(C)$ when every cluster has a host. The bound follows from
Theorem~\ref{thm:case3}, since $d_\Sigma=2k$.
\end{proof}

Soft Tree Containment on binary tree-child networks was known to be polynomial. Bentert, Mal\'{\i}k, and
Weller decide it on reticulation-visible networks of reticulation indegree at most two in $O(n^{3})$
time~\cite[Corollary~25]{bmw2018}, by a reduction to soft containment on $2$-labelled
trees~\cite[Theorem~24]{bmw2018}. The journal version~\cite{bmw2021} solves that problem through $O(n)$
separate $2$-SAT instances of size $O(n^{2})$, whose variables encode an embedding of $T$ into
$N$~\cite[Theorem~4 and Construction~3, at label-multiplicity two]{bmw2021}, and it states polynomial-time
solvability at indegree two, but not as a numbered result. Their notion of soft display coincides with
ours here, because $N$ is binary (Section~\ref{sec:sym}). The $2$-SAT instance of
Corollary~\ref{cor:binarycase} has one variable per reticulation and $O(k^{2})$ distinct clauses. Since
distinct reticulations have distinct minimum taxa, none of which is $\pi_1$ (Lemma~\ref{lem:dichotomy}),
$k\le n-1$. Thus the bound $O(|N|+|T|+nk^{2})$ is never larger than $O(n^{3})$, and it is smaller when
$k=o(n)$.

On a positive answer, Algorithms~\ref{alg:case1} and~\ref{alg:case3} return a switching $f$, which is the
output that an application such as PhyloFusion needs: $T(N,f)$, computed in $O(|N|)$ time, refines $T$ and
is firmly displayed by $N$, and the edges kept by $f$ give the embedding.

\section{The symmetric variant}\label{sec:sym}

The soft display of the previous sections is asymmetric: the polytomies of $T$ may be resolved by $N$,
but those of $N$ may not (Section~\ref{sec:prelim}). Bentert et al.~\cite{bmw2018,bmw2021} study the
symmetric variant, in which the polytomies of both $T$ and $N$ may be resolved. In this section, we show
that the method of Section~\ref{sec:case3} also decides the symmetric variant on tree-child networks in
polynomial time, for reticulations of any indegree. The switchings that satisfy the condition of one
cluster have a simpler description than in Section~\ref{sec:case3}, and Lemma~\ref{cor:hostrow} is not
needed. The proofs omitted from this section are given in
Appendix~\ref{app:resolutions}.

\subsection{Binary resolutions}

A {\it binary resolution} of an RPN $M$ on $X$ is an RPN $M_B$ on $X$ whose tree nodes and reticulations
are binary, together with a set of edges of $M_B$ whose contraction, that is, the identification of the
two endpoints of each of these edges, yields $M$. The remaining edges of $M_B$ are in bijection with the
edges of $M$, and each leaf of $M_B$ contracts to the leaf of $M$ with the same label. For a vertex $u$ of
$M_B$, let $\kappa(u)$ be the vertex of $M$ that $u$ contracts to. For a vertex $v$ of $M$, the set
$\kappa^{-1}(v)$ is called the {\it block} of $v$. It induces a weakly connected subgraph of $M_B$, and
the edges entering and leaving it correspond, under the bijection, to the incoming and outgoing edges of
$v$. Bentert et al.\ define resolutions through the map $\kappa$ alone and observe that no block contains
both a reticulation and a tree node~\cite[Observation~1]{bmw2018}. We take this observation as part of
their definition; Appendix~\ref{app:resolutions} explains why their results need it. The network $M_B$
need not be tree-child even when $M$ is. Hence, for a switching $f_B$ of $M_B$, some vertices of
$f_B(M_B)$ may reach no leaf; the definition of $T(M_B,f_B)$ in Section~\ref{sec:prelim} deletes them,
and $\mathcal{C}(T(M_B,f_B))$ is the set of nonempty sets $L_{f_B}(u)$, $u\in V(M_B)$.

An RPN $N$ {\it symmetrically softly displays} a phylogenetic tree $T$ on $X$ if $T(N_B,f_B)\cong T_B$
for some binary resolutions $N_B$ of $N$ and $T_B$ of $T$ and some switching $f_B$ of $N_B$. The next
lemma passes from a switching of a resolution of $N$ to a switching of $N$.

\begin{lemma}\label{lem:resolveswitch}
Let $N$ be an RPN on $X$, let $N_B$ be a binary resolution of $N$, and let $f_B$ be a switching of
$N_B$. Then some switching $f$ of $N$ satisfies $\mathcal{C}(T(N,f))\subseteq\mathcal{C}(T(N_B,f_B))$.
\end{lemma}

If all tree nodes of $N$ are binary, then symmetric soft display coincides with the condition of
Theorem~\ref{thm:star}, that is, with soft display. The next proposition uses this to transfer the
hardness result of Bentert et al.~\cite[Corollary~20]{bmw2018}, \cite[Corollary~3]{bmw2021}.

\begin{proposition}[Hardness beyond tree-childness]\label{prop:hardness}
Deciding whether some switching $f$ satisfies $\mathcal{C}(T)\subseteq\mathcal{C}(T(N,f))$, that is, whether $N$ softly
displays $T$ (Theorem~\ref{thm:star}), is NP-hard on single-labelled
reticulation-visible networks with binary tree nodes and maximum reticulation indegree three.
\end{proposition}

\subsection{Reduction to clusters}

In the rest of this section, $N$ is a tree-child network on $X$, $T$ is a phylogenetic tree on $X$, and
$n=|X|$. Let $f$ be a switching of $N$. Every vertex $v$ reaches a leaf along tree edges
(Lemma~\ref{lem:committed}), and $f(N)$ contains all tree edges, so $L_f(v)\neq\varnothing$. By
Lemma~\ref{thm:decomp}, the cluster set of $T(N,f)$ is $\{L_f(v):v\in V(N)\}$.

Lemma~\ref{lem:resolveswitch} passes from a resolution of $N$ to $N$. The next lemma passes from $N$ to a
resolution.

\begin{lemma}\label{lem:symresolve}
Let $N$ be a tree-child network on $X$, let $f$ be a switching of $N$, and let $R$ be a binary
phylogenetic tree on $X$ with $\mathcal{C}(T(N,f))\subseteq\mathcal{C}(R)$. Then $T(N_B,f_B)=R$ for some
binary resolution $N_B$ of $N$ and some switching $f_B$ of $N_B$.
\end{lemma}

\begin{proposition}[Symmetric cluster characterization]\label{prop:sym}
A tree-child network $N$ symmetrically softly displays a phylogenetic tree $T$ on $X$ if and only if
$\mathcal{C}(T)\cup\mathcal{C}(T(N,f))$ is laminar for some switching $f$ of $N$.
\end{proposition}

Proposition~\ref{prop:sym} also holds for the resolutions of Bentert et al.: Lemma~\ref{lem:resolveswitch}
applies to their resolutions, and the resolution constructed in the proof of Lemma~\ref{lem:symresolve}
satisfies their observation and is minimal in their sense (Lemma~\ref{lem:minimal} in
Appendix~\ref{app:resolutions}), so it is one of their resolutions. Hence the two definitions of symmetric soft display agree on tree-child networks.

A set $A\subseteq X$ is {\it compatible} with a phylogenetic tree $R$ on $X$ if every cluster of $R$ is
contained in $A$, contains $A$, or is disjoint from $A$. If $\mathcal{C}(T)\subseteq\mathcal{C}(T(N,f))$,
then every cluster of $T$ is compatible with $T(N,f)$, since the clusters of a tree are laminar. The
symmetric variant requires only the latter, and this requirement splits into one condition per cluster
of $T$.

\begin{corollary}\label{cor:symdecomp}
For $A\subseteq X$, let $\mathrm{compat}(A)$ be the set of switchings $f$ of $N$ such that $A$ is
compatible with $T(N,f)$. Then $N$ symmetrically softly displays $T$ if and only if
$\bigcap_{A}\mathrm{compat}(A)\neq\varnothing$, where $A$ ranges over the clusters of $T$ with
$2\le|A|\le n-1$.
\end{corollary}

\begin{proof}
Let $f$ be a switching of $N$. The families $\mathcal{C}(T)$ and $\mathcal{C}(T(N,f))$ are laminar. Hence
their union is laminar if and only if every cluster of $T$ is compatible with $T(N,f)$, that is, if and
only if $f\in\mathrm{compat}(A)$ for every cluster $A$ of $T$. The clusters $X$ and $\{x\}$, $x\in X$, of
$T$ are compatible with every phylogenetic tree on $X$. The claim now follows from
Proposition~\ref{prop:sym}.
\end{proof}

\subsection{Hosts of a cluster}

Fix $A\subseteq X$ with $2\le|A|\le n-1$. For a vertex $v$, let $K_0(v)$ and $L(K_0(v))$ be as in
Section~\ref{sec:case3}: $K_0(v)$ is the set of vertices reachable from $v$ along tree edges, and
$L(K_0(v))$ is the set of taxa of the leaves in $K_0(v)$. The set $L(K_0(v))$ is nonempty
(Lemma~\ref{lem:committed}), and $L(K_0(v))\subseteq L_f(v)$ for every switching $f$. We give every vertex
$v$ the status of $K_0(v)$, defined as in Section~\ref{sec:case3} with $A$ in place of $C$: $v$ is
{\it must-in} if $L(K_0(v))\subseteq A$, {\it must-out} if $L(K_0(v))\cap A=\varnothing$, and {\it mixed}
otherwise. For a reticulation $r$, $K_0(r)=\mathrm{comp}(r)$, so this agrees with the status of $r$ in
Section~\ref{sec:case3}. We write $\mathcal{I}$ and $\mathcal{O}$ for the sets of must-in and must-out
vertices. For a vertex $u$, let $\mathrm{tp}[u]$ be the set of vertices of the path of tree edges from the
root of $\mathrm{comp}(u)$ to $u$, including both ends, and let $\mathrm{tp}(u)=\mathrm{tp}[u]\setminus\{u\}$.
A vertex of indegree one lies in $\mathrm{tp}[u]$ only if its parent does. Let
\[
  \mathcal{I}_u=\mathcal{I}\setminus \mathrm{tp}[u].
\]
For a reticulation $r$, let
\[
  E_u(r)=\begin{cases}
    \mathrm{par}(r)\cap(\mathcal{I}_u\cup\{u\}), & r\in\mathcal{I}_u,\\
    \mathrm{par}(r)\setminus\mathcal{I}_u, & r\in\mathcal{O},\ r\neq u,\\
    \mathrm{par}(r), & \text{otherwise.}
  \end{cases}
\]
In the third case, $r=u$, or $r$ is mixed, or $r$ is must-in and is the root of $\mathrm{comp}(u)$.
Let $B_u=\prod_r E_u(r)$. We call $u$ a {\it host} of $A$, and write $u\in U(A)$, if the following three
conditions hold.
\begin{itemize}
\item[(s1)] Every mixed vertex lies in $\mathrm{tp}[u]$.
\item[(s2)] No vertex of $\mathcal{I}_u$ is the root, and the parent of every vertex of $\mathcal{I}_u$
of indegree one lies in $\mathcal{I}_u\cup\{u\}$.
\item[(s3)] $E_u(r)\neq\varnothing$ for every reticulation $r$.
\end{itemize}
Condition (s1) follows from (s2) (Lemma~\ref{lem:s1froms2} in Appendix~\ref{app:resolutions}); it is
listed separately because the proofs and the algorithm use it directly.

\begin{lemma}\label{lem:symreach}
Let $u\in U(A)$. Then $x\prec u$ for every $x\in\mathcal{I}_u$, and
$\mathrm{par}(u)\cap\mathcal{I}_u=\varnothing$. Moreover, if $u$ is not must-in, then $u$ is a tree node.
\end{lemma}

\begin{theorem}[Symmetric carve]\label{thm:symcarve}
Let $A\subseteq X$ with $2\le|A|\le n-1$. Then $\mathrm{compat}(A)=\bigcup_{u\in U(A)}B_u$.
\end{theorem}

Theorem~\ref{thm:symcarve} differs from Lemma~\ref{lem:carve} in two ways. First, the sets $E_u(r)$ are
determined by $N$ and $A$ alone, whereas the sets $A_w(r)$ of Section~\ref{sec:case3} depend on the set
$R(w)$, which is defined through reachability in $N$. Second, the hosts of $A$ need not be comparable.
For example, let $N$ be the tree-child network with root $\rho$, tree nodes $t$, $p$ and $q$,
reticulations $r_1$ and $r_2$, leaves $x_p$, $x_q$, $a_1$ and $a_2$, and edges $(\rho,t)$, $(t,p)$,
$(t,q)$, $(p,x_p)$, $(q,x_q)$, $(p,r_i)$, $(q,r_i)$ and $(r_i,a_i)$ for $i=1,2$. Then $p$ and $q$ are
incomparable, and both are hosts of $A=\{a_1,a_2\}$. Hence Lemma~\ref{cor:hostrow} has no counterpart
here, and the algorithm relies on min-closure alone.

\subsection{Min-closure and the algorithm}

Let $\wedge$ be the operation on switchings defined in Section~\ref{sec:case3}: $(f\wedge g)(r)$ is the
deeper of $f(r)$ and $g(r)$ for every reticulation $r$.

\begin{lemma}\label{lem:symminclosed}
Let $A\subseteq X$ with $2\le|A|\le n-1$. For all $f,g\in\mathrm{compat}(A)$, $f\wedge g\in\mathrm{compat}(A)$.
\end{lemma}

The constraint network $\mathcal{N}'$ has a variable for each reticulation $r$, with domain
$\mathrm{par}(r)$, and, for each cluster $A$ of $T$ with $2\le|A|\le n-1$, the constraint
$\mathrm{compat}(A)$, which involves all variables. A {\it solution} of $\mathcal{N}'$ is a switching that
lies in $\mathrm{compat}(A)$ for every such $A$. By Corollary~\ref{cor:symdecomp}, $N$ symmetrically softly
displays $T$ if and only if $\mathcal{N}'$ has a solution. {\it Generalized arc consistency} is enforced
on $\mathcal{N}'$ by removing from the domain of a variable $r$ every value $p$ for which some constraint
$\mathrm{compat}(A)$ contains no switching $g$ with $g(r)=p$ and $g(r')$ in the current domain of $r'$
for every reticulation $r'$, until no value can be removed. A removed value belongs to no solution, so
the solutions do not change. By Theorem~\ref{thm:symcarve}, such a switching $g$ exists if and only if
some host $u\in U(A)$ satisfies $p\in E_u(r)$ and $E_u(r')$ meets the current domain of $r'$ for every
$r'\neq r$. Algorithm~\ref{alg:sym} decides symmetric soft display.

\begin{algorithm}[H]
\caption{Symmetric Soft Tree Containment for a tree-child network and a phylogenetic tree.}
\label{alg:sym}
\begin{algorithmic}%
\State {\bf Input}: A tree-child network $N$ and a phylogenetic tree $T$ on the same taxon set $X$.
\State {\bf Output}: `Yes' and a switching $f$ such that $\mathcal{C}(T)\cup\mathcal{C}(T(N,f))$ is laminar if $N$ symmetrically softly displays $T$, and `No' otherwise.\\
\State compute a topological order of $V(N)$, the tree components of $N$, and the intervals $J(v)$ of Lemma~\ref{lem:preproc}(T1);
\State $\mathcal{N}'\gets$ the constraint network with a variable $r$ of domain $\mathrm{par}(r)$ for each reticulation $r$, and no constraints;
\For{each cluster $A$ of $T$ with $2\le|A|\le n-1$}
  \State compute the status of every vertex, $U(A)$, and the sets $E_u(r)$ for $u\in U(A)$; \Comment{{\tt Appendix~\ref{app:resolutions}}}
  \State {\bf if} $U(A)=\varnothing$ {\bf then} return `No'; \Comment{{\tt $\mathrm{compat}(A)=\varnothing$ by Theorem~\ref{thm:symcarve}}}
  \State add the constraint $\mathrm{compat}(A)=\bigcup_{u\in U(A)}B_u$ to $\mathcal{N}'$;
\EndFor
\State enforce generalized arc consistency on $\mathcal{N}'$;
\State {\bf if} some domain is empty {\bf then} return `No'; \Comment{{\tt $\bigcap_A\mathrm{compat}(A)=\varnothing$}}
\State $f(r)\gets$ the deepest value in the domain of $r$, for each reticulation $r$; \Comment{{\tt Lemma~\ref{lem:symminclosed}}}
\State compute $T(N,f)$; {\bf if} $\mathcal{C}(T)\cup\mathcal{C}(T(N,f))$ is not laminar {\bf then} return `No'; \Comment{{\tt certifies $f$; never happens}}
\State return `Yes' and $f$.
\end{algorithmic}
\end{algorithm}

\begin{theorem}[Correctness of Algorithm~\ref{alg:sym}]\label{thm:symcase}
For a tree-child network $N$ and a phylogenetic tree $T$ on $X$, Algorithm~\ref{alg:sym} returns `Yes'
if and only if $N$ symmetrically softly displays $T$, and then $\mathcal{C}(T)\cup\mathcal{C}(T(N,f))$ is
laminar for the switching $f$ it returns. It runs in $O(|N|+|T|+n\,|N|\,(d_{\Sigma}{+}1))$ time.
\end{theorem}

\begin{proof}
If the algorithm returns `Yes' and $f$, then it has verified that $\mathcal{C}(T)\cup\mathcal{C}(T(N,f))$
is laminar, and $N$ symmetrically softly displays $T$ by Proposition~\ref{prop:sym}. Conversely, suppose
that $N$ symmetrically softly displays $T$. By Corollary~\ref{cor:symdecomp}, some switching $g$ lies in
$\mathrm{compat}(A)$ for every cluster $A$ of $T$ with $2\le|A|\le n-1$. By Theorem~\ref{thm:symcarve},
every such $A$ has a host, so the algorithm does not stop in the loop, and $g$ is a solution of
$\mathcal{N}'$. Generalized arc consistency removes no value of a solution, so no domain becomes empty.
Let $D(r)$ be the domain of $r$ after generalized arc consistency has been enforced, and let $f$ be the
switching chosen by the algorithm, so that $f(r)$ is the deepest value of $D(r)$. We show that
$f\in\mathrm{compat}(A)$ for every cluster $A$ of $T$ with $2\le|A|\le n-1$. Fix such an $A$. If $N$ has
no reticulation, then $f=g$. Otherwise, for each reticulation $r$, the value $f(r)$ was not removed from
the domain of $r$, so some $g_r\in\mathrm{compat}(A)$ satisfies $g_r(r)=f(r)$ and $g_r(r')\in D(r')$ for
every $r'$. Let $h$ be the switching such that $h(r')$ is the deepest of the values $g_r(r')$, $r\in R(N)$,
for every $r'$. It is obtained from the switchings $g_r$ by repeated application of $\wedge$, so
$h\in\mathrm{compat}(A)$ by Lemma~\ref{lem:symminclosed}. For every $r'$, the values $g_r(r')$ lie in
$D(r')$, and one of them, $g_{r'}(r')=f(r')$, is the deepest value of $D(r')$; hence $h(r')=f(r')$. So
$f=h\in\mathrm{compat}(A)$. This is the argument of Theorem~\ref{thm:minclosed}, for constraints on all
variables~\cite{jeavonscooper1995,dalmau1999}. Thus every cluster of $T$ is compatible with $T(N,f)$,
the clusters $X$ and $\{x\}$ being compatible with every tree, and $\mathcal{C}(T)\cup\mathcal{C}(T(N,f))$
is laminar, as in the proof of Corollary~\ref{cor:symdecomp}. Hence the algorithm returns `Yes'. The
running time is analysed in Appendix~\ref{app:resolutions}.
\end{proof}

Counting the edges of $N$ by their heads and by their tails gives $n+t+d_{\Sigma}=|E(N)|\ge 1+k+2t$,
where $t$ is the number of tree nodes. Hence $|N|=O(n+d_{\Sigma})$, and the bound of
Theorem~\ref{thm:symcase} is $O(n^{2}(d_{\Sigma}{+}1)+n\,d_{\Sigma}^{2})$. The second term is the bound
of Theorem~\ref{thm:case3}; the first comes from testing every vertex as a host of every cluster, whereas
Algorithm~\ref{alg:case3} tests only the candidate hosts of Lemma~\ref{lem:signature}.

\begin{remark}[Single labelling]\label{rem:symsingle}
The proof of Theorem~\ref{thm:symcarve} uses that each taxon labels exactly one leaf
(Section~\ref{sec:prelim}): from $a\in A\subseteq L_f(u)$ it passes to the path from $u$ to $\ell_a$ in
$f(N)$. This assumption cannot be dropped: symmetric soft Tree Containment is NP-hard on binary
tree-child networks in which a taxon may label two leaves~\cite[Corollary~5]{bmw2021}.
\end{remark}

\section{Discussion}\label{sec:discussion}

Since the cluster characterization of Theorem~\ref{thm:star} holds on all networks, the tractability shown
here comes from the tree-childness property: the canonical order places the vertices that can host a given cluster on a
single principal path (Lemma~\ref{cor:hostrow}), and the interval and min-closure arguments of
Section~\ref{sec:case3} turn this into an algorithm. The networks built by
PhyloFusion are tree-child~\cite[Supplementary Proposition~C.7]{phylofusion}, so
Section~\ref{sec:case3} gives an exact soft-display test for all networks this method produces.
Reference implementations in Python of the algorithms of Sections~\ref{sec:case1}--\ref{sec:sym} are
available at \url{https://github.com/ErdemUgurlu/Soft-TC-Containment}.

Bruchhold and Weller~\cite{bruchholdweller} give a fixed-parameter algorithm for soft Tree Containment
on arbitrary networks (Theorem~1), in $2^{O(\Delta_T s\log s)}n^{O(1)}$ time. A tree extension of $N$ is given
as part of the input, $s$ is its scanwidth plus the maximum outdegree of $N$, and $\Delta_T$ is the
maximum outdegree of $T$.
Their notion is symmetric, but built on a split-based notion of binary resolution. Their algorithm
applies to all networks and is exponential in the scanwidth and in the outdegrees. Ours is confined to
tree-child networks and runs in polynomial time with no parameter. Removing the superpolynomial
dependence on the outdegrees is left open in~\cite{bruchholdweller}.  On tree-child networks our
algorithms achieve it. Whether soft Tree Containment is polynomial-time solvable for any superclass of tree-child networks is an interesting problem for future research.

\section*{Acknowledgment}
The authors thank Yufeng Wu and Daniel Huson for helpful discussions on the tree containment problem. L. X. Zhang was supported by the Singapore Ministry of Education (MOE) Academic Research Fund (AcRF) Tier 1 grant A-8001951-00-00.

\FloatBarrier

\appendix

\section{Proofs for Section~\ref{sec:prelim}}\label{app:prelim}

\begin{proof}[Proof of Lemma~\ref{lem:faithful}]
A phylogenetic tree $T$ on $X$ is a tree-child network, every order $\pi$ of $X$ is canonical for $T$,
and the LTH encoding of $T$ is its encoding as a tree-child network (Section~\ref{sec:prelim}); so
Lemma~\ref{lem:dichotomy} applies to $T$ and $\pi$. Since a phylogenetic tree is determined by its
cluster set (Section~\ref{sec:prelim}), it suffices to show that the encoding determines $\mathcal{C}(T)$.

Let $\tau\in X$, and let $v_1,\dots,v_p$ be the tree nodes on $\mathrm{path}(\tau)$, from top to bottom,
so that $\mathrm{LTH}(\tau)=(\lambda_\pi(v_1),\dots,\lambda_\pi(v_p))$. As $T$ has no reticulations, the
vertices of $\mathrm{path}(\tau)$ below $v_1$ are $v_2,\dots,v_p,\ell_\tau$. Let $1\le i\le p$. The child
of $v_i$ on $\mathrm{path}(\tau)$ has minimum taxon $\tau$ (Lemma~\ref{lem:dichotomy}(i)); the other
children of $v_i$ have distinct minimum taxa, as their clusters are disjoint, and these are the elements
of $\lambda_\pi(v_i)$. For $a\in\lambda_\pi(v_i)$, the only occurrence of $a$ in the encoding is in the
label of the parent of $\mathrm{top}(a)$ (Lemma~\ref{lem:dichotomy}(ii) and (iv)), so $\mathrm{top}(a)$ is the child of
$v_i$ with minimum taxon $a$. Since the cluster of a vertex is the union of the clusters of its children,
induction on $p-i$ gives
\begin{equation}\label{eq:decode}
  L(v_i)=\{\tau\}\cup\bigcup\bigl\{\,L(\mathrm{top}(a)) : a\in\lambda_\pi(v_i)\cup\dots\cup\lambda_\pi(v_p)\,\bigr\}.
\end{equation}
Here $L(\mathrm{top}(a))=\{a\}$ if $\mathrm{LTH}(a)$ is empty, since then $\mathrm{top}(a)=\ell_a$; otherwise $\mathrm{top}(a)$
is the first tree node of $\mathrm{path}(a)$, and $L(\mathrm{top}(a))$ is given by (\ref{eq:decode}) for $a$ and
$i=1$. Moreover, $\tau<_\pi a$ for every $a\in\lambda_\pi(v_i)$, as $\tau=m_\pi(v_i)$ is the least
minimum taxon of a child of $v_i$. Hence, processing the taxa in the order $\pi_n,\dots,\pi_1$, all sets
$L(v_i)$ are computed from the encoding by (\ref{eq:decode}). Every tree node $v$ of $T$ lies on
$\mathrm{path}(m_\pi(v))$ (Lemma~\ref{lem:dichotomy}(i)), so these sets are the clusters of the tree
nodes of $T$. The other clusters of $T$ are $X$ and the singletons. Hence the encoding determines
$\mathcal{C}(T)$.
\end{proof}

\begin{proof}[Proof of Lemma~\ref{lem:dichotomy}]
Let $V_\tau$ denote the set of vertices of $N$ with minimum taxon $\tau$.

(i) Every vertex of $\mathrm{path}(\tau)$ lies in $V_\tau$: the leaf $\ell_\tau$ does, and the process that
defines $\mathrm{path}(\tau)$ moves from the current vertex to its parent only if the parent has minimum taxon
$\tau$. Conversely, let $v\in V_\tau$. Starting at $x_0=v$, as long as the current vertex $x_i$ is not
$\ell_\tau$, let $x_{i+1}$ be a child of $x_i$ with $\tau\in L(x_{i+1})$; such a child exists, because
$\tau\in L(x_i)$ and $L(x_i)$ is the union of the clusters of the children of $x_i$. Since
$L(x_{i+1})\subseteq L(x_i)$ and $\tau\in L(x_{i+1})$, we have $m_\pi(x_{i+1})=m_\pi(x_i)=\tau$. This yields a
directed path $v=x_0\to x_1\to\dots\to x_s=\ell_\tau$ whose vertices all lie in $V_\tau$. No $x_i$ with
$i\ge1$ is a reticulation, since otherwise Lemma~\ref{lem:canonical}(ii) would give
$m_\pi(x_{i-1})<_\pi m_\pi(x_i)$. So each $x_i$ with $i\ge1$ has indegree one, and its parent is $x_{i-1}$,
which has minimum taxon $\tau$. Hence the process that defines $\mathrm{path}(\tau)$, which starts at
$x_s=\ell_\tau$, moves from $x_i$ to $x_{i-1}$ for $i=s,s-1,\dots,1$, and $v=x_0$ lies on
$\mathrm{path}(\tau)$. Thus $V_\tau$ is the vertex set of $\mathrm{path}(\tau)$. Finally, the process moves
from a vertex to its parent only if the vertex has indegree one, so every vertex of $\mathrm{path}(\tau)$
other than $\mathrm{top}(\tau)$ has indegree one, and only $\mathrm{top}(\tau)$ can be a reticulation.

(ii)--(iv) We first show that the tree nodes whose labels contain $\tau$ are exactly the parents of
$\mathrm{top}(\tau)$. By the definition of the labels, $\tau\in\lambda_\pi(v)$ for a tree node $v$ if and only if $v$
has a child $c$ with $m_\pi(c)=\tau$ and $m_\pi(v)\neq\tau$. Let $v$ be such a tree node and $c$ such a child.
By (i), $c$ lies on $\mathrm{path}(\tau)$ and $v$ does not. If $c\neq\mathrm{top}(\tau)$, then $c$ has indegree one,
and its parent is the vertex preceding $c$ on $\mathrm{path}(\tau)$; but $v$ is a parent of $c$ and does not
lie on $\mathrm{path}(\tau)$. Hence $c=\mathrm{top}(\tau)$, and $v$ is a parent of $\mathrm{top}(\tau)$. Conversely, let $v$
be a parent of $\mathrm{top}(\tau)$. Then $m_\pi(v)\neq\tau$: if $\mathrm{top}(\tau)$ is a reticulation, this is
Lemma~\ref{lem:canonical}(ii); otherwise $\mathrm{top}(\tau)$ is a tree node or a leaf, and the process that defines
$\mathrm{path}(\tau)$ stopped at $\mathrm{top}(\tau)$ because its parent $v$ does not have minimum taxon $\tau$.
Moreover, $v$ is a tree node: the root and the reticulations have a single child $c$, for which $L(v)=L(c)$
and hence $m_\pi(v)=m_\pi(c)$, whereas here $m_\pi(v)\neq\tau=m_\pi(\mathrm{top}(\tau))$. Since $\mathrm{top}(\tau)$ is a
child of $v$ with minimum taxon $\tau\neq m_\pi(v)$, we get $\tau\in\lambda_\pi(v)$.

Next, the label of a tree node $v$ appears in the encoding of $N$ exactly once, in
$\mathrm{LTH}_{N,\pi}(m_\pi(v))$: by (i), $v$ lies on $\mathrm{path}(m_\pi(v))$ and on no other principal
path, as the vertices of $\mathrm{path}(\tau')$ have minimum taxon $\tau'$. Therefore $\tau$ occurs in the
encoding of $N$ once for each parent of $\mathrm{top}(\tau)$, in the label of that parent, and nowhere else. If
$\mathrm{top}(\tau)$ is a tree node or a leaf, it has exactly one parent, which proves (ii). If $\mathrm{top}(\tau)$ is a
reticulation $r$, its parents are the parents of $r$, which proves (iii). If $\mathrm{top}(\tau)$ is the root, it has
no parent, and $\tau$ does not occur in the encoding. For (iv) it remains to show that $\mathrm{top}(\tau)$ is the
root if and only if $\tau=\pi_1$. The root has minimum taxon $\pi_1$, so it lies on $\mathrm{path}(\pi_1)$ by
(i), and since it has no parent, it is $\mathrm{top}(\pi_1)$. Conversely, if $\mathrm{top}(\tau)$ is the root, then
$\tau=m_\pi(\mathrm{root})=\pi_1$.

Finally, if two distinct reticulations had the same minimum taxon $\tau$, both would lie on
$\mathrm{path}(\tau)$ and both would be $\mathrm{top}(\tau)$ by (i). Since $\mathrm{top}(\pi_1)$ is the root, no reticulation
has minimum taxon $\pi_1$.
\end{proof}

\section{The choice of contractible edges}\label{app:reading}

The main text of PhyloFusion~\cite{phylofusion} states the relation as follows: ``$T$ can be obtained from
$N$ by deleting nodes and edges, and/or contracting edges''. It does not say which edges may be contracted.
We read it as the internal tree edges, which means that a contracted edge $u\to v$ must have a head $v$ that is not
a leaf and has indegree $1$ in the current graph. Both restrictions are needed. If $v$ is the leaf
labelled by a taxon $\tau$, the contraction merges $u$ and $v$ into a vertex that carries $\tau$ and keeps
the other children of $u$, so $\tau$ labels an internal vertex and the result is not a phylogenetic tree on
$X$. If $v$ is a reticulation $r$ of indegree $d_r\ge 2$, the contraction merges $r$ into $u$, and the
merged vertex inherits the other $d_r-1$ incoming edges of $r$ on top of $u$'s own, which makes its indegree
exceed that of $u$. Hence the reticulation moves upward without disappearing, and the remaining parents of
$r$ become ancestors of all descendants of $u$. The proof of Lemma~\ref{lem:expansion} needs both
restrictions: the leaf one keeps all labels on leaves, and the indegree one makes the vertices merged into
a single vertex form a tree hanging from one top, leaving each reticulation with at most one incoming edge that
was not deleted, which is the parent a switching records. Without them a sequence does not have to correspond to any switching, and
the example below shows that the cluster characterization then fails.

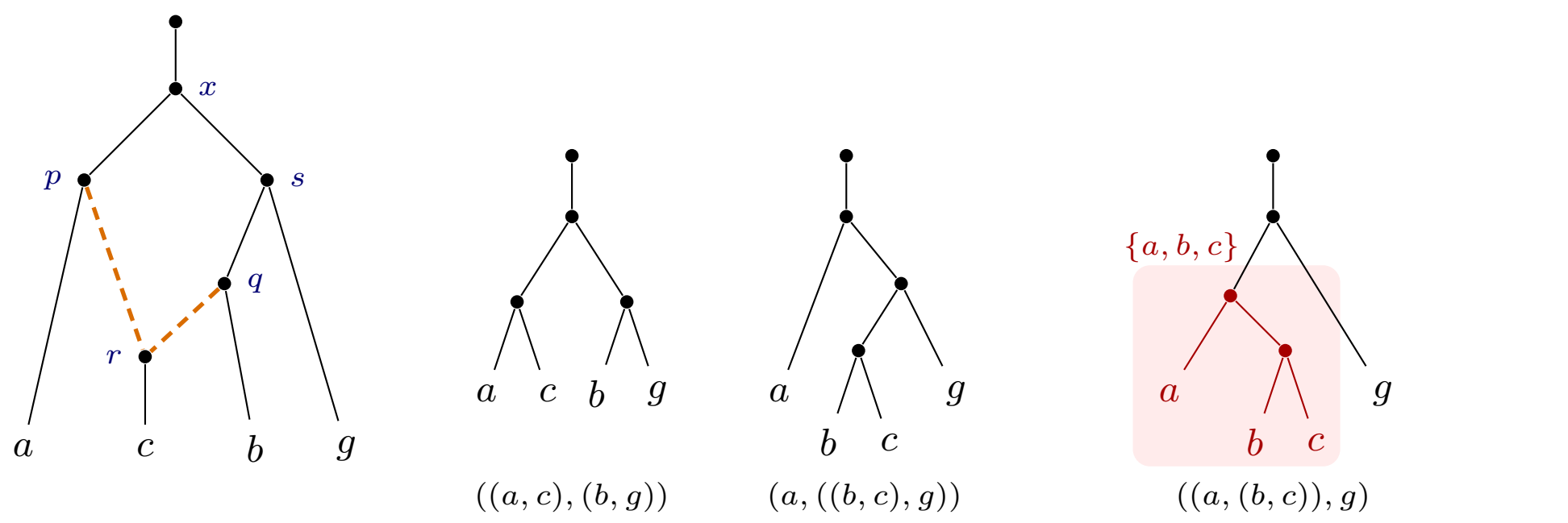
\begin{figure}[H]
\centering
\begin{tikzpicture}[
  int/.style={circle, fill, inner sep=1.1pt},
  lf/.style={font=\small},
  nm/.style={font=\scriptsize, text=blue!45!black},
  mo/.style={red!65!black},
  re/.style={line width=1.0pt, orange!85!black, densely dashed},
  cc/.style={font=\scriptsize}
]
\begin{scope}
  \node[int] (rr) at (0.9,4.5) {};
  \node[int, label={[nm]right:$x$}] (x) at (0.9,3.95) {};
  \node[int, label={[nm]left:$p$}] (p) at (0.15,3.2) {};
  \node[int, label={[nm]right:$s$}] (s) at (1.65,3.2) {};
  \node[int, label={[nm]right:$q$}] (q) at (1.3,2.35) {};
  \node[int, label={[nm]left:$r$}] (r) at (0.65,1.75) {};
  \node[lf] (la) at (-0.35,1.0) {$a$};
  \node[lf] (lc) at (0.65,1.0) {$c$};
  \node[lf] (lb) at (1.55,1.0) {$b$};
  \node[lf] (lg) at (2.3,1.0) {$g$};
  \draw (rr)--(x); \draw (x)--(p); \draw (x)--(s);
  \draw (p)--(la); \draw (s)--(q); \draw (s)--(lg);
  \draw (q)--(lb); \draw (r)--(lc);
  \draw[re] (p)--(r); \draw[re] (q)--(r);
  \node[font=\small] at (0.9,0.1) {(a) the network $N$};
\end{scope}
\begin{scope}[xshift=3.6cm]
  \node[int] (b1r) at (0.55,3.4) {};
  \node[int] (b1v) at (0.55,2.9) {};
  \node[int] (b1m) at (0.1,2.2) {};
  \node[int] (b1n) at (1.0,2.2) {};
  \node[lf] (b1a) at (-0.15,1.45) {$a$};
  \node[lf] (b1c) at (0.35,1.45) {$c$};
  \node[lf] (b1b) at (0.75,1.45) {$b$};
  \node[lf] (b1g) at (1.25,1.45) {$g$};
  \draw (b1r)--(b1v); \draw (b1v)--(b1m); \draw (b1v)--(b1n);
  \draw (b1m)--(b1a); \draw (b1m)--(b1c); \draw (b1n)--(b1b); \draw (b1n)--(b1g);
  \node[cc] at (0.55,0.6) {$((a,c),(b,g))$};
\end{scope}
\begin{scope}[xshift=5.8cm]
  \node[int] (b2r) at (0.6,3.4) {};
  \node[int] (b2v) at (0.6,2.9) {};
  \node[lf] (b2a) at (0.05,1.45) {$a$};
  \node[int] (b2u) at (1.05,2.35) {};
  \node[int] (b2w) at (0.7,1.8) {};
  \node[lf] (b2b) at (0.45,1.05) {$b$};
  \node[lf] (b2c) at (0.95,1.05) {$c$};
  \node[lf] (b2g) at (1.5,1.45) {$g$};
  \draw (b2r)--(b2v); \draw (b2v)--(b2a); \draw (b2v)--(b2u);
  \draw (b2u)--(b2w); \draw (b2u)--(b2g);
  \draw (b2w)--(b2b); \draw (b2w)--(b2c);
  \node[cc] at (0.75,0.6) {$(a,((b,c),g))$};
\end{scope}
\node[font=\small] at (4.95,0.1) {(b) the two displayed trees};
\begin{scope}[xshift=9.3cm]
  \fill[red!8, rounded corners] (-0.55,0.85) rectangle (1.15,2.5);
  \node[int] (cr) at (0.6,3.4) {};
  \node[int] (cv) at (0.6,2.9) {};
  \node[int, mo] (cw1) at (0.25,2.25) {};
  \node[lf, mo] (ca) at (-0.25,1.45) {$a$};
  \node[int, mo] (cw2) at (0.7,1.8) {};
  \node[lf, mo] (cb) at (0.45,1.05) {$b$};
  \node[lf, mo] (cc2) at (0.95,1.05) {$c$};
  \node[lf] (cg) at (1.5,1.45) {$g$};
  \draw (cr)--(cv); \draw (cv)--(cw1); \draw (cv)--(cg);
  \draw[mo] (cw1)--(ca); \draw[mo] (cw1)--(cw2); \draw[mo] (cw2)--(cb); \draw[mo] (cw2)--(cc2);
  \node[cc, mo] at (-0.15,2.65) {$\{a,b,c\}$};
  \node[cc] at (0.6,0.6) {$((a,(b,c)),g)$};
  \node[font=\small] at (0.6,0.1) {(c) the unrestricted contraction};
\end{scope}
\end{tikzpicture}
\caption{The two readings on a tree-child network $N$ over $\{a,b,c,g\}$ with binary tree nodes and
one reticulation $r$ (parents $p,q$). (a)~$N$; the dashed edges are the incoming edges of $r$, of which
a switching keeps exactly one. (b)~The two switchings display $((a,c),(b,g))$ and
$(a,((b,c),g))$. (c)~The tree $((a,(b,c)),g)$ that the unrestricted reading reaches by the three
steps given in the text. Its cluster $\{a,b,c\}$ (shaded) is a cluster of neither displayed tree. That
reading thus accepts strictly more trees. The tree
$((a,(b,c)),g)$ is binary, so it is not soft-displayed under the refinement-based definition of
PhyloFusion's Supplementary Section~B either.}
\label{fig:reading}
\end{figure}

The unrestricted reading accepts more trees. This happens even on networks
in which all tree nodes are binary, where our notion and the symmetric one of
Section~\ref{sec:sym} agree (Figure~\ref{fig:reading}). Take the tree-child
network over $X=\{a,b,c,g\}$ given by $\mathrm{root}\to x$, $x\to p,s$,
$p\to r,a$, $s\to q,g$, $q\to r,b$ and $r\to c$. Its tree nodes are binary and $r$ is its only
reticulation, and its two switchings display $((a,c),(b,g))$ and $(a,((b,c),g))$. The unrestricted
reading, however, also allows this: contract $q\to r$, delete the edge from $s$ to the merged vertex, and
suppress $s$. The result is $((a,(b,c)),g)$, and its cluster $\{a,b,c\}$ is a cluster of neither
displayed tree.

The appendix of PhyloFusion resolves the same ambiguity from the other side. Its
Supplementary Section~B defines $T$ to be soft-displayed in $N$ when a refinement of $T$ is
displayed in $N$. Our definition agrees with it on all networks: by Theorem~\ref{thm:star}, a
refinement of $T$ is displayed if and only if some switching $f$ has
$\mathcal{C}(T)\subseteq\mathcal{C}(T(N,f))$. The unrestricted reading, on the other hand, accepts more trees. As the tree
$((a,(b,c)),g)$ is binary, it is soft-displayed in that sense only if some switching displays it
exactly, and none does.

\section{Proofs for Section~\ref{sec:char}}\label{app:char}

\begin{proof}[Proof of Lemma~\ref{lem:expansion}]
($\Leftarrow$) Delete the reticulation incoming edges that $f$ discards. Delete the leaves outside $X'$ and
then, repeatedly, any non-leaf vertex left with outdegree zero: this is the discard step of
Section~\ref{sec:prelim} together with the pruning of the restriction, all of it vertex deletions.
Suppress each vertex of indegree and outdegree one by contracting the edge into it, an internal tree edge,
as its head has indegree one and is not a leaf. The result is $T(N,f)|_{X'}$, and the prescribed
internal-edge contractions then produce $T$. Together these steps form a soft-display sequence.

($\Rightarrow$) Here a soft-display sequence $N=G_0,\dots,G_m\cong T$ is given (Section~\ref{sec:prelim}),
and our goal is to read a switching off its final tree. A vertex of $N$ is deleted when the
vertex of the current graph containing it is deleted. It then loses all its edges, contracted ones
included. Call an edge of $N$ deleted if an edge deletion or a vertex deletion removed it,
contracted if it was contracted, and surviving otherwise. We prove three claims.

Claim 1 (blocks). Because contractions merge vertices, a vertex of $N$ does not have to
survive as a vertex of $G_m$. We keep track of where it goes. Each vertex of $G_m$ is the merger of a set of vertices of
$N$, its block, joined by the contracted edges; a deleted vertex of $N$ lies in no block, and each
edge of $G_m$ is a surviving edge of $N$ joining two blocks. No edge of the current graph joins two
vertices of one block, since the graphs stay acyclic and a contraction creates no loop. An edge is
contracted only when its head has indegree one at that moment, in the multigraph count, and hence no vertex of
$N$ is the head of two contracted edges. The contracted edges of a block therefore form a tree directed
away from one vertex, the top of the block, which reaches all vertices of the block along
contracted edges. Moreover, when an edge $x\to y$ is contracted, $y$ has no other incoming edge left. Thus each
incoming edge of $y$ other than $x\to y$ was deleted earlier. A vertex that is not the top of its block has
exactly one non-deleted incoming edge, the contracted one. Hence a surviving edge into a block enters it at
the top, and the non-deleted incoming edges of a top are the incoming edges of its block in $G_m$: one, or none when
the block is the root of $G_m$.

Claim 2 (the switching). At a reticulation $r$ of $N$ with a non-deleted incoming edge, which is unique by
Claim~1, let $f(r)$ be the tail of that edge; at the other reticulations choose $f(r)$ arbitrarily. Then
$f$ retains every non-deleted edge of $N$: such an edge is a tree edge, or the unique non-deleted incoming edge
of a reticulation.

Claim 3 (the clusters). For every block $y$ with top $t$, $L_f(t)\cap X'=L_{G_m}(y)$. Each leaf of
$G_m$ carries a taxon of $X'$; the leaf of $N$ with that label survives, its label being lost otherwise,
and is a block by itself, since no edge into a leaf is contracted.
($\supseteq$) Let $x\in L_{G_m}(y)$ and let $y=y_0,y_1,\dots,y_j=\{x\}$ be the path of $G_m$ from $y$ to
the leaf $x$. The edge $y_i\to y_{i+1}$ is an edge of $N$ from a vertex $u_i$ of $y_i$ to the top of
$y_{i+1}$, and $u_i$ is reached from the top of $y_i$ along contracted edges (Claim~1). All these edges
are non-deleted, hence retained by $f$ (Claim~2), and together they form a path from $t$ to $x$ along
edges $f$ retains. Hence $x\in L_f(t)$.
($\subseteq$) Let $x\in L_f(t)\cap X'$, via a path $P$ from $t$ to $x$ along edges $f$ retains, and suppose
some edge of $P$ is deleted. Let $s$ be the first vertex of $P$ that survives and has its incoming edge on $P$
deleted; it exists, because $x$ survives and a deleted vertex loses its edges. Then $s\neq t$, and $s$ has
no non-deleted incoming edge at all: if $s$ has indegree one, its incoming edge is the deleted one; if $s$ is a
reticulation, $f(s)$ is its predecessor on $P$, which is not the tail of a non-deleted incoming edge of $s$;
by Claim~2, no such incoming edge exists. By Claim~1, $s$ is then the top of the root block of $G_m$. The block
of $t$ is a different one, since a block has one top and $t\neq s$. Thus it lies strictly below the root
block in the tree $G_m$. The argument of ($\supseteq$) turns the path of $G_m$ between them into a path of
$N$ from $s$ to $t$, which with the segment of $P$ from $t$ to $s$ closes a directed cycle in $N$, which
is impossible. This contradiction shows that no edge of $P$ is deleted. Every edge of $P$ is then contracted or an edge of $G_m$, and all vertices
of $P$ survive; read block by block, $P$ is a walk of $G_m$ from $y$ to the leaf $x$, giving
$x\in L_{G_m}(y)$.

By Claim~3 each cluster of $G_m\cong T$ is $L_f(t)\cap X'$ for the top $t$ of some block, and so a
cluster of $T(N,f)|_{X'}$, whose clusters are the nonempty sets $L_f(v)\cap X'$ (Section~\ref{sec:prelim}).
That is, $\mathcal{C}(T)\subseteq\mathcal{C}(T(N,f)|_{X'})$, which by the refinement characterization of
Section~\ref{sec:prelim} says that $T$ is a contraction of $T(N,f)|_{X'}$ by internal edges.
\end{proof}

\section{Proofs for Section~\ref{sec:case3}}\label{app:case3}

\begin{proof}[Proof of Lemma~\ref{lem:carve}]
We first record two facts about reachability. Let $f$ be a switching.

(F1) Every vertex of $K_0(w)$ is reachable from $w$ in $f(N)$, since tree edges are kept. Let $r$ be a
reticulation other than $w$. Every vertex of $\mathrm{comp}(r)$ other than $r$ has indegree one, and its
parent lies in $\mathrm{comp}(r)$, so a path that enters $\mathrm{comp}(r)$ from outside enters it at $r$.
Hence the vertices of $\mathrm{comp}(r)$ are reachable from $w$ in $f(N)$ if and only if $r$ is, and $r$
is reachable from $w$ in $f(N)$ if and only if $f(r)$ is, because $(f(r),r)$ is the only incoming edge of
$r$ in $f(N)$. Likewise, the vertices of $\mathrm{comp}(r)$ lie in $R(w)$ if and only if $r$ does, and
$r\in R(w)$ if and only if $r$ is must-in and some parent of $r$ lies in $R(w)$, because the incoming edges
of $r$ in $G_{\mathrm{in}}(C)$ are all of them if $r$ is must-in and none otherwise.

(F2) Every vertex $v\preceq w$ lies in $K_0(w)$ or in $\mathrm{comp}(r)$ for a reticulation $r$
constrained at $w$. Indeed, take a path from $w$ to $v$. If all its edges are tree edges, then
$v\in K_0(w)$. Otherwise let $(p,r)$ be its last reticulation edge; then $r$ is a strict descendant of $w$,
and the path continues from $r$ to $v$ along tree edges, so $v\in\mathrm{comp}(r)$.

(i) Suppose that $K_0(w)$ is not must-in. Then some leaf of $K_0(w)$ carries a taxon outside $C$, and this
taxon lies in $L_f(w)$ for every $f$ by (F1). Hence $L_f(w)\neq C$ for every $f$.

(ii) Suppose that $K_0(w)$ is must-in. We show that $f\in S_w(C)$ if and only if $f(r)\in A_w(r)$ for every
reticulation $r$.

Let $f\in S_w(C)$, that is, $L_f(w)=C$, and let $r$ be a reticulation. If $r$ is free at $w$, then
$f(r)\in\mathrm{par}(r)=A_w(r)$. Let $r$ be constrained at $w$. If $r$ were mixed, then $\mathrm{comp}(r)$
would have a leaf with taxon in $C$ and a leaf with taxon outside $C$, and by (F1) both or neither of these
taxa would lie in $L_f(w)$, contradicting $L_f(w)=C$. Suppose that $r$ is must-in. Then a leaf of
$\mathrm{comp}(r)$ has its taxon in $C=L_f(w)$, so $r$ and $f(r)$ are reachable from $w$ in $f(N)$ by (F1).
Let $P$ be a path from $w$ to $f(r)$ in $f(N)$. For every reticulation edge $(q,r')$ on $P$, the
reticulation $r'$ is reachable from $w$ in $f(N)$, so all taxa of $L(\mathrm{comp}(r'))$ lie in $L_f(w)=C$
by (F1); that is, $r'$ is must-in, and $(q,r')$ is an edge of $G_{\mathrm{in}}(C)$. Hence $P$ is a path of
$G_{\mathrm{in}}(C)$, $f(r)\in R(w)$, and $f(r)\in A_w(r)$. Suppose that $r$ is must-out. Then no leaf of
$\mathrm{comp}(r)$ has its taxon in $C=L_f(w)$, so $r$ and $f(r)$ are not reachable from $w$ in $f(N)$ by
(F1). Assume, for a contradiction, that $f(r)\in R(w)$. By (F2), $f(r)\in K_0(w)$ or
$f(r)\in\mathrm{comp}(r')$ for a reticulation $r'$ constrained at $w$. The first case is impossible by
(F1). In the second case, a path from $w$ to $f(r)$ in $G_{\mathrm{in}}(C)$ enters $\mathrm{comp}(r')$ at
$r'$ through an incoming edge of $r'$, so $r'$ is must-in. Then a leaf of $\mathrm{comp}(r')$ has its taxon
in $C=L_f(w)$, so $r'$ is reachable from $w$ in $f(N)$, and so is $f(r)\in\mathrm{comp}(r')$ by (F1), a
contradiction. Hence $f(r)\notin R(w)$, and $f(r)\in A_w(r)$.

Conversely, let $f(r)\in A_w(r)$ for every reticulation $r$. Then no reticulation constrained at $w$ is
mixed. We claim that a reticulation $r$ constrained at $w$ is reachable from $w$ in $f(N)$ if and only if
$r$ is must-in. We prove the claim by induction along a topological order of $V(N)$. Let $r$ be constrained
at $w$, and assume the claim for the reticulations constrained at $w$ that precede $r$. We first show that
$f(r)$ is reachable from $w$ in $f(N)$ if and only if $f(r)\in R(w)$. If $f(r)\not\preceq w$,
neither holds. If $f(r)\in K_0(w)$, both hold by (F1). Otherwise, by (F2), $f(r)\in\mathrm{comp}(r')$ for a
reticulation $r'$ constrained at $w$, and $r'$ precedes $r$, being an ancestor of $f(r)$. By (F1), $f(r)$
is reachable from $w$ in $f(N)$ if and only if $r'$ is, that is, by induction, if and only if $r'$ is
must-in. Also by (F1), $f(r)\in R(w)$ if and only if $r'\in R(w)$, which requires $r'$ to be must-in; and if
$r'$ is must-in, then $f(r')\in A_w(r')=\mathrm{par}(r')\cap R(w)$, so $r'\in R(w)$ by (F1). Hence
$f(r)\in R(w)$ if and only if $r'$ is must-in, which proves the equivalence. Now, if $r$ is must-in, then
$f(r)\in A_w(r)\subseteq R(w)$, so $f(r)$ and $r$ are reachable from $w$ in $f(N)$. If $r$ is must-out,
then $f(r)\in A_w(r)$ gives $f(r)\notin R(w)$, so $f(r)$ and $r$ are not reachable. This proves the claim.

By (F2), every leaf reachable from $w$ in $f(N)$ lies in $K_0(w)$ or in $\mathrm{comp}(r)$ for a
reticulation $r$ constrained at $w$, and by (F1) and the claim, the leaves of $\mathrm{comp}(r)$ are
reachable if and only if $r$ is must-in. So the taxa in $L_f(w)$ are those of $L(K_0(w))$ and of
$L(\mathrm{comp}(r))$ for the must-in reticulations $r$ constrained at $w$, and all of them lie in $C$.
Conversely, let $x\in C$. Since $C\subseteq L(w)$, we have $\ell_x\preceq w$, so by (F2)
it lies in $K_0(w)$ or in $\mathrm{comp}(r)$ for a reticulation $r$ constrained at $w$. In the second case,
$\mathrm{comp}(r)$ has a leaf with taxon in $C$ and is not mixed, so it is must-in, and $\ell_x$ is
reachable from $w$ in $f(N)$. In both cases $x\in L_f(w)$. Hence $L_f(w)=C$, and $f\in S_w(C)$.

Finally, $\prod_rA_w(r)$ is a box by definition, and a product of sets is nonempty if and only if every
factor is nonempty.
\end{proof}

\begin{proof}[Proof of Lemma~\ref{lem:signature}]
(i) Let $w$ be a host of $C$ and $f\in S_w(C)$. By Lemma~\ref{cor:hostrow}, $w$ lies on
$\mathrm{path}(\tau)$, so $w\in\mathrm{comp}(\tau)$ and $K_0(w)\subseteq\mathrm{comp}(\tau)$. The taxon of
a leaf of $K_0(w)$ lies in $L_f(w)=C$ by fact (F1) in the proof of Lemma~\ref{lem:carve}; hence
$L(K_0(w))\subseteq C\cap L(\mathrm{comp}(\tau))$. Conversely, let $x\in C\cap L(\mathrm{comp}(\tau))$.
Since $x\in L_f(w)$, we have $\ell_x\preceq w$, so by fact (F2) $\ell_x$ lies in $K_0(w)$ or in
$\mathrm{comp}(r)$ for a reticulation $r$ constrained at $w$. The second case is impossible: the root of
$\mathrm{comp}(w)$ is an ancestor of $w$ and hence not $r$, so $\mathrm{comp}(r)\neq\mathrm{comp}(w)=\mathrm{comp}(\tau)$,
while $\ell_x\in\mathrm{comp}(\tau)$. Hence
$\ell_x\in K_0(w)$, and $x\in L(K_0(w))$. This proves $L(K_0(w))=C\cap L(\mathrm{comp}(\tau))$.
Consequently, $I(w)$ is the set of ranks of the taxa in $C\cap L(\mathrm{comp}(\tau))$. As $I(w)$ is an
interval, it is the interval from the least to the greatest of these ranks, that is, $I(w)=I_C$, and
$|C\cap L(\mathrm{comp}(\tau))|=|I(w)|=|I_C|$. Since $w\in\mathrm{path}(\tau)$, $w$ is a candidate
host.

(ii) Let $v$ be a candidate host, so that $I(v)=I_C$. Since distinct taxa have distinct ranks, the taxa
whose ranks lie in $I(v)$ are exactly those of $L(K_0(v))$. Every taxon of $C\cap L(\mathrm{comp}(\tau))$
has its rank in $I_C=I(v)$, so $C\cap L(\mathrm{comp}(\tau))\subseteq L(K_0(v))$, and
$|C\cap L(\mathrm{comp}(\tau))|=|I_C|=|L(K_0(v))|$. Hence
$L(K_0(v))=C\cap L(\mathrm{comp}(\tau))\subseteq C$, and $K_0(v)$ is must-in.

(iii) Let $v_0\succ v_1\succ\dots\succ v_t$ be the vertices of $\mathrm{path}(\tau)$, from $\mathrm{top}(\tau)$
down to $\ell_\tau$. For $i<t$, the vertex $v_{i+1}$ is a child of $v_i$ joined to it by a tree edge, so
$K_0(v_{i+1})\subseteq K_0(v_i)$ and $I(v_{i+1})\subseteq I(v_i)$. If $v_i$ is a tree node with a child
$c\neq v_{i+1}$ joined to it by a tree edge, then a leaf reachable from $c$ along tree edges
(Lemma~\ref{lem:committed}) lies in $K_0(v_i)$ but not in $K_0(v_{i+1})$, because $K_0(v_i)$ is a tree in
which $c$ and $v_{i+1}$ are distinct children of $v_i$; so $I(v_{i+1})\neq I(v_i)$. Hence the candidate
hosts, which have the same interval $I_C$, are consecutive vertices $v_a,\dots,v_b$ of the path, and for
$a\le i<b$ the vertex $v_i$ is the root, a reticulation, or a tree node whose children other than
$v_{i+1}$ are all reticulations. Only $v_0$ can be the root or a reticulation. Each tree node $v_i$ with
$a\le i<b$ has a reticulation child $r_i$, and the edges $(v_i,r_i)$ are distinct reticulation edges, of
which $N$ has $d_\Sigma$. Hence there are at most $d_\Sigma$ such tree nodes, and at most $d_\Sigma+3$
candidate hosts.
\end{proof}

\begin{proof}[Proof of Lemma~\ref{lem:preproc}]
Algorithm~\ref{alg:preproc} builds the tables. For (T1), Algorithm~\ref{alg:canonical} performs a
depth-first search of each tree component along tree edges. Number the vertices in the order in which
these searches visit them, and let $J(v)$ be the interval from the number of $v$ to the largest number of a
vertex visited below $v$. Since the searches of distinct components use disjoint ranges of numbers,
$u\in K_0(v)$ if and only if $J(u)\subseteq J(v)$. The intervals $I(v)$ and the numbers $|L(K)|$ are
computed in the same searches. For the dictionary, note that for $v\in\mathrm{path}(\tau)$ the interval
$I(v)$ begins at $\mathrm{rank}(\tau)$: the leaf $\ell_\tau$ lies in $K_0(v)$, because the part of
$\mathrm{path}(\tau)$ below $v$ consists of tree edges, and $\tau=m_\pi(v)$ is the $\pi$-minimum of
$L(v)\supseteq L(K_0(v))$. So the vertices $v$ of $\mathrm{path}(\tau)$ with $I(v)=I$ are determined by the
pair of integers $(\tau,\max I)$, and by Lemma~\ref{lem:signature}(iii) they are consecutive on
$\mathrm{path}(\tau)$. The keys $(m_\pi(v),\max I(v))$ of all vertices $v$ and the keys
$(\min_\pi C,\max I_C)$ of all clusters $C$, available once (T2) is built, are radix-sorted together in
$O(|N|+|T|)$ time, and one pass over the sorted list attaches to each cluster the topmost vertex with the
same key and the number of such vertices. The lookup of $\widehat W(C)$ is then an $O(1)$ read.

For (T2), one bottom-up pass over $T$ computes, for each vertex $u$ of $T$, a vector with one entry per
tree component $K$, holding $|L(u)\cap L(K)|$ and the least and the greatest rank of a taxon in
$L(u)\cap L(K)$. At a leaf with taxon $x$, only the entry of the component containing $\ell_x$ is nonzero,
with count $1$ and both ranks equal to $\mathrm{rank}(x)$. At an internal vertex, the vectors of the
children are combined entry by entry: the counts are added, and the least and the greatest ranks are taken
over the children, which is correct because the clusters of the children are disjoint. The network has
$k+1$ tree components and $T$ has at most $2n$ vertices, so this takes $O(n(k+1))$ time. The interval $I_C$
is read from the entry of $\mathrm{comp}(\tau)$.

With the tables, a tree component $K$ is must-in with respect to $C$ if and only if
$|C\cap L(K)|=|L(K)|$, and must-out if and only if $C\cap L(K)=\varnothing$, that is, if and only if $K$
has no entry for $C$; the condition of Lemma~\ref{lem:signature}(ii) compares
$|C\cap L(\mathrm{comp}(\tau))|$ with $|I_C|$; and $\widehat W(C)$ is returned by the dictionary.
\end{proof}

\begin{algorithm}[H]
\caption{\textsc{Preprocess}$(N,T)$: the tables of Lemma~\ref{lem:preproc}.}
\label{alg:preproc}
\begin{algorithmic}[1]
\Procedure{Preprocess}{$N,T$} \Comment{$N$: a tree-child network on $X$; $T$: a phylogenetic tree on $X$}
  \State $\pi\gets$ \Call{CanonicalOrder}{$N$}; number the vertices of $N$ in the order in which its depth-first searches visit them
  \State compute $m_\pi(v)$ for every vertex $v$ \Comment{as in Algorithm~\ref{alg:encode}}
  \State for every vertex $v$, record $J(v)$ and $I(v)$; for every tree component $K$, record $|L(K)|$ \Comment{table (T1)}
  \For{each vertex $u$ of $T$, in postorder} \Comment{table (T2)}
    \If{$u$ is a leaf with taxon $x$}
      \State set the entry of the tree component containing $\ell_x$ to $(1,\mathrm{rank}(x),\mathrm{rank}(x))$, and the other entries to $(0,-,-)$
    \Else
      \State combine the vectors of the children of $u$ entry by entry: add the counts, and take the least and the greatest rank
    \EndIf
  \EndFor
  \State for each cluster $C$ of $T$ with $\tau=\min_\pi C$, read $I_C$ from the entry of $\mathrm{comp}(\tau)$
  \State radix-sort the keys $(m_\pi(v),\max I(v))$ of the vertices $v$ of $N$ together with the keys $(\min_\pi C,\max I_C)$ of the clusters $C$ of $T$, and attach to each cluster the topmost vertex with the same key and the number of such vertices \Comment{the dictionary}
  \State \Return the tables (T1) and (T2)
\EndProcedure
\end{algorithmic}
\end{algorithm}

\begin{proof}[Proof of Theorem~\ref{thm:2decomp}]
Let $w_0\succ w_1\succ\dots\succ w_m$ be the hosts of $C$. We show that each $H_{r,p}$ consists of
consecutive hosts.

Suppose first that $r$ is must-in. If $r$ is free at every host, then $H_{r,p}=W(C)$. Otherwise $r$ is
bound (Lemma~\ref{lem:bound}), $A_w(r)=\mathrm{par}(r)\cap R(w)$ at every host $w$, and
$H_{r,p}=\{w\in W(C): p\in R(w)\}$ is an initial segment of the chain by Lemma~\ref{lem:monoreach}.

Suppose next that $r$ is must-out. The hosts at which $r$ is constrained are the hosts that are strict
ancestors of $r$. They form an initial segment $\{w_0,\dots,w_\gamma\}$ of the chain, possibly empty, since
an ancestor of a strict ancestor of $r$ is a strict ancestor of $r$. At every other host $w$,
$p\in A_w(r)=\mathrm{par}(r)$. At a host $w$ with $w\succeq w_\gamma$, $p\in A_w(r)$ if and only if
$p\notin R(w)$. By Lemma~\ref{lem:monoreach}, the hosts $w$ with $p\in R(w)$ form an initial segment
$\{w_0,\dots,w_\alpha\}$. Each of them is a strict ancestor of $r$, since $p\preceq w$ for $p\in R(w)$
and $(p,r)$ is an edge, while $w=r$ would put the parent $p$ of $w$ below $w$. Hence $\alpha\le\gamma$, and
$H_{r,p}=\{w_{\alpha+1},\dots,w_m\}$ is a final segment of the chain.

Suppose finally that $r$ is mixed. At a host $w$, every reticulation constrained at $w$ has
$A_w(r)\neq\varnothing$ (Lemma~\ref{lem:carve}(ii)), so $r$ is free at every host, and $H_{r,p}=W(C)$.

Now let $f$ be a switching with $f(r)\in\mathrm{proj}_r(\mathrm{realized}(C))$ for every reticulation $r$
and $(f(r),f(r'))\in\mathrm{proj}_{r,r'}(\mathrm{realized}(C))$ for all reticulations $r\neq r'$. We show
that $f\in\mathrm{realized}(C)$; the converse is clear. A switching $g$ lies in $\mathrm{realized}(C)$ if
and only if $g\in S_w(C)$ for some host $w$, that is, if and only if some host lies in $H_{r,g(r)}$ for
every $r$. Since $S_w(C)$ is a box at every host $w$, the condition on $f(r)$ says that
$H_{r,f(r)}\neq\varnothing$, and the condition on $(f(r),f(r'))$ says that
$H_{r,f(r)}\cap H_{r',f(r')}\neq\varnothing$. Write $H_{r,f(r)}=\{w_{a_r},\dots,w_{b_r}\}$ with
$a_r\le b_r$, and let $b=\min_r b_r$. For every $r$, the set $H_{r,f(r)}$ meets the set attaining this
minimum, so $a_r\le b$, and $b\le b_r$; hence $w_b\in H_{r,f(r)}$. Thus $f\in S_{w_b}(C)\subseteq
\mathrm{realized}(C)$.

The hypothesis that $C$ has a host is needed only when $k=0$. Then there is no reticulation, the conditions
on $f$ are vacuous, and $f$ is the unique switching, while $\mathrm{realized}(C)$ is empty if $C$ has no
host. When $k\ge1$ and $C$ has no host, $\mathrm{proj}_r(\mathrm{realized}(C))=\varnothing$, and no switching
satisfies the conditions.
\end{proof}

For a bound must-in reticulation $r$ and a parent $p$ of $r$, let $Q(p)=\{w\in W(C): p\in R(w)\}$, an
initial segment of the host chain (Lemma~\ref{lem:monoreach}). If $p$ is reachable from $p'$ in
$G_{\mathrm{in}}(C)$, then $Q(p)\supseteq Q(p')$, since a host that reaches $p'$ in $G_{\mathrm{in}}(C)$
also reaches $p$. The next two lemmas show that $Q(p)=Q(p')$ when $p$ and $p'$ are {\it incomparable} in
$G_{\mathrm{in}}(C)$, that is, when neither is reachable from the other in $G_{\mathrm{in}}(C)$.

\begin{lemma}\label{lem:stableleaf}
Let $w$ be a host of $C$ and $p\in R(w)$. Then there is a path $p=x_0\to x_1\to\dots\to x_t$ in $N$ such
that $x_t$ is a leaf whose taxon lies in $C$ and $x_1,\dots,x_t$ have indegree one.
\end{lemma}

\begin{proof}
Descending from $p$ along tree edges reaches a leaf $x_t$ of $\mathrm{comp}(p)$ (Lemma~\ref{lem:committed}),
and the vertices after $p$ on this path have indegree one. It remains to show that the taxon of $x_t$ lies
in $C$. If $w\notin\mathrm{comp}(p)$, then a path from $w$ to $p$ in $G_{\mathrm{in}}(C)$ enters
$\mathrm{comp}(p)$ at its root through an incoming edge of the root; the root is not the root of $N$, which
has no incoming edge, so it is a must-in reticulation, and $L(\mathrm{comp}(p))\subseteq C$. If
$w\in\mathrm{comp}(p)$, then a path from $w$ to $p$ in $G_{\mathrm{in}}(C)$ cannot leave $\mathrm{comp}(p)$,
since it could re-enter it only at its root, an ancestor of $w$, closing a cycle. So the path consists of
tree edges, $x_t\in K_0(w)$, and the taxon of $x_t$ lies in $L_f(w)=C$ for $f\in S_w(C)$.
\end{proof}

\begin{lemma}\label{lem:incomp}
Let $r$ be a bound must-in reticulation, and let $p,q$ be parents of $r$ with $Q(p)\neq\varnothing$ and
$Q(q)\neq\varnothing$ that are incomparable in $G_{\mathrm{in}}(C)$. Then $Q(p)=Q(q)$.
\end{lemma}

\begin{proof}
Suppose that $Q(p)\neq Q(q)$. Both are nonempty initial segments of the host chain, so one is a proper
subset of the other, say $Q(p)\subsetneq Q(q)$. Let $w$ be the last host in $Q(q)$; then $q\in R(w)$ and
$p\notin R(w)$. Let $f\in S_w(C)$. Since $Q(p)\neq\varnothing$, Lemma~\ref{lem:stableleaf} gives a path
$p=x_0\to x_1\to\dots\to x_t$ in which $x_t$ is a leaf whose taxon lies in $C$ and $x_1,\dots,x_t$ have
indegree one. As this taxon lies in $C=L_f(w)$, the vertex $w$ is an ancestor of $x_t$. Since
$x_1,\dots,x_t$ have indegree one, the ancestors of $x_t$ are $x_1,\dots,x_{t-1}$, $p$, and the ancestors of
$p$.

If $w=x_i$ with $1\le i\le t-1$, then $w$ is reachable from $p$ along tree edges, and $q\in R(w)$; so $q$ is
reachable from $p$ in $G_{\mathrm{in}}(C)$, contradicting incomparability. If $w=p$, then $p\in R(w)$, a
contradiction. So $w$ is a strict ancestor of $p$. The path from $w$ to $x_t$ in $f(N)$ passes through $p$,
since the only parent of $x_i$ is $x_{i-1}$ for $1\le i\le t$. Let $(y,r')$ be a reticulation edge on the
part of this path from $w$ to $p$. Then $r'$ is reachable from $w$ in $f(N)$, so every leaf of
$\mathrm{comp}(r')$ is, and $L(\mathrm{comp}(r'))\subseteq L_f(w)=C$; that is, $r'$ is must-in, and
$(y,r')$ belongs to $G_{\mathrm{in}}(C)$. Hence the part from $w$ to $p$ is a path of $G_{\mathrm{in}}(C)$,
and $p\in R(w)$, a contradiction.
\end{proof}

\begin{proof}[Proof of Lemma~\ref{lem:minclosed}]
Let $f\in S_{w_1}(C)$ and $g\in S_{w_2}(C)$ for hosts $w_1\succeq w_2$ (Lemma~\ref{cor:hostrow}), and let
$h=f\wedge g$. We show that $h\in S_{w_2}(C)$, that is, $h(r)\in A_{w_2}(r)$ for every reticulation $r$
(Lemma~\ref{lem:carve}).

If $r$ is free at $w_2$, then $A_{w_2}(r)=\mathrm{par}(r)\ni h(r)$. Let $r$ be constrained at $w_2$. Then
$r$ is not mixed, and $r$ is also constrained at $w_1$, since $w_1\succeq w_2$ is a strict ancestor of $r$ as
well. If $r$ is must-out, then $f(r)\notin R(w_1)$ and $g(r)\notin R(w_2)$, and
$R(w_2)\subseteq R(w_1)$ (Lemma~\ref{lem:monoreach}) gives $f(r),g(r)\notin R(w_2)$; so
$h(r)\in\mathrm{par}(r)\setminus R(w_2)=A_{w_2}(r)$. If $r$ is must-in, then $r$ is bound,
$f(r)\in R(w_1)$ and $g(r)\in R(w_2)$. If $h(r)=g(r)$, we are done. Otherwise $h(r)=f(r)$ is deeper than
$g(r)$. Then $f(r)$ is not a strict ancestor of $g(r)$, since an ancestor precedes its descendants in the
topological order. So either $f(r)$ is reachable from $g(r)$ in $G_{\mathrm{in}}(C)$, and then
$Q(f(r))\supseteq Q(g(r))$, or $f(r)$ and $g(r)$ are incomparable in $G_{\mathrm{in}}(C)$, and then
$Q(f(r))=Q(g(r))$ by Lemma~\ref{lem:incomp}, as $w_1\in Q(f(r))$ and $w_2\in Q(g(r))$. In both cases
$w_2\in Q(f(r))$, that is, $f(r)\in R(w_2)$, and $h(r)=f(r)\in\mathrm{par}(r)\cap R(w_2)=A_{w_2}(r)$.
\end{proof}

\begin{proof}[Proof of Theorem~\ref{thm:minclosed}]
The operation $\wedge$ is idempotent, commutative and associative, and each domain $\mathrm{par}(r)$ is
closed under it. By Lemma~\ref{lem:minclosed}, every $\mathrm{realized}(C)$ is closed under $\wedge$, and
so are its projections: if $(a,b)$ and $(a',b')$ in $\mathrm{proj}_{r,r'}(\mathrm{realized}(C))$ come from
$f$ and $g$ in $\mathrm{realized}(C)$, then $(a\wedge a',b\wedge b')$ comes from $f\wedge g$. Hence every
constraint of $\mathcal{N}$ is closed under $\wedge$. A network of this kind has width
one~\cite{jeavonscooper1995,dalmau1999}; we include the argument. Enforcing arc consistency removes only
values that belong to no solution, so if some domain becomes empty, then $\mathcal{N}$ has no solution.
Suppose that all domains are nonempty after arc consistency, and let $m_r$ be the deepest value of the
domain of $r$. Each $m_r$ satisfies the unary constraints on $r$, since arc consistency removes the values
that violate them. Consider a binary constraint $\mathcal{R}$ on $r$ and $r'$. By arc consistency, there
are a value $b$ in the domain of $r'$ with $(m_r,b)\in\mathcal{R}$ and a value $a$ in the domain of $r$ with
$(a,m_{r'})\in\mathcal{R}$. As $\mathcal{R}$ is closed under $\wedge$,
$(m_r\wedge a,\,b\wedge m_{r'})\in\mathcal{R}$, and $m_r\wedge a=m_r$ and $b\wedge m_{r'}=m_{r'}$, because
$m_r$ and $m_{r'}$ are the deepest values of their domains. Hence $(m_r,m_{r'})\in\mathcal{R}$, and the
switching $r\mapsto m_r$ is a solution.
\end{proof}

\begin{proof}[Proof of Lemma~\ref{lem:eligsets}]
Let $\tau=\min_\pi C$, and let $w_0\succ\dots\succ w_m$ be the candidate hosts of $C$, obtained from the
tables in $O(1)$ time plus $O(1)$ per candidate (Lemma~\ref{lem:preproc}); there are $O(d_\Sigma)$ of them
(Lemma~\ref{lem:signature}(iii)). If $|C\cap L(\mathrm{comp}(\tau))|\neq|I_C|$, then $C$ has no host
(Lemma~\ref{lem:signature}(i)). Otherwise $K_0(w)$ is must-in for every candidate host $w$
(Lemma~\ref{lem:signature}(ii)), and by Lemma~\ref{lem:carve} a candidate host $w$ is a host if and only if
$C\subseteq L(w)$ and $A_w(r)\neq\varnothing$ for every reticulation $r$. The status of each tree component,
and of each reticulation, is an $O(1)$ lookup (Lemma~\ref{lem:preproc}); it is computed once per cluster,
in $O(k)$ time.

The tests use two reachability relations: whether a reticulation is constrained at a candidate host, and
whether a vertex lies in $R(w)$. For candidate hosts $w\succeq w'$ and a vertex $p$, $p\preceq w'$ implies
$p\preceq w$, and $p\in R(w')$ implies $p\in R(w)$ (Lemma~\ref{lem:monoreach}). So the candidate hosts $w$
with $p\preceq w$ form an initial segment
of the chain, and so do those with $p\in R(w)$. Let $\rho(p)$ and $\lambda(p)$ be the last candidate hosts
of these two segments, with $\rho(p)=\bot$ or $\lambda(p)=\bot$ if the segment is empty. Likewise, for a
reticulation $r$, the candidate hosts at which $r$ is constrained form an initial segment, and so do those
$w$ with $r\in R(w)$; let $\nu(r)$ and $\beta(r)$ be their last elements, or $\bot$. Thus $r$ is constrained
at $w$ if and only if $w\succeq\nu(r)$, where $w\succeq\bot$ is false; $p\in R(w)$ if and only if
$w\succeq\lambda(p)$; and $r\in R(w)$ if and only if $w\succeq\beta(r)$.

These thresholds are computed as follows. A reticulation $r$ is constrained at $w$ if and only if $p\preceq w$
for some parent $p$ of $r$, and by fact (F1) in the proof of Lemma~\ref{lem:carve},
$r\in R(w)$ if and only if $r$ is must-in and some parent of $r$ lies in $R(w)$. Hence
\[
  \nu(r)=\text{the deepest of }\{\rho(p): p\in\mathrm{par}(r)\},
  \qquad
  \beta(r)=
  \begin{cases}
    \text{the deepest of }\{\lambda(p): p\in\mathrm{par}(r)\} & \text{if $r$ is must-in},\\[2pt]
    \bot & \text{otherwise},
  \end{cases}
\]
where the deepest of a family of candidate hosts and $\bot$'s is $\bot$ if all members are $\bot$. For a
vertex $p$, there are three cases. If $p\in\mathrm{comp}(\tau)$, then $p\preceq w$ if and only if
$p\in K_0(w)$, that is, $J(p)\subseteq J(w)$, because a path from $w$ to $p$ cannot leave
$\mathrm{comp}(\tau)$ and re-enter it, and the same holds for $p\in R(w)$. The intervals $J(w)$ of the
candidate hosts are nested, so $\rho(p)=\lambda(p)$ is found by binary search in $O(\log d_\Sigma)$ time. If
$p\in\mathrm{comp}(r')$ for a reticulation $r'$, then $p\preceq w$ if and only if $r'$ is
constrained at $w$, and $p\in R(w)$ if and only if $r'\in R(w)$, by (F1) and because $r'\neq w$; so
$\rho(p)=\nu(r')$ and $\lambda(p)=\beta(r')$. If $p$ lies in the tree component of the root of $N$ and this
component is not $\mathrm{comp}(\tau)$, then no candidate host reaches $p$, and $\rho(p)=\lambda(p)=\bot$.
Processing the reticulations in a topological order of $V(N)$, every value on the right-hand side is
available when it is needed, since the root of the component containing a parent of $r$ is an ancestor of
$r$. All reticulations are processed, at a cost of $O(1)$ or $O(\log d_\Sigma)$ per parent, that is, in
$O(d_\Sigma\log d_\Sigma)$ time in all.

A candidate host $w$ is a host if and only if the following four conditions hold.
\begin{enumerate}
\item[(a)] For every tree component $K\neq\mathrm{comp}(\tau)$ with $C\cap L(K)\neq\varnothing$, the root of $K$
is a reticulation $r_K$ with $w\succeq\nu(r_K)$. This is the condition $C\subseteq L(w)$: the taxa of $C$
in $\mathrm{comp}(\tau)$ lie in $L(K_0(w))\subseteq L(w)$ by Lemma~\ref{lem:signature}(ii), and a taxon of
$C$ in another component $K$ lies in $L(w)$ if and only if the root of $K$ is a reticulation constrained at
$w$, as $w\in\mathrm{comp}(\tau)$ and $K$ is entered only at its root.
\item[(b)] No mixed reticulation $r$ has $w\succeq\nu(r)$. This says that $A_w(r)\neq\varnothing$ for the
mixed reticulations.
\item[(c)] Every must-in reticulation $r$ with $w\succeq\nu(r)$ has $w\succeq\beta(r)$. This says that
$\mathrm{par}(r)\cap R(w)\neq\varnothing$, that is, $r\in R(w)$, for the constrained must-in
reticulations.
\item[(d)] Every must-out reticulation $r$ with $w\succeq\nu(r)$ has a parent $p$ with
$w\not\succeq\lambda(p)$. This says that $\mathrm{par}(r)\setminus R(w)\neq\varnothing$ for the constrained
must-out reticulations.
\end{enumerate}
The tests range over the reticulations and, in (d), over the parents of a reticulation, so they take
$O(d_\Sigma)$ time per candidate host and $O(d_\Sigma^{2})$ time for all candidate hosts.

Now let $C$ have hosts, and let $r$ be a reticulation and $p$ a parent of $r$. If $r$ is free at the first
host, then $r$ is free at every host, because the hosts at which $r$ is constrained form an initial
segment; then $H_{r,p}=W(C)$. If $r$ is must-in and constrained at the first host, then $r$ is bound and
$H_{r,p}=\{w\in W(C): w\succeq\lambda(p)\}$. If $r$ is must-out and constrained at the first host, then
$H_{r,p}$ consists of the hosts $w$ at which $r$ is free or $p\notin R(w)$; as $p\in R(w)$ implies that $r$
is constrained at $w$, this is $\{w\in W(C): w\not\succeq\lambda(p)\}$. A mixed $r$ is free at every host.
In each case, $H_{r,p}$ is an initial or a final segment of $W(C)$ determined by one threshold. Counting,
along the chain of candidate hosts, the hosts at or above each candidate, which takes $O(d_\Sigma)$ time
once, turns each threshold into a position in the chain $W(C)$ in $O(1)$ time. So the first and the last
host of every $H_{r,p}$ are found in $O(1)$ time, and $O(d_\Sigma)$ time in all.

Finally, $p\in\mathrm{proj}_r(\mathrm{realized}(C))$ if and only if $H_{r,p}\neq\varnothing$, and
$(p,p')\in\mathrm{proj}_{r,r'}(\mathrm{realized}(C))$ if and only if $H_{r,p}\cap H_{r',p'}\neq\varnothing$,
because $S_w(C)$ is a nonempty box at every host $w$, so that some $f\in S_w(C)$ has $f(r)=p$ and
$f(r')=p'$ whenever $p\in A_w(r)$ and $p'\in A_w(r')$. Each test is an $O(1)$ comparison of intervals, and
there are $\sum_rd_r+\sum_{r\neq r'}d_rd_{r'}\le d_\Sigma+d_\Sigma^{2}$ of them.
\end{proof}

\begin{algorithm}[H]
\caption{\textsc{Carve}$(C)$: the computation of Lemma~\ref{lem:eligsets}, in $O(d_\Sigma^{2})$ time.}
\label{alg:carve}
\begin{algorithmic}[1]
\Procedure{Carve}{$C$} \Comment{$C$: a cluster of $T$ with $2\le|C|\le n-1$; the tables of Lemma~\ref{lem:preproc} are available}
  \State $\tau\gets\min_\pi C$; \quad $w_0\succ\dots\succ w_m\gets$ the candidate hosts of $C$ \Comment{the dictionary of (T1)}
  \State {\bf if} $|C\cap L(\mathrm{comp}(\tau))|\neq|I_C|$ {\bf then return} {\bf no host} \Comment{Lemma~\ref{lem:signature}(i)}
  \State determine the status of every reticulation \Comment{table (T2)}
  \For{each reticulation $r$, in a topological order of $V(N)$} \Comment{the thresholds}
    \State for each parent $p$ of $r$: $\rho(p),\lambda(p)\gets$ the deepest $w_i$ with $J(p)\subseteq J(w_i)$, by binary search, if $p\in\mathrm{comp}(\tau)$; $\rho(p)\gets\nu(r')$ and $\lambda(p)\gets\beta(r')$ if $p\in\mathrm{comp}(r')$ for a reticulation $r'$; and $\rho(p),\lambda(p)\gets\bot$ otherwise
    \State $\nu(r)\gets$ the deepest of $\{\rho(p): p\in\mathrm{par}(r)\}$
    \State $\beta(r)\gets$ the deepest of $\{\lambda(p): p\in\mathrm{par}(r)\}$ if $r$ is must-in, and $\bot$ otherwise
  \EndFor
  \State $W\gets\varnothing$
  \For{$i=0,\dots,m$} \Comment{which candidate hosts are hosts}
    \State {\bf if} conditions (a)--(d) of the proof of Lemma~\ref{lem:eligsets} hold for $w_i$ {\bf then} add $w_i$ to $W$
  \EndFor
  \State {\bf if} $W=\varnothing$ {\bf then return} {\bf no host}
  \State $c_i\gets$ the number of hosts among $w_0,\dots,w_i$, for $i=0,\dots,m$
  \For{each reticulation $r$ and each parent $p$ of $r$} \Comment{the sets $H_{r,p}$, as positions in $W$}
    \State $H_{r,p}\gets W$ if $r$ is free at the first host; the first $c_i$ hosts, where $w_i=\lambda(p)$, if $r$ is must-in and constrained at the first host, and $\varnothing$ if $\lambda(p)=\bot$; the hosts after the first $c_i$ hosts if $r$ is must-out and constrained at the first host, and $W$ if $\lambda(p)=\bot$
  \EndFor
  \State $\mathrm{proj}_r\gets\{p: H_{r,p}\neq\varnothing\}$ for each $r$, and $\mathrm{proj}_{r,r'}\gets\{(p,p'): H_{r,p}\cap H_{r',p'}\neq\varnothing\}$ for each pair $r\neq r'$
  \State \Return $W$, the sets $H_{r,p}$, and the projections
\EndProcedure
\end{algorithmic}
\end{algorithm}

\section{Binary resolutions, and proofs for Section~\ref{sec:sym}}\label{app:resolutions}

Our definition of a binary resolution in Section~\ref{sec:sym} allows any
block. Bentert et al.\ define the notion through the contraction map alone, without the bijection of edges,
ask that the resolution be minimal, with no biconnected component having exactly one incoming and one
outgoing edge, in the journal version none lying within a single block~\cite[Preliminaries]{bmw2018,bmw2021}, and record as an observation that no block contains
both a reticulation and a tree node~\cite[Observation~1]{bmw2018}. The definition,
minimality included, does not imply the observation: without the observation, even a binary tree has
resolutions other than itself, and some of them display trees that the tree does not refine.
We therefore take the observation as part of their definition. Under this definition, their subgraph
characterization holds~\cite[Lemma~11(d)]{bmw2018}, \cite[Lemma~3(d)]{bmw2021}. Their hardness
proof also needs it: the network there
is binary, and the proof immediately reads soft display as the containment of a binary tree in $N$~\cite[proof of Lemma~17]{bmw2018}, which requires that resolving a
binary network adds no displayed tree. Proposition~\ref{prop:hardness} transfers their
hardness result under this definition, and Lemma~\ref{lem:resolveswitch} applies to their resolutions.\footnote{Their condition on edges asks only that the edge $xy$ of $N$ exist
if and only if $N_B$ has an edge from $\chi^{-1}(x)$ to $\chi^{-1}(y)$~\cite[Preliminaries]{bmw2021} (the conference version asks for an edge between the two sets), so several edges of
$N_B$ may stand for one edge of $N$. Of the bijection the proof of
Lemma~\ref{lem:resolveswitch} uses only that a reticulation's block has one leaving edge and a tree node's
block one entering edge, which the observation guarantees; at most one representative of an edge lies on
the walk it follows; and a leaf's block may receive several representatives of its one incoming edge, but the walk
up from the leaf still leaves the block through an edge of $f_B(N_B)$.} Bruchhold and Weller build the same condition into their
construction: they resolve a tree node of outdegree at least three and a reticulation of indegree at
least three by separate splitting operations~\cite[Definitions~1--3]{bruchholdweller}, so that each
block consists of tree nodes alone or of reticulations alone.

\begin{proof}[Proof of Lemma~\ref{lem:resolveswitch}]
In $f_B(N_B)$, every vertex other than the root has exactly one incoming edge. As a reticulation $r$ of $N$
has outdegree one, its block $\kappa^{-1}(r)$ has a single leaving edge; write $s_r$ for its tail. Follow
incoming edges of $f_B(N_B)$ upward from $s_r$. The walk visits distinct vertices and hence, by acyclicity,
ends at the root of $N_B$, which lies in the block of the root of $N$ (the only block with no entering edge)
and not in $\kappa^{-1}(r)$. Thus the walk leaves $\kappa^{-1}(r)$, and it can do so only through an entering
edge of the block. That edge corresponds to an incoming edge of $r$; let $f(r)$ be the tail of that incoming
edge. This defines a switching $f$ of $N$.

For a tree node or leaf $v$ of $N$, its block has a single entering edge; write $t_v$ for its head, and let
$t_v$ be the root of $N_B$ when $v$ is the root of $N$. For a reticulation $r$ put $t_r=s_r$. Let $v$ be a
tree node, a leaf, or the root. If $v$ is not the root, then the entering edge of $\kappa^{-1}(v)$ is an edge
of $f_B(N_B)$: any other incoming edge of $t_v$ lies inside $\kappa^{-1}(v)$, and a walk up incoming edges
of $f_B(N_B)$ from its tail would stay inside the block until it left through the only entering edge,
returning to $t_v$ and closing a cycle. Every other vertex of $\kappa^{-1}(v)$ has all its incoming edges
inside $\kappa^{-1}(v)$, and so its incoming edge in $f_B(N_B)$ lies inside $\kappa^{-1}(v)$. Hence all
vertices of $\kappa^{-1}(v)$ are descendants of $t_v$ in $f_B(N_B)$. No block of a non-leaf contains a leaf.

We claim $L_{f_B}(t_v)=L_f(v)$ for every vertex $v$ of $N$, both sides possibly empty, by induction along a
reverse topological order of $N$. For a leaf both sides are its taxon. Let $v$ be a tree node or the root,
$c$ a child of $v$, $e$ the edge of $N_B$ corresponding to $(v,c)$, and $y$ the head of $e$. The edge $e$
leaves $\kappa^{-1}(v)$. As all vertices of that block are descendants of $t_v$ in $f_B(N_B)$, it
contributes to $L_{f_B}(t_v)$ the set $L_{f_B}(y)$ if it is an edge of $f_B(N_B)$, and nothing otherwise.
If $c$ is a tree node or a leaf, then $e$ enters $\kappa^{-1}(c)$ at $y=t_c$, so $e$ is an edge of
$f_B(N_B)$, and the contribution is $L_{f_B}(t_c)=L_f(c)$. If $c$ is a reticulation, then, since the only
edge leaving $\kappa^{-1}(c)$ starts at $s_c$, a descendant of $y$ in $f_B(N_B)$ that does not reach $s_c$
stays inside $\kappa^{-1}(c)$ and reaches no leaf. The contribution is therefore $L_{f_B}(s_c)$ if $e$ is an
edge of $f_B(N_B)$ and $f_B(N_B)$ contains a path from $y$ to $s_c$, and empty otherwise. This condition
holds if and only if $e$ lies on the upward walk from $s_c$: such a path reversed is an initial segment of
the walk, and all edges of the walk are edges of $f_B(N_B)$. By construction this happens if and only if
$f(c)=v$, and the contribution is then $L_{f_B}(s_c)=L_f(c)$. Thus $L_{f_B}(t_v)$ is the union of $L_f(c)$
over the children $c$ of $v$ in $f(N)$, which is $L_f(v)$. Finally let $v$ be a reticulation with child $c$,
and apply the same case analysis to the edge leaving $\kappa^{-1}(v)$, which starts at $s_v=t_v$. Its
contribution is $L_f(c)$ if $c$ is a tree node or a leaf; if $c$ is a reticulation, it is $L_f(c)$ when
$f(c)=v$ and empty otherwise. In each case this is $L_f(v)$.

The clusters of $T(N,f)$ are the nonempty sets $L_f(v)$, each equal to $L_{f_B}(t_v)$ and so a cluster of
$T(N_B,f_B)$.
\end{proof}

\begin{proof}[Proof of Proposition~\ref{prop:hardness}]
Bentert et al.\ prove soft Tree Containment NP-hard on reticulation-visible networks with maximum
reticulation indegree three and maximum outdegree two~\cite[Corollary~20]{bmw2018}, \cite[Corollary~3]{bmw2021}, hence on networks with binary tree nodes. These networks are single-labelled: in their
networks no label occurs twice unless said otherwise~\cite[Preliminaries]{bmw2021}, and the reduction
behind the corollary merges the leaves of each label of a binary $3$-labelled tree into one reticulation
with a single new leaf of that label~\cite[the paragraph preceding Corollary~20]{bmw2018}. The merged
network has no parallel edges: the leaves of one label lie in different parts of the construction, one
part for each vertex of that colour, since no path of the source instance carries a colour twice~\cite[the paragraph preceding Construction~2]{bmw2021}. Consequently no two of these leaves are siblings.
Their notion of soft display is the symmetric one: $N$ softly
displays $T$ if and only if some binary resolutions $N_B$ of $N$ and $T_B$ of $T$ satisfy $T(N_B,f_B)\cong T_B$
for a switching $f_B$ of $N_B$, with resolutions as defined above. We show that on a network with binary
tree nodes, this notion coincides with the switching condition of the statement.

Suppose first that $N_B$, $T_B$ and $f_B$ witness the symmetric display. Lemma~\ref{lem:resolveswitch}
gives a switching $f$ of $N$ with $\mathcal{C}(T(N,f))\subseteq\mathcal{C}(T(N_B,f_B))=\mathcal{C}(T_B)$.
Since the tree nodes of $N$ are binary and every reticulation has one incoming edge in $f(N)$, every
vertex of $f(N)$ has outdegree at most two, so $T(N,f)$ is a binary phylogenetic tree on $X$. Its cluster set is thus
a maximal laminar family on $X$. Being contained in the laminar family $\mathcal{C}(T_B)$, it equals it, and
$\mathcal{C}(T)\subseteq\mathcal{C}(T_B)=\mathcal{C}(T(N,f))$.

Conversely let $f$ satisfy $\mathcal{C}(T)\subseteq\mathcal{C}(T(N,f))$. Resolve each reticulation $r$ of
$N$, with parents $p_1,\dots,p_{d_r}$, into a {\it cascade} of $d_r-1$ binary reticulations
$h_1,\dots,h_{d_r-1}$: the incoming edges of $h_1$ come from $p_1$ and $p_2$, those of $h_j$ from $h_{j-1}$
and $p_{j+1}$, and the outgoing edge of $h_{d_r-1}$ goes to the child of $r$. Contracting the edges
$(h_{j-1},h_j)$ recovers $r$, and the tree nodes, which are binary, are left as they are. Thus the result $N_B$ is a binary
resolution of $N$. To realize $f(r)=p_1$, let $f_B$ retain $(p_1,h_1)$ and every $(h_{j-1},h_j)$; to
realize $f(r)=p_{j+1}$, let $f_B$ retain $(p_{j+1},h_j)$ and every $(h_i,h_{i+1})$ with $i\ge j$, together with
an arbitrary incoming edge at each of $h_1,\dots,h_{j-1}$, which then reach no leaf and are deleted. The switching $f_B$ so
defined has $T(N_B,f_B)=T(N,f)$, a binary tree, and a binary tree contracts to $T$ if and only if its cluster
set contains $\mathcal{C}(T)$ (Section~\ref{sec:prelim}). Therefore $T(N,f)$ is a binary resolution $T_B$ of $T$
and $N_B$ firmly displays it. The symmetric problem and the switching condition thus coincide on these
networks, and the NP-hardness of one is the NP-hardness of the other.
\end{proof}

\begin{proof}[Proof of Lemma~\ref{lem:symresolve}]
Resolve each reticulation $r$, with parents $p_1,\dots,p_{d_r}$, into a cascade of $d_r-1$ binary
reticulations $h_1,\dots,h_{d_r-1}$: the incoming edges of $h_1$ come from $p_1$ and $p_2$, those of $h_j$
from $h_{j-1}$ and $p_{j+1}$, and the outgoing edge of $h_{d_r-1}$ goes to the child of $r$. Contracting
the edges $(h_{j-1},h_j)$ recovers $r$. To realize $f(r)=p_1$, let $f_B$ retain $(p_1,h_1)$ and every
$(h_{j-1},h_j)$; to realize $f(r)=p_{j+1}$, let $f_B$ retain $(p_{j+1},h_j)$ and every $(h_{i},h_{i+1})$
with $i\ge j$.

Resolve each tree node $v$ as follows. Let $Q$ be the set of children of $v$ in $f(N)$; $Q$ is nonempty,
since by tree-childness $v$ has a child of indegree one, which is a child of $v$ in $f(N)$. The sets
$L_f(c)$, $c\in Q$, partition $L_f(v)$, and each is a cluster of $T(N,f)$, and so of $R$. Let $\beta_v$ be
the minimal subtree of $R$ spanning the vertices carrying those clusters, with each of them contracted to
a leaf. No internal vertex of that subtree has a single child in it: a child subtree of $R$ carrying none
of the $L_f(c)$ would hold taxa of $L_f(v)$ lying in none of them, and the $L_f(c)$ partition $L_f(v)$.
Hence $\beta_v$ is a binary tree with $|Q|$ leaves, and its root carries $L_f(v)$. Replace $v$ by a copy of $\beta_v$,
identifying its leaves with the children in $Q$. Attach each remaining child, that is, each reticulation
child $c$ with $f(c)\neq v$, by subdividing an edge of the copy once and hanging the child on the new
vertex. When $|Q|=1$, the copy is a single vertex, identified with the one child in $Q$, and has no edge
to subdivide; subdivide then the edge entering this vertex, once per remaining child. The block of $v$
consists of the internal vertices of the copy and the subdivision vertices; it is nonempty, as $v$ has at
least two children. Each vertex of the block has indegree one and outdegree two, and contracting the
edges inside the block recovers $v$ with all its children. Let
$f_B$ retain the cascade edges prescribed above and, at each cascade vertex that has no retained incoming
edge yet, an arbitrary incoming edge.

The resulting $N_B$ is a binary resolution of $N$, and $f_B$ induces $f$ in the sense of
Lemma~\ref{lem:resolveswitch}. The proof of that lemma gives $L_{f_B}(t_v)=L_f(v)$ for all $v$. In $f_B(N_B)$, the
edges between vertices of the block of $v$ and from the block to the children in $Q$ form the copy of
$\beta_v$ with some of its edges subdivided or, when $|Q|=1$, a path ending at the child in $Q$. The internal vertices of $\beta_v$ carry the
clusters of $R$ lying strictly between $L_f(v)$ and the $L_f(c)$, $c\in Q$. In $f_B(N_B)$, the cascade
vertices at which an arbitrary incoming edge was retained reach no leaf, and the definition of
$T(N_B,f_B)$ deletes them; the subdivision vertices and the other cascade vertices then have one child
each and are suppressed. All clusters of $R$ arise in this way, and no other nonempty cluster does. Thus
$\mathcal{C}(T(N_B,f_B))=\mathcal{C}(R)$, which gives $T(N_B,f_B)=R$.
\end{proof}

\begin{lemma}\label{lem:minimal}
The resolution $N_B$ constructed in the proof of Lemma~\ref{lem:symresolve} is minimal in the sense of
Bentert et al., in both versions of their definition: no biconnected component of $N_B$ that contains a
cycle lies within a block, and every such component has at least two outgoing edges.
\end{lemma}

\begin{proof}
The blocks of $N_B$ are trees, so no biconnected component that contains a cycle lies within a block.
Let $Z$ be a biconnected component of $N_B$ that contains a cycle. On a cycle of $Z$, some vertex $x$ has
both of its cycle edges leaving it. Then $x$ has outdegree two, so $x$ lies in the block of a tree node
$v$ of $N$. Following either outgoing edge of $x$ downwards inside the block of $v$, one reaches an edge
from the block of $v$ into the block of a child of $v$. Let $c_1$ and $c_2$ be the children reached from
the two outgoing edges. They are distinct, since the block of $v$ induces an out-tree and each child of
$v$ is attached to it by one edge. Extend the two routes to leaves: through the cascade of $c_i$ when
$c_i$ is a reticulation, which ends at the child of $c_i$, a tree node or a leaf, and from a tree node on
along children of indegree one, which exist because $N$ is tree-child, each time inside the block of the
current tree node. On the $i$-th path, all vertices after $x$ have indegree one, except those in the
cascade of $c_i$. If the paths met after $x$, their first common vertex could not have indegree one: its
unique parent would be an earlier common vertex, or $x$, which the two paths leave by different edges.
Nor could it be a cascade vertex, as the cascades of $c_1$ and $c_2$ are disjoint. Both paths end at
leaves, which lie on no cycle and hence outside $Z$, and the first edges by which they leave $Z$ are
distinct, as their tails are distinct or both equal to $x$. Thus $Z$ has at least two outgoing edges.
\end{proof}

\begin{proof}[Proof of Proposition~\ref{prop:sym}]
For two trees on $X$ this laminarity condition is due to Bentert, Mal\'{\i}k, and Weller~\cite[Lemma~7]{bmw2018}, \cite[Lemma~1]{bmw2021}: a tree
$N$ symmetrically softly displays $T$ if and only if each cluster of $N$ and each cluster of $T$ are nested
or disjoint. The proposition reads the same condition on the switchings of a network, as
Lemma~\ref{lem:expansion} does for Lemma~3(d) of~\cite{bmw2021}.

Suppose $T(N_B,f_B)\cong T_B$ with $T_B$ a binary resolution of $T$. Lemma~\ref{lem:resolveswitch} gives a
switching $f$ of $N$ with $\mathcal{C}(T(N,f))\subseteq\mathcal{C}(T(N_B,f_B))=\mathcal{C}(T_B)$, and
$\mathcal{C}(T)\subseteq\mathcal{C}(T_B)$ because $T_B$ resolves $T$. The clusters of a tree are laminar,
so $\mathcal{C}(T)\cup\mathcal{C}(T(N,f))$ is laminar.

Conversely let $\mathcal{C}(T)\cup\mathcal{C}(T(N,f))$ be laminar. A laminar family on $X$ extends to a
maximal one, which is the cluster set of a binary phylogenetic tree $R$ on $X$~\cite{semplesteel2003}. Then $R$ refines
both $T$ and $T(N,f)$. By Lemma~\ref{lem:symresolve} there are a binary resolution $N_B$ and a
switching $f_B$ with $T(N_B,f_B)=R$, and $R$ is a binary resolution of $T$.
\end{proof}

\begin{lemma}\label{lem:s1froms2}
Every vertex $u$ of $N$ that satisfies (s2) satisfies (s1).
\end{lemma}

\begin{proof}
Suppose that $u$ satisfies (s2) and that some mixed vertex $y$ lies outside $\mathrm{tp}[u]$. Pick
$a\in L(K_0(y))\cap A$, and let $P$ be a path of tree edges from $y$ to $\ell_a$. Let $z$ be the first
must-in vertex on $P$, which exists because $\ell_a$ is must-in, and let $y'$ be the vertex preceding
$z$ on $P$; as $z$ has indegree one, $y'$ is the parent of $z$. The set $L(K_0(y'))$ contains $a$, and
$y'$ is not must-in, so $y'$ is mixed. Neither $y'$ nor $z$ lies in $\mathrm{tp}[u]$: a vertex of
indegree one lies in $\mathrm{tp}[u]$ only if its parent does, and $y\notin\mathrm{tp}[u]$. Hence
$z\in\mathcal{I}_u$, $z$ has indegree one, and its parent $y'$ lies outside $\mathcal{I}_u\cup\{u\}$,
since $y'$ is not must-in and $y'\neq u$, as $u\in\mathrm{tp}[u]$. This contradicts (s2).
\end{proof}

\begin{proof}[Proof of Lemma~\ref{lem:symreach}]
Each $x\in\mathcal{I}_u$ has a parent in $\mathcal{I}_u\cup\{u\}$: its parent by (s2) if its indegree is
one, and otherwise any $p\in E_u(x)$, which is nonempty by (s3). Walking up such parents from $x$ visits
distinct vertices. By acyclicity it therefore ends, and it can end only at $u$. This gives $x\prec u$ for
every $x\in\mathcal{I}_u$, and then a parent of $u$ in $\mathcal{I}_u$ would close a directed cycle. For
the last claim, since $|A|\ge2$ there is a taxon $b\in A$ with $\ell_b\neq u$. The leaf $\ell_b$ is must-in,
and it does not lie in $\mathrm{tp}[u]$, since a leaf other than $u$ is not an ancestor of $u$; so
$\ell_b\in\mathcal{I}_u$ and $u$ has a strict descendant. If $u$ had a single child $c$, the walk just
described would enter $u$ through $c$, so $c\in\mathcal{I}_u$ and $L(K_0(u))=L(K_0(c))\subseteq A$, making
$u$ must-in. Hence, if $u$ is not must-in, then $u$ has at least two children, and is a tree node.
\end{proof}

\begin{proof}[Proof of Theorem~\ref{thm:symcarve}]
We use the following fact. (F) Let $y$ be a vertex and $a$ a taxon such that $\ell_a$ is reachable from
$y$ along tree edges, and let $P$ be such a path. Every vertex of $P$ other than $y$ has indegree one,
since it is the head of a tree edge. Hence every directed path ending at $\ell_a$ contains $y$ or
starts at a vertex of $P$ below $y$.

We say that a set of taxa is pure if it is contained in $A$ or disjoint from $A$. Let $f$ be a switching.
We first show that $A$ is compatible with $T(N,f)$ if and only if some vertex $u$ satisfies
$A\subseteq L_f(u)$ and $L_f(c)$ is pure for every child $c$ of $u$ in $f(N)$. Suppose that $A$ is
compatible with $T(N,f)$. Fix $a\in A$. The vertices $v$ with $A\subseteq L_f(v)$ are ancestors of
$\ell_a$ in $f(N)$, and $f(N)$ is a tree, so they form a chain. Let $u$ be the deepest of them. No child
$c$ of $u$ in $f(N)$ has $L_f(c)\supseteq A$, so compatibility makes each $L_f(c)$ pure. Conversely,
suppose that $u$ satisfies the two conditions, and let $v$ be any vertex. If $v$ is $u$ or an ancestor of
$u$ in $f(N)$, then $L_f(v)\supseteq A$. If $v$ is a strict descendant of $u$ in $f(N)$, then
$L_f(v)\subseteq L_f(c)$ for the child $c$ of $u$ in $f(N)$ above $v$, so $L_f(v)$ is pure. Otherwise
$L_f(v)$ is disjoint from $L_f(u)$, and so from $A$. Hence $A$ is compatible with $T(N,f)$.

For a vertex $u$, let $\Sigma_u$ be the set of switchings $f$ such that $A\subseteq L_f(u)$ and $L_f(c)$
is pure for every child $c$ of $u$ in $f(N)$. We prove that $\Sigma_u=B_u$ for $u\in U(A)$ and
$\Sigma_u=\varnothing$ otherwise. Note first that for $f\in\Sigma_u$, every strict descendant $v$ of $u$
in $f(N)$ has $L_f(v)$ pure, and $\varnothing\neq L(K_0(v))\subseteq L_f(v)$; so $v$ is must-in if
$L_f(v)\subseteq A$ and must-out if $L_f(v)\cap A=\varnothing$, and $v$ is not mixed. $(\dagger)$

Let $f\in\Sigma_u$. If $y$ is mixed, pick $a\in L(K_0(y))\cap A$. As $a\in A\subseteq L_f(u)$, there is a
path from $u$ to $\ell_a$ in $f(N)$, and by (F) that path contains $y$, or else $u$ lies strictly below
$y$ on a path of tree edges from $y$ to $\ell_a$. In the first case $y$ is a descendant of $u$ in $f(N)$,
which forces $y=u$ by $(\dagger)$; in the second $y\in\mathrm{tp}(u)$. This is (s1). The same argument
applied to $x\in\mathcal{I}_u$ and $a\in L(K_0(x))\subseteq A$ gives that $x$ is a descendant of $u$ in
$f(N)$, the alternative $x\in\mathrm{tp}(u)$ being excluded; and $x\neq u$. That is, $\mathcal{I}_u$
consists of strict descendants of $u$ in $f(N)$. Such an $x$ is not the root, and if its indegree is one,
its parent $p$ lies on the path from $u$ to $x$ in $f(N)$, so $p=u$ or $p$ is a strict descendant of $u$
in $f(N)$. In the latter case $p$ is not mixed by $(\dagger)$, and $L(K_0(p))\supseteq L(K_0(x))$ meets
$A$, so $p$ is must-in, and, as $p\prec u$, $p$ is not in $\mathrm{tp}[u]$. This is (s2). For
$r\in\mathcal{I}_u$, the parent $f(r)$ of $r$ in $f(N)$ is $u$ or a strict descendant of $u$ in $f(N)$.
The argument just given does not apply, since $(f(r),r)$ is a reticulation edge and $L(K_0(f(r)))$ need
not contain $L(K_0(r))$; but $L_f(f(r))\supseteq L_f(r)\supseteq L(K_0(r))$ meets $A$ and is pure by
$(\dagger)$. It follows that $L_f(f(r))\subseteq A$, so $f(r)$ is must-in, and again
$f(r)\notin\mathrm{tp}[u]$. In both cases $f(r)\in\mathcal{I}_u\cup\{u\}$. For $r\in\mathcal{O}$ with
$r\neq u$, a parent $f(r)\in\mathcal{I}_u$ would be a strict descendant of $u$ in $f(N)$, so $L_f(f(r))$
would be pure by $(\dagger)$ and, as it contains the nonempty set $L(K_0(f(r)))\subseteq A$, contained in
$A$; but $L_f(f(r))\supseteq L(K_0(r))$ is nonempty and disjoint from $A$. Thus $f\in B_u$, and (s3)
holds.

Conversely, let $u\in U(A)$ and $f\in B_u$. Each $x\in\mathcal{I}_u$ has exactly one incoming edge in
$f(N)$, and its tail lies in $\mathcal{I}_u\cup\{u\}$: if the indegree of $x$ is one, the tail is the
parent of $x$, which lies there by (s2); otherwise it is $f(x)\in E_u(x)$. Walking up from $x$ along
incoming edges of $f(N)$ therefore stays in $\mathcal{I}_u\cup\{u\}$ and, by acyclicity, ends at $u$.
Hence every $x\in\mathcal{I}_u$ is a strict descendant of $u$ in $f(N)$. As in the proof of
Lemma~\ref{lem:symreach}, some taxon of $A$ has its leaf distinct from $u$, and this leaf lies in
$\mathcal{I}_u$. Hence $u$ is not a leaf, and no taxon of $A$ labels $u$. Every leaf $\ell_a$ with $a\in A$
is thus must-in and outside $\mathrm{tp}[u]$, that is, $\ell_a\in\mathcal{I}_u$, and so
$A\subseteq L_f(u)$. Now let $v$ be a strict descendant of $u$ in $f(N)$. Then $v\notin\mathrm{tp}[u]$;
by (s1), $v$ is not mixed, so $v\in\mathcal{I}_u$ or $v\in\mathcal{O}$. Every child $c$ of $v$ in $f(N)$
lies in the same one of the two sets. If $(v,c)$ is a tree edge, this holds because $L(K_0(c))$ is a
nonempty subset of $L(K_0(v))$ and $c$ again lies outside $\mathrm{tp}[u]$. If $c$ is a reticulation with
$f(c)=v$, then $c$ is a strict descendant of $u$ in $f(N)$, so $c\notin\mathrm{tp}[u]$ and $c\neq u$, and
$c$ is not mixed by (s1); if $c\in\mathcal{I}_u$, then $v\in\mathcal{I}_u\cup\{u\}$ by the first case of
$E_u$, so $v\in\mathcal{I}_u$; if $c\in\mathcal{O}$, then $v\notin\mathcal{I}_u$ by the second case, so
$v\in\mathcal{O}$. Descending $f(N)$ from the children of $u$, we obtain $L_f(v)\subseteq A$ for
$v\in\mathcal{I}_u$ and $L_f(v)\cap A=\varnothing$ for $v\in\mathcal{O}$. Applied to the children of $u$
in $f(N)$, this gives $f\in\Sigma_u$.
\end{proof}

\begin{proof}[Proof of Lemma~\ref{lem:symminclosed}]
By Theorem~\ref{thm:symcarve}, it suffices to show that $f\wedge g\in\mathrm{compat}(A)$ for $f\in B_u$
and $g\in B_{u'}$ with $u,u'\in U(A)$. Put $h=f\wedge g$, which is again a switching. We use
Lemma~\ref{lem:symreach} throughout, in particular $x\prec u$ for every $x\in\mathcal{I}_u$, and
$\mathrm{par}(u)\cap\mathcal{I}_u=\varnothing$.

Suppose first $u\preceq u'$. We show $h\in B_u$. Let $r\in\mathcal{O}$, $r\neq u$; then
$f(r)\notin\mathcal{I}_u$. If $h(r)=g(r)\in\mathcal{I}_u$, then $g(r)\notin\mathcal{I}_{u'}$, by the
box $B_{u'}$ if $r\neq u'$ and by Lemma~\ref{lem:symreach} if $r=u'$. Hence $g(r)$ is must-in and lies in
$\mathrm{tp}[u']$, giving $g(r)\succeq u'\succeq u$, which contradicts $g(r)\prec u$, as
$g(r)\in\mathcal{I}_u$. Let now $r\in\mathcal{I}_u$ and $h(r)=g(r)\neq f(r)$. Then $g(r)$ is the deeper
of the two. First, $r\in\mathcal{I}_{u'}$: otherwise $r$ is must-in and lies in $\mathrm{tp}[u']$, giving
$r\succeq u'\succeq u$, again contradicting $r\prec u$, as $r\in\mathcal{I}_u$. Thus
$g(r)\in\mathcal{I}_{u'}\cup\{u'\}$. If $g(r)=u'$, then either $u'=u$, and we are done, or $u'\succ u$,
and $f(r)\in\mathcal{I}_u\cup\{u\}$ is a strict descendant of $u'$, and so deeper than $g(r)$, a
contradiction. If $g(r)\in\mathcal{I}_{u'}$ but $g(r)\notin\mathcal{I}_u$, then $g(r)$ is must-in and
lies in $\mathrm{tp}[u]$; for $g(r)=u$ this makes $f(r)\in\mathcal{I}_u$ deeper than $g(r)$, and for
$g(r)\in\mathrm{tp}(u)$ it makes $g(r)$ a strict ancestor of $u\succeq f(r)$, both contradicting the
choice of $g(r)$ as the deeper parent. Therefore $g(r)\in\mathcal{I}_u\cup\{u\}$. The remaining
coordinates are unconstrained in $B_u$.

Suppose now that $u$ and $u'$ are incomparable, and let $u$ be the deeper of the two. Neither is must-in:
a must-in $u$ lies outside $\mathrm{tp}[u']$, so $u\in\mathcal{I}_{u'}$ and hence $u\prec u'$, contrary to
incomparability. No vertex of $\mathrm{tp}[u]$ is must-in either: if $x\in\mathrm{tp}[u]$ were must-in,
then $u$, which lies below $x$ along tree edges, would be must-in. Thus
$\mathcal{I}_u=\mathcal{I}_{u'}=\mathcal{I}$, and both boxes read: $\mathrm{par}(r)\cap
(\mathcal{I}\cup\{\text{host}\})$ at a must-in $r$, $\mathrm{par}(r)\setminus\mathcal{I}$ at a must-out
$r$ other than the host, and $\mathrm{par}(r)$ at the host and at a mixed $r$. At $r=u$, the box $B_{u'}$
reads $\mathrm{par}(u)\setminus\mathcal{I}$ or $\mathrm{par}(u)$, and $B_u$ reads $\mathrm{par}(u)$;
these agree, since $\mathrm{par}(u)\cap\mathcal{I}=\mathrm{par}(u)\cap\mathcal{I}_u=\varnothing$ by
Lemma~\ref{lem:symreach}. The same holds at $r=u'$. All elements of $\mathcal{I}\cup\{u\}$ are deeper
than $u'$, because $x\prec u$ for every $x\in\mathcal{I}=\mathcal{I}_u$, and $u$ is deeper than $u'$.
Hence at a must-in $r$ we have $h(r)\neq u'$, and $h(r)=u$ only when $g(r)=u'$; in each case
$h(r)\in\mathcal{I}\cup\{u\}$. At a must-out $r\neq u$ we have $f(r)\notin\mathcal{I}$, and
$g(r)\notin\mathcal{I}$ by the box if $r\neq u'$ and by Lemma~\ref{lem:symreach} if $r=u'$. Therefore
$h\in B_u$.
\end{proof}

\begin{proof}[Proof of the running-time bound of Theorem~\ref{thm:symcase}]
A topological order of $V(N)$, the tree components, and the intervals $J(v)$ of
Lemma~\ref{lem:preproc}(T1) are computed in $O(|N|)$ time; then whether $v\in\mathrm{tp}[u]$, that is,
whether $u\in K_0(v)$, is tested in $O(1)$ time. Fix a cluster $A$. One bottom-up pass over the forest
of tree edges gives $|L(K_0(v))\cap A|$ and $|L(K_0(v))|$ for all $v$, and so the status of every vertex,
in $O(|N|)$ time. Condition (s1) needs one preliminary test. If a mixed vertex $x$ has indegree one, then
its parent $p$ is mixed, since $L(K_0(p))\supseteq L(K_0(x))$ meets $A$ and is not contained in $A$.
Hence, if $v_A$ is the deepest mixed vertex, then every vertex of $\mathrm{tp}[v_A]$ is mixed, and the
mixed vertices lie in $\mathrm{tp}[u]$ for some $u$ if and only if they are exactly the vertices of
$\mathrm{tp}[v_A]$, that is, if and only if their number is $|\mathrm{tp}[v_A]|$. This is one comparison,
and if it fails, $A$ has no host. Otherwise (s1) holds at $u$ if and only if $v_A\in\mathrm{tp}[u]$, which
is one ancestor test. Condition (s2) fails at a vertex $u$ that satisfies (s1) if and only if the root of
$N$ is must-in and $u$ lies outside the tree component of the root, or some $p\in\mathrm{tp}(u)$ has a
must-in child, joined to $p$ by a tree edge, other than the next vertex of $\mathrm{tp}[u]$: for if $x$
violates (s2) and has a parent, then its parent lies in $\mathrm{tp}(u)$, either because that parent is
mixed, and then (s1) places it in $\mathrm{tp}[u]$, or because it is must-in and outside $\mathcal{I}_u$.
Counting, for each vertex $p$, the must-in children joined to $p$ by tree edges, and sweeping each tree
component from its root, thus settles (s2) at every $u$ in $O(|N|)$ time. Condition (s3) and the sets
$E_u(r)$ cost $O(d_{\Sigma})$ time per vertex $u$. Since the passes above are needed whatever the value of
$d_{\Sigma}$, one cluster is processed in $O(|N|\,(d_{\Sigma}{+}1))$ time, and all clusters in
$O(n\,|N|\,(d_{\Sigma}{+}1))$ time.

The constraint $\mathrm{compat}(A)$ is stored as the boxes $B_u$, $u\in U(A)$, each given by its sets
$E_u(r)$, so the constraints have total size $O(n\,|N|\,d_{\Sigma})$. Generalized arc consistency is
enforced in time linear in that size by an adaptation of the counter scheme of Mohr and
Henderson~\cite{mohrhenderson1986}. By the characterization of supports in Section~\ref{sec:sym}, it
suffices to keep, for each box $B_u$ and each reticulation $r$, the number of values of $E_u(r)$ still in
the domain of $r$, and, for each triple $(A,r,p)$, the number of boxes $B_u$ of $\mathrm{compat}(A)$ with
$p\in E_u(r)$ whose sets $E_u(r')$, $r'\neq r$, all meet the current domains. A box is discarded once,
when one of its counts reaches zero, and a value is deleted once, when its count reaches zero. The final
test computes $T(N,f)$ in $O(|N|)$ time and tests each of the $O(n)$ clusters of $T$ for compatibility
with $T(N,f)$ in $O(n)$ time, by one bottom-up pass over $T(N,f)$ comparing $|L_{T(N,f)}(w)\cap A|$ with
$|L_{T(N,f)}(w)|$. This is $O(n^{2})$ time in all, absorbed by the term $n\,|N|$ since $|N|\ge n$.
Adding the $O(|N|+|T|)$ time to read the input gives the bound.
\end{proof}


\begin{thebibliography}{99}

\bibitem{kanj2008}
I.~A. Kanj, L.~Nakhleh, C.~Than, G.~Xia.
Seeing the trees and their branches in the network is hard.
\emph{Theoretical Computer Science} 401(1--3):153--164, 2008.

\bibitem{bmw2018}
M.~Bentert, J.~Mal\'{\i}k, M.~Weller.
Tree Containment With Soft Polytomies.
In \emph{Proc.\ 16th Scandinavian Symposium and Workshops on Algorithm Theory (SWAT 2018)},
LIPIcs vol.~101, pp.~9:1--9:14, 2018.

\bibitem{bmw2021}
M.~Bentert, M.~Weller.
Tree Containment With Soft Polytomies.
\emph{Journal of Graph Algorithms and Applications} 25(1):417--436, 2021.

\bibitem{huson2010}
D.~H. Huson, R.~Rupp, C.~Scornavacca.
\emph{Phylogenetic Networks: Concepts, Algorithms and Applications}.
Cambridge University Press, 2010.

\bibitem{crv2009}
G.~Cardona, F.~Rossell\'{o}, G.~Valiente.
Comparison of tree-child phylogenetic networks.
\emph{IEEE/ACM Transactions on Computational Biology and Bioinformatics} 6(4):552--569, 2009.

\bibitem{willson2010}
S.~J. Willson.
Properties of normal phylogenetic networks.
\emph{Bulletin of Mathematical Biology} 72(2):340--358, 2010.

\bibitem{semplesteel2003}
C.~Semple, M.~Steel.
\emph{Phylogenetics}.
Oxford Lecture Series in Mathematics and its Applications, vol.~24. Oxford University Press, 2003.

\bibitem{viss2010}
L.~van Iersel, C.~Semple, M.~Steel.
Locating a tree in a phylogenetic network.
\emph{Information Processing Letters} 110(23):1037--1043, 2010.

\bibitem{gunawan2018}
A.~D.~M. Gunawan.
Solving the Tree Containment Problem for Reticulation-Visible Networks in Linear Time.
In \emph{Algorithms for Computational Biology (AlCoB 2018)}, LNCS vol.~10849, pp.~24--36, 2018.

\bibitem{weller2018}
M.~Weller.
Linear-Time Tree Containment in Phylogenetic Networks.
In \emph{Comparative Genomics (RECOMB-CG 2018)}, LNCS vol.~11183, pp.~309--323, 2018.

\bibitem{jm2020}
R.~Janssen, Y.~Murakami.
Linear time algorithm for tree-child network containment.
In \emph{Algorithms for Computational Biology (AlCoB 2020)}, LNBI vol.~12099, pp.~93--107, 2020.

\bibitem{jm2021}
R.~Janssen, Y.~Murakami.
On Cherry-Picking and Network Containment.
\emph{Theoretical Computer Science} 856:121--150, 2021.

\bibitem{bruchholdweller}
S.~Bruchhold, M.~Weller.
Exploiting Low Scanwidth to Resolve Soft Polytomies.
In \emph{Proc.\ SOFSEM 2026}, LNCS vol.~16448, pp.~332--346, 2026. Full version: arXiv:2511.20771v3;
results are cited here by the numbering of that version.

\bibitem{vijwscanwidth}
L.~van Iersel, M.~Jones, M.~Weller.
Tree Containment Parameterized by Scanwidth.
arXiv:2605.31071, 2026.

\bibitem{vijwtreewidth}
L.~van Iersel, M.~Jones, M.~Weller.
Embedding phylogenetic trees in networks of low treewidth.
\emph{Discrete Mathematics \& Theoretical Computer Science} 25(2):\#4, 2023.

\bibitem{gunawanLuZhang2016}
A.~D.~M. Gunawan, B.~Lu, L.~Zhang.
A program for verification of phylogenetic network models.
\emph{Bioinformatics} 32(17):i503--i510, 2016.

\bibitem{chan2013}
J.~M. Chan, G.~Carlsson, R.~Rabad\'{a}n.
Topology of viral evolution.
\emph{Proceedings of the National Academy of Sciences} 110(46):18566--18571, 2013.

\bibitem{treangen2011}
T.~J. Treangen, E.~P.~C. Rocha.
Horizontal transfer, not duplication, drives the expansion of protein families in prokaryotes.
\emph{PLoS Genetics} 7(1):e1001284, 2011.

\bibitem{zhang2023}
L.~Zhang, N.~Abhari, C.~Colijn, Y.~Wu.
A fast and scalable method for inferring phylogenetic networks from trees by aligning lineage taxon
strings.
\emph{Genome Research} 33:1053--1060, 2023.
Preprint version: arXiv:2301.00992. The definitions and lemmas of the Supplemental Methods are cited here
by the numbering of that version.

\bibitem{phylofusion}
L.~Zhang, B.~Cetinkaya, D.~H. Huson.
PhyloFusion---Fast and Easy Fusion of Rooted Phylogenetic Trees into Rooted Phylogenetic Networks.
\emph{Systematic Biology} 75(1):88--99, 2026.

\bibitem{jeavons1998}
P.~Jeavons, D.~Cohen, M.~C. Cooper.
Constraints, consistency and closure.
\emph{Artificial Intelligence} 101(1--2):251--265, 1998.

\bibitem{schaefer1978}
T.~J. Schaefer.
The complexity of satisfiability problems.
In \emph{Proc.\ 10th ACM Symposium on Theory of Computing (STOC 1978)}, pp.~216--226, 1978.

\bibitem{jeavonscooper1995}
P.~G. Jeavons, M.~C. Cooper.
Tractable constraints on ordered domains.
\emph{Artificial Intelligence} 79(2):327--339, 1995.

\bibitem{dalmau1999}
V.~Dalmau, J.~Pearson.
Closure functions and width 1 problems.
In \emph{Principles and Practice of Constraint Programming (CP~1999)}, LNCS vol.~1713, pp.~159--173.
Springer, 1999.

\bibitem{apt1979}
B.~Aspvall, M.~F. Plass, R.~E. Tarjan.
A linear-time algorithm for testing the truth of certain quantified boolean formulas.
\emph{Information Processing Letters} 8(3):121--123, 1979.

\bibitem{mohrhenderson1986}
R.~Mohr, T.~C. Henderson.
Arc and path consistency revisited.
\emph{Artificial Intelligence} 28(2):225--233, 1986.

\bibitem{companion}
E.~Ugurlu, L.~Zhang.
Soft Tree Containment with Missing Taxa on Tree-Child Networks.
Manuscript in preparation, 2026.

\end{thebibliography}
\end{document}